\documentclass[aps,prb,amssymb,twocolumn]{revtex4}

\usepackage{graphicx}
\usepackage{docs}
\def\be{\begin{equation}}
\def\ee{\end{equation}}
\def\ba#1{\begin{array}{#1}}
\def\ea{\end{array}}
\def\bn{\begin{enumerate}}
\def\en{\end{enumerate}}
\def\r{\right}
\def\l{\left}
\def\H{{\cal H}}

\begin{document}
\title{Dissipation and quantum phase transitions of a pair of Josephson junctions}
\author{{\sc Gil Refael${}^1$, Eugene Demler${}^1$, Yuval Oreg${}^2$, Daniel S. Fisher${}^1$}\\ 
{\small ${}^1$\em Dept.\ of Physics, Harvard University, Cambridge MA, 02138}\\
{\small ${}^2$\em Department of Condensed Matter Physics, Weizmann Institute of Science, Rehovot, 76100, ISRAEL }}

%--------------------------------paper

\date{\today}

\begin{abstract}
 A model system consisting of
a mesoscopic superconducting grain coupled by Josephson junctions to two
macroscopic superconducting electrodes is studied. We focus on the effects of ohmic dissipation caused by resistive shunts and superconducting-normal charge relaxation within the grain.
As the temperature is lowered, the behavior crosses over from uncoupled
Josephson junctions, similar to situations analyzed previously,  to strongly interacting
junctions. The crossover temperature is related to the energy-level spacing of the grain and is of the order of the inverse escape time from the grain.  In the limit of zero temperature, the two-junction system exhibits five
distinct quantum phases, including a novel superconducting state with
localized Cooper pairs on the grain but phase coherence between the
leads due to Cooper pair cotunneling processes. In contrast to a single junction, the transition from the fully superconducting to fully normal phases is found to be controlled by an intermediate coupling fixed point whose critical exponents vary continuously as the resistances are changed. The model is analyzed via   two component sine-Gordon models and related Coulomb gases that
provide effective low temperature descriptions  in
both the weak and the strong Josephson coupling limits. The complicated phase diagram is
consistent with symmetries of the two component sine-Gordon models,
which include weak to strong coupling duality and permutation
triality. Experimental
consequences of the results and potential implications for
superconductor to normal transitions in thin wires and films are discussed briefly.
\end{abstract}
\pacs{PACS numbers:}

\maketitle

%------------------------------------------------------------introduction---------------------------------------------------------
\section{Introduction \label{int}}

Understanding the effects of dissipation on quantum phase transitions
has proved to be a challenging problem in many contexts including
quantum Hall transitions, \cite{Auerbach} and quantum critical points
in antiferromagnets. \cite{Sachdev} Transitions from superconductor to ``normal" metal or insulator in 
thin wires and films have been extensively studied, \cite{Goldman1,dynes,Gantmakher1,Kapitulnik1,mason}
as well as in Josephson
junction arrays \cite{fazio,haviland,rimberg,japan,haviland2,haviland3} and superconducting nanowires.
\cite{bezryadin,tinkham,lau,giordano,dynes0} One of the most intriguing
aspects of these transitions is the role of dissipation. \cite{mason2,wagenblast,voelker,larkin,phillips,spivak}
%Such destruction of superconductivity by quantum fluctuations is the %subject of this psper.
Theoretically, there has been extensive work on the effects of dissipation on a single resistively 
shunted Josephson junction (RSJJ). The resistor can be modeled
theoretically as a Caldeira-Leggett ohmic heat bath, 
\cite{c-leggett,leggett2,chakravarty,schmid,bulga,korshunov0,simanek,weiss}
and precise predictions
for the transport properties can be worked out (see [\onlinecite{schoen-zaikin}]
for a review). The system undergoes a  superconductor--to--normal transition at zero temperature 
when the shunt resistance increases through a critical value equal to the quantum of resistance 
$R_Q=h/4e^2=6.53\,k\Omega$. Recent experiments by Penttil\"a {\it et.al.}
\cite{paalanen} showed good agreement with the theoretical analysis.

\begin{figure}
\includegraphics[width=8.5cm]{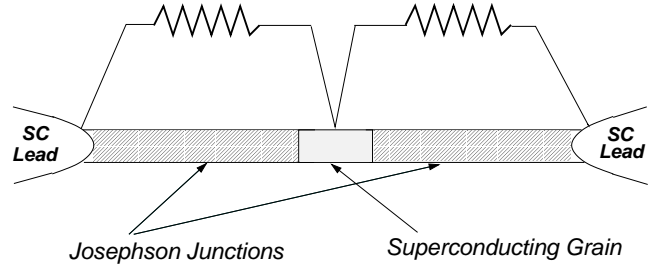}
\caption{A mesoscopic superconducting grain connected to 
superconducting leads via Josephson junctions and resistive shunts.}
  \label{Figure1}
\end{figure}

Arrays of RSJJ have been studied in the same framework in terms of the {\it local} 
physics of the individual junctions.  \cite{schoen-zaikin,pz,fisher,chakravarty2,bobbert,korshunov,schoen} 
   By percolation arguments, this local physics has been argued to apply to granular 
films and wires with the superconductor--to--normal transition in these extended systems 
occuring when the {\it individual} shunting resistances along a critical percolation path 
become equal to $R_Q$.\cite{fisher} 
   
    The prediction for destruction of superconductivity via this local mechanism is 
in striking contrast to what one would expect in the absence of dissipation: domination 
near to the quantum phase transition by  {\it collective} long-wavelength quantum fluctuations 
rather than local physics. In addition to the nature of the transition, {\it where} it 
would be expected to occur as parameters of the system are varied is strikingly 
different for the two pictures.  The long-wavelength quantum fluctuations should 
be controlled by the interplay between the Josephson couplings among grains and 
the Coulomb interactions, the former acting  to decrease the phase fluctuations 
and the latter to decrease the charge fluctuations.  If long-wavelength physics 
dominates, the location of the transition would thus be expected to depend markedly 
on the strength of the Josephson couplings. In contrast,  for a single junction, and 
by naive extension for a network of junctions, the location of the dissipation induced 
transition would be entirely determined by the shunting resistances, independent of the 
Josephson couplings. 
    
    The primary purpose of this paper is to begin to reconcile these two approaches by 
studying a deceptively simple system: 
two resistively shunted Josephson junctions coupled in series through a superconducting grain. 
This system, in addition to its intrinsic
interest, \cite{2JJ} provides a simple paradigm for the competing effects of dissipation and 
quantum fluctuations on superconductivity.

An important simplification in all previous theoretical
studies of JJ arrays is the assumption that 
the superconducting grains are sufficiently large that they can effectively be treated as 
macroscopic. In the case of several junctions
in series, such
an assumption leads to a result that the superconductor-to-normal transition
occurs on each junction separately and takes place
when the values of the individual shunting resistances are equal to the
quantum of resistance $R_Q=h/(2e)^2$.  
In this paper we take into account the effects of finite size 
grains, specifically by considering two bulk superconducting leads connected by a pair of 
Josephson junctions in series 
through a {\it mesoscopic} grain.  We show that the quantum dynamics of the two junctions, 
which are independent over a wide range of temperatures, become strongly coupled below a  
characteristic 
crossover temperature. In the low temperature regime, this simple  system exhibits surprisingly 
rich behavior, including two distinct superconducting
phases. In some regimes of parameter space, the superconductor-to-normal transition between the 
two macroscopic leads is
determined by the {\it total} shunting resistance of the system,
rather than individual resistances of the junctions; while in other regimes its location depends 
on the strengths of the 
 Josephson couplings as well as the shunting resistances. In this latter case, the corresponding critical behavior becomes very different from the 
single junction case.  

%Our analysis is based on a
%effective two fluid model in which we introduce a phenomenolgical
%parameter to describe charge relaxation between the normal and the
%super conducting fluids on the grain.

\begin{figure}
\includegraphics[width=8.5cm]{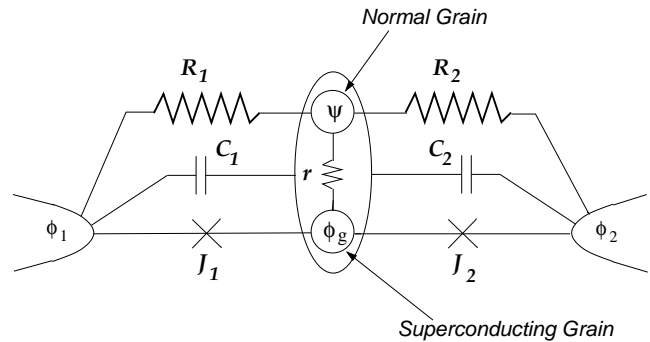}
\caption{Effective circuit consisting of two Josephson-junctions ($J_1,\,J_2$)
connecting the macroscopic electrodes ($\phi_1,\,\phi_2$) to a 
mesoscopic grain. The
grain is modeled in a two-fluid manner, as a superconducting grain
($\phi_g$) connected through a phenomenological resistance $r$ to a
normal-fluid grain ($\psi$). $R_1,\,R_2$ are the shunt resistors
connecting the normal-fluid of the
grain to the superconducting contacts, in which the normal-superconducting relaxation is fast. }  \label{Figure2}
\end{figure}

The basic system is shown in Fig. \ref{Figure1}.
Dissipation occurs in ohmic shunts
between the superconducting contacts and the grain. Such systems may
be understood in terms of  a two-fluid model in which Cooper
pairs tunneling across Josephson junctions represent the superfluid, and electrons flowing through the 
shunt resistors represent the normal fluid. \cite{sz, AES, bruder} The
presence of two fluids in the middle grain suggests considering it as a double grain with a 
superconducting part and a normal part as shown in
Fig. \ref{Figure2}.  We  assume for simplicity that the normal and superconducting 
charges of the two parts experience the same electrostatic
potential as they overlap in space. The chemical potentials of the two parts, however, do {\it not} have
to be the same. When these differ, the resulting electrochemical potential difference
can cause charge relaxation within the grain that will act to equilibrate its normal and superfluid components.  
In this paper we assume a simple ohmic model of this
relaxation with the conversion current
\be 
I_{ns} = \frac{V_n - V_s}{r} 
\ee 
where $V_n$ and $V_s$ are the electrochemical
potentials of the normal and the superconducting fluids on the grain.  The
coefficient $r$ is a phenomenological parameter of our model that we
will call the conversion resistance.  Decoupling of the two chemical
potentials is similar to the nonequlibrium state of the
superconducting and normal fluids, as discussed for phase slip centers
at finite current. \cite{beasley,tinkham2,ivlev}  We assume that the
two leads are macroscopic, so that there is perfect coupling
between the superconducting and normal fluids in each of them (this corresponds to the conversion resistances in the leads being negligible).

The model we arrive at using the arguments above is quite general. 
One could also obtain it by considering the electromagnetic modes that Cooper-pair 
tunneling events excite as discussed in Appendix \ref{altdiss}. 
This alternative approach does not require a two-fluid picture. 

It is worth pointing out that our system bears some resemblance to the
Cooper pair box systems studied recently in the context of quantum
computing and mesoscopic qubits.\cite{nakamura,cottet,lehnert} The
charge on the grain could be used as a the quantum number of a qubit.
The biggest obstacle to quantum computation is then the limited
lifetime of the quantum state of the qubit.  Quantum fluctuations and
interactions with the environment limit the life time of such a
state, so practical realizations of qubits require
systems with low dissipation. In this paper, in contrast, we study the Cooper
pair box system in a highly dissipative environment.

This paper is organized as follows. In Section II we present a
microscopic Hamiltonian  and derive the quantum
action. To ascertain the consistency of this derivation we demonstrate
in Appendix A that the classical equations of motion obtained from
the action correspond to the electrodynamics of the circuit
in Fig. \ref{Figure2}.  From the analysis of the quantum model we show
the existence of a new temperature scale $T^*$ set by the level
spacing in the grain.  At temperatures higher than $T^*$ the two
junctions are decoupled and can be considered separately. If the grain
is macroscopic, $T^* \rightarrow 0$ and the system is always in the decoupled
regime. This is the case considered in the literature thus far.
\cite{fisher,chakravarty2,bobbert,korshunov,schoen} 
For temperatures below $T^*$ one
cannot neglect interactions between the junctions, and the effective
low temperature description is given by two coupled quantum
sine-Gordon models. 

In section III we use renormalization group (RG) methods
to analyze the two component sine-Gordon theory in the limit of weak
Josephson coupling and obtain its phase diagram.  We show that the
system can have five distinct phases: fully supercondicting, FSC, where both junctions are
superconducting; normal, NOR, where both junctions are normal and there is
no phase coherence between the leads; N1-S2, where junction one is
normal and junction two is superconducting; S1-N2, where junction
one is superconducting and junction two is normal; and  $SC^{\star}$, in which
Cooper pairs are localized on the grain, so individual junctions are
insulating, but there is superconducting coherence between the leads due to
cotunneling processes.  We provide simple arguments for
the phase boundaries based on electrical circuit considerations of
the effective shunting resistances for various Cooper pair tunneling
events. 

In Section IV we analyze the system in the opposite regime of
strong Josephson couplings using a dual two component sine-Gordon
model and considerations of quantum phase slips.   The RG analysis is again supplemented by 
effective shunting resistance arguments which  determine the action of the various quantum phase slip
processes.  It is found that the phase diagrams obtained in the weak and strong coupling limits
differ in the location of the NOR to FSC phase boundary.  

In Section V we show that the difference between strong and weak coupling
phase diagrams signals the existence of a novel regime with the fully normal to fully superconducting transition controlled by
a critical fixed point at intermediate Josephson coupling. We analyze the appropriate fixed
point, whose properties depend continuously on the resistances, and discuss the RG flow in its vicinity.

In Section VI we
explore the surprisingly rich symmetries of the two-junction system. In addition to a weak-to-strong duality, the system also exhibits a 
permutation triality that implies that  aspects of the phase diagram are invariant under interchange of any of the three resistances involved in the dissipative transport.

In Section VII we review some experimental implications of our work and discuss
such questions as observation of the crossover temperature scale
$T^*$, the experimental identification of the novel superconducting phase
$SC^{\star}$, and the universality of the resistance at the superconductor to normal
transition.  We also suggest that our results may be relevant for
understanding some puzzling experimental results on superconductor to
normal transitions in thin wires and films.  

Finally, in Section VIII
we summarize the main results.  To maintain the coherence of the presentation we delegate most of the
technical calculations to appendices. In particular, the renormalization group analysis of the two component
sine-Gordon model and the relations to classical
Coulomb gasses  are given in Appendices D (weak coupling), and E
(strong coupling).

\section{Microscopic Model}

\subsection{Hamiltonian of the two-junction system}

\label{TwoJSHamiltonian}

The system we wish to describe consists of a mesoscopic
superconducting grain situated between two macroscopic superconducting
leads (Fig. \ref{Figure2}).  The grain interacts with the leads
both electrostatically and through a weak link. The electostaticl interaction is capacitative while the weak link allows the flow of both Cooper-pairs and normal electrons. Cooper
pairs flow through a Josephson junction from the superconducting part
of the grain to the leads. Normal electrons flow from the normal part of the grain
to the leads
through
what we model as a shunt resistor.

In order to understand the quantum dynamics of this system  
we must first obtain an appropriate low-energy
effective Hamiltonian.  This should include the charging energy for
the grain and leads, the Josephson coupling energies for the
junctions, and appropriate Hamiltonians for the shunt resistors which
can be approximated by heat baths. \cite{c-leggett}

The charging energy of the system includes both electrostatic and
electrochemical capacitances. All the islands (here we use the term
island to denote either the electrodes or the grain) have part of
their charge, $Q_{Si}$, in the form of superconducting Cooper pairs
and part of their charge, $Q_{Ni}$, in the form of normal fluid. Both
kinds of charge contribute to the electrostatic potential and have
their own compressibility. The electrochemical potentials for the
superconducting and normal electrons on island $i$ are
\be
\ba{c}
V_{Si}=\varphi_i+ D_{Si} Q_{Si}\vspace{2mm}\\
V_{Ni}=\varphi_i+ D_{Ni} Q_{Ni}.
\label{potentials} \ea 
\ee
The index $i$ is summed over electrodes $1$, $2$, and the grain $g$;
$\varphi_i$ is electric potential; $D_{i}$'s are the inverse of
the compressibilities of the fluids $S$ and $N$, in a non-interacting approximation,  $e^2 D_{Ni}$ is the level spacings of the normal electrons in
the island $i$. \cite{buttiker93}  The electrostatic potential on 
island $i$ is related to the charges on all the islands via the
capacitance matrix $C_{ij}$:
\be
\varphi_i = \sum_j C^{-1}_{ij} (Q_{Sj}+Q_{Nj}).
\ee
Hence, for the electrochemical potentials we have
\be
\ba{rcl}
V_{Si} &=&  \sum_j ( \kappa^{-1}_{Sij} Q_{Sj} + C^{-1}_{ij} Q_{Nj} )\vspace{2mm}\\
V_{Ni} &=& \sum_j ( C^{-1}_{ij} Q_{Sj} + \kappa^{-1}_{Nij} Q_{Nj} ),
\label{pot2}
\ea
\ee
where we defined
\be
\ba{rcl}
\kappa^{-1}_{Sij} &=& C^{-1}_{ij} +D_{Si} \delta_{ij},\vspace{2mm}\\
\kappa^{-1}_{Nij} &=& C^{-1}_{ij} +D_{Ni} \delta_{ij}
\label{NetworkKappa}
\ea
\ee
with $\delta_{ij}$ a Kronecker delta. By integrating out the electro-chemical potentials in (\ref{pot2})
we find the charging part of the Hamiltonian
\be
\ba{c}
{\cal H}_Q = \frac{1}{2} \sum_{ij} \kappa^{-1}_{Sij} Q_{Si} Q_{Sj}
+ \frac{1}{2} \sum_{ij} \kappa^{-1}_{Nij} Q_{Ni} Q_{Nj}\vspace{2mm}\\
+ \sum_{ij} C^{-1}_{ij} Q_{Si} Q_{Nj}.
\label{ChargingHamiltonian}
\ea
\ee

At this point we  introduce superconducting phases, $\phi_{i}$, on the islands
and ``normal phases" $\psi_{i}$ that we define formally to
be conjugate to $Q_{Ni}$: \cite{schoen-zaikin}
\be
\ba{rcl}
\l[ Q_{Ni},\psi_j \r]=-ie \delta_{ij}	&&	\l[Q_{Si},\phi_i \r]=
-2ie \delta_{ij} \vspace{2mm}\\
\l[Q_{Ni},\phi_i \r]=0	&&	\l[Q_{Si},\psi_j \r]=0.
\label{QPhiCommutation_relations}
\ea
\ee
By using (\ref{ChargingHamiltonian}) and
(\ref{QPhiCommutation_relations}), it is easy to verify that the Heisenberg equations of motion for the
two phases give the correct Josephson relations:
\be
\ba{rcl}
\frac{\hbar}{2e} \frac{d \phi_i}{d t} &=&
\frac{i}{2e}\l[\H_Q,\phi_i \r]= V_{Si}
\vspace{2mm}\\
\frac{\hbar}{e} \frac{d \psi_i}{d t} &=&
\frac{i}{e}\l[\H_Q,\psi_i \r]= V_{Ni}.
\label{NetworkJosphson_relations}
\ea
\ee

The other important energies involving the superconducting degrees of freedom are the Cooper pair tunnelings, with
\be
{\cal H}_J=
-J_1 \cos(\phi_g-\phi_1)- J_2 \cos(\phi_2-\phi_g).
\ee
The dissipation in the ohmic shunts, $R_1$, $R_2$, and the internal
charge relaxation, $r$, are modeled following Caldeira and Leggett (see
refs.  \cite{leggett2,schoen-zaikin,weiss} for a review).  In this
approach, the shunting resistances are replaced by collections of harmonic
oscillators (heat baths), with appropriately chosen spectral functions:
\be
\ba{cc} {\cal H}_{dis} = {\cal H}_{bath}(R_1,2\psi_1-2\psi_g) +
{\cal H}_{bath}(R_2, 2\psi_2-2\psi_g)\vspace{2mm}\\ +{\cal H}_{bath}(r,
\phi_g- 2\psi_g).  \ea 
\ee 
We will not give the explicit form of the appropriate Hamiltonians
here, but in the next subsection we give the effective actions
obtained after integrating out the heat-bath degrees of freedom. The
heat bath model is the simplest quantum model that gives the correct
classical equations of motion for systems with dissipation. Later in
this paper we will discuss some of its drawbacks, however, we believe
that it gives a qualitatively correct picture for a
general mechanism of dissipation.

Collecting all the terms, we obtain an effective Hamiltonian
that describes the system shown in Fig. \ref{Figure2}:
\be
{\cal H}(Q_{Ni}, Q_{Si}, \phi_{i}, \psi_{i}) 
= {\cal H}_Q + {\cal H}_J + {\cal H}_{dis}.
\label{Hamiltonian}
\ee
\vspace{3mm}

\subsection{Imaginary time action \label{ita}\label{ActionSection}}

From the Hamiltonian (\ref{Hamiltonian}) and commutation relations
(\ref{QPhiCommutation_relations}), we can construct the imaginary time 
action and partition function for the system 
in Fig. \ref{Figure2}: 
\be
\ba{rcl}
Z&=& \int {\cal D} Q_{Ni} {\cal D} Q_{Si} {\cal D} \phi_i 
{\cal D} \psi_i \exp \left( -S \right)
\vspace{2mm}\\
S&=&
-\frac{i}{2e} \sum_i \int_0^\beta d\tau\, Q_{Si}\, \dot{\phi}_i 
- \frac{i}{e} \sum_i \int_0^\beta d\tau\, Q_{Ni}\, \dot{\psi}_i \vspace{2mm}\\&&
+ \int_0^\beta d\tau {\cal H}(Q_{Ni}, Q_{Si}, \phi_{i}, \psi_{i}). 
\label{ZwithQ} 
\ea
\ee 
It is important to point out that in the presence of ohmic
dissipation the phase variables $\phi_i$ and $\psi_i$ should be
periodic at $\tau=0$ and $\tau=\beta$ with {\it no phase twists} by
multiples of $2\pi$ allowed. This follows from the fact that a $2\pi$
phase twist causes dissipation and is thus measurable. The ohmic
dissipation allows continuous charge transfer (as opposed to transfer
of multiples of $e$) from the shunting resistors to the
grain. Therefore any non-integer charge induced by the gate voltage
can be screened out. (For a more detailed discussion see
\cite{likharev,schoen-zaikin}). This potential drawback of the Caldeira-Leggett
model of dissipation may be overcome if one introduces a more
complicated form of dissipation, such as via quasiparticle
tunneling (see e.g. \cite{schoen-zaikin}). 

The quantum action in (\ref{ZwithQ}) is quadratic in $Q_{Si}$ and
$Q_{Ni}$, so they may be integrated out (for details, see
Appendix \ref{AppendixA1}). The electrochemical contribution is (in terms of the electrochemical potentials)
\be
\ba{c}
S_Q = \int_0^\beta d\tau\; \frac{1}{2(2e)^2}\;
\left( \sum_{i} 
C_{Qi} ( V_{Si} - V_{Ni} )^2
\r.\vspace{2mm}\\

\l.+  \sum_{ij} (s_i V_{Si} 
+\eta_i V_{Ni})\; C_{ij}\; (s_j V_{Sj} 
+\eta_j V_{Nj})  \right). 
\label{SQ1} 
\ea
\ee
This is very easy to interpret. The level spacings give rise to the first term in the brackets
making a potential difference between the two fluids on one island energetically costly. 
The second term in the brackets is the charging energy one would expect from a conventional system of islands, but 
the potential on each island is replaced by a weighted average of the normal-fluid potential and the superfluid potential:
$
\overline{V}_i=s_i V_{Si} +\eta_i V_{Ni}.
$

In terms of the phase variables, the full action can be written as
\begin{widetext}
\be
\ba{rcl}
Z &=& \int 
{\cal D} \phi_i {\cal D} \psi_i 
\exp \left( -  S_Q - S_J - S_{dis} \right)
\label{Z1}
\vspace{2mm}\\
S_Q &=& \int_0^\beta d\tau\; \frac{1}{2(2e)^2}\;
\left( \sum_{i} 
C_{Qi} ( \dot{\phi}_i - 2 \dot{\psi}_i )^2
+  \sum_{ij} (s_i \dot{\phi}_i 
+\eta_i 2 \dot{\psi}_i)\; C_{ij}\; (s_j \dot{\phi}_j 
+\eta_j 2 \dot{\psi}_j)  \right) 
\vspace{2mm}\\
S_J&=&\int_0^\beta d \tau (-J_1 \cos(\phi_g-\phi_1)- 
J_2 \cos(\phi_2-\phi_g))\vspace{2mm}
\label{SJ1} 
\vspace{2mm}\\
S_{dis}&=&\beta \sum_{\omega_n}\; \frac{R_Q}{4\pi}\;
\l( \frac{|\omega_n|}{R_1}|2 \psi_{1,\,(\omega_n)}-2 \psi_{g,\,(\omega_n)}|^2+
\frac{|\omega_n|}{R_2}|2 \psi_{2,\,(\omega_n)}-2 \psi_{g,\,(\omega_n)}|^2 +
\frac{|\omega_n|}{r}|\phi_{g,\,(\omega_n)} - 2 \psi_{g,\,(\omega_n)} |^2
\r)
\label{Sdis1}
\ea
\ee
\end{widetext}
where the Matsubara frequencies are $\omega_n = 2 \pi T n$, and we have defined  
\be
C_{Qi} = (D_{Si}+D_{Ni})^{-1}
\ee
\be
s_i=D_{Ni}/(D_{Si}+D_{Ni})
\ee
 and 
 \be
 \eta_i=D_{Si}/(D_{Si}+D_{Ni}) \ . 
\ee
An important consequence of the domain of the phase fields $\phi_g$
and $\psi$ being the real line rather than a circle, is that the Berry 
phase has no effect on the behavior of the system. 
A  Berry phase could arise if we included
the gate voltage effects in (\ref{potentials}) and
(\ref{ChargingHamiltonian}) by shifting $Q_{Sg} \rightarrow
Q_{Sg}-Q_0$, which would lead to additional terms in the action
(\ref{SQ1}) of the form $i Q_0 \int_0^\beta \dot{\phi}_g$. But
because $2\pi$ phase twists are not allowed, the additional action
vanishes due to the periodic boundary conditions in imaginary time.

As a consistency check on the action (\ref{Z1}), we demonstrate
in Appendix \ref{AppendixA2} 
that its real time equivalent gives rise to equations of
motion that coincide exactly with the basic electrodynamic equations
for the circuit in Figure \ref{Figure2}.

In this paper we consider the limit of macroscopic electrodes, so we can set the corresponding $D_1=D_2=0$ on these.  The first term
in (\ref{SQ1}) then imposes perfect coupling between the
superconducting and normal fluids in the electrodes, i.e., $\phi_1=2\psi_1$ and
$\phi_2=2\psi_2$. Note that this assumption does not restrict us to
taking an infinite capacitance for the electrodes: the inverse of the
level spacing grows as the volume of the grains, whereas capacitances
increase only linearly with the dimensions. We restrict our
discussion to the case when the largest capacitances in the system are the
{\it mutual capacitances} between the electrodes and the grain, $C_1$ and
$C_2$, for electrodes one and two respectively. In Appendix \ref{AppendixA3} we show
that in this case the charging part can be simplified if we introduce
the phase difference variables
\be
\ba{c}
\Delta_1=\phi_g-\phi_1\vspace{2mm}\\
\Delta_2=\phi_2-\phi_g\vspace{2mm}\\
\Delta_g=\phi_g - 2 \psi_g,
\ea
\label{delt1}
\ee
and the center of mass variable, $\Phi$,
\be
\ba{c}
\Phi = \frac{C_{11}+C_{12}+C_{1g}}{C_{tot}} \phi_1
+  \frac{C_{22}+C_{12}+C_{2g}}{C_{tot}} \phi_2\vspace{2mm}\\
+  \frac{C_{1g}+C_{2g}+C_{gg}}{C_{tot}} \; s_g \; \phi_g
+  \frac{C_{1g}+C_{2g}+C_{gg}}{C_{tot}} \: \eta_g\; 2 \psi_g,
\ea
\ee
where 
\be
C_{tot}=\sum_{ij} C_{ij}
\ee
 (note that $C_{tot}$ is not affected
by the mutual capacitances $C_1$ and $C_2$ but is determined by the 
capacitance of the system to the ground). We thus have 
\begin{eqnarray}
S_Q = \frac{1}{2(2e)^2} \int_0^\beta d\tau \left(
C_1 ( - \dot{\Delta}_1 + \eta_g \dot{\Delta}_g )^2
\r.\nonumber\\\l.
+ C_2 ( \dot{\Delta}_2 + \eta_g \dot{\Delta}_g )^2
+ C_Q \dot{\Delta}^2_g +C_{tot}\dot{\Phi}^2 \right)
\label{SQ2}
\end{eqnarray}
The center of mass
coordinate, $\Phi$, completely decouples from the phase differences
in the charging part of the action, and it is not present in $S_J$
and $S_{dis}$; these can be written as
\begin{widetext}
\be
\ba{rcl}
S_J&=&\int_0^\beta d \tau (-J_1 \cos(\Delta_1)- 
J_2 \cos(\Delta_2))\vspace{2mm}\\
\label{SJ2} 
S_{dis}&=&\beta \sum_{\omega_n}\; \frac{R_Q}{2\pi}\;
\l( \frac{|\omega_n|}{R_1}|\Delta_{1,\,(\omega_n)}+\Delta_{g,\,(\omega_n)}|^2+
\frac{|\omega_n|}{R_2}|\Delta_{2,\,(\omega_n)}
+\Delta_{g,\,(\omega_n)}|^2 +
\frac{|\omega_n|}{r}|\Delta_{g,\,(\omega_n)} |^2
\r).
\label{Sdis2}
\ea
\ee
Therefore the center of mass coordinate, $\Phi$, factors out in the partition
function.
From Eqs. (\ref{SQ2})-(\ref{Sdis2}) we see that $\Delta_g$
appears quadratically in the action and can be integrated out.
After this integration, and also after neglecting terms involving $C_1/C_Q,\,C_2/C_Q\ll 1$, we obtain
\begin{eqnarray}
\begin{array}{c}
S=\frac{R_Q}{2\pi}\beta \sum\limits_{\omega_n}\l
(\l|\Delta_{1,(\omega_n)}\r|^2
\left(\frac{|\omega_n|}{2R_1} \; \frac{\l[\frac{\hbar}{C_Q}\l(\frac{1}{r}+\frac{1}{R_2}\r)
+|\omega_n|
+C_1 R_1 \omega_n^2/\hbar
\r]}{\l[\frac{\hbar}{C_Q}\l(\frac{1}{R_1}+\frac{1}{R_2}+\frac{1}{r}\r)+|\omega_n|\r]}\r)
\right.\vspace{2mm}\\\left.+
\l|\Delta_{2,(\omega_n)}\r|^2
\left(\frac{|\omega_n|}{2R_2} \; \frac{\l[\frac{\hbar}{C_Q}\l(\frac{1}{r}+\frac{1}{R_1}\r)
+|\omega_n|+C_2 R_2 \omega_n^2/\hbar\r]}{\l[\frac{\hbar}{C_Q}\l(\frac{1}{R_1}+\frac{1}{R_2}+\frac{1}{r}\r)+|\omega_n|\r]}\r)
\right.\vspace{2mm}\\\left.+
\Delta_{1,(\omega_n)}\Delta_{2,(-\omega_n)}\frac{|\omega_n|}{R_1R_2}
\frac{\hbar/C_Q (1+ |\omega_n| \eta_g C_1 R_1/\hbar) 
(1+ |\omega_n| \eta_g C_2 R_2/\hbar)}
{ \l[\frac{\hbar}{C_Q}\l(\frac{1}{R_1}+\frac{1}{R_2}+\frac{1}{r}\r)+
|\omega_n|\r] } \r)+S_J,
\end{array}
\label{action1}
\end{eqnarray}
\end{widetext}
Looking at the ubiquitous denominators of Eq. (\ref{action1}) we
notice the expression:
\[
\frac{\hbar}{C_Q}\left(\frac{1}{r}+\frac{1}{R_1}+\frac{1}{R_2}\right)+|\omega_n|.
\]
The scale for the Matsubara frequencies, 
$\omega_n$, is set by temperature, hence  a
new temperature scale emerges from (\ref{action1}):
\be
T^*=
%\frac{1}{2\pi}
(2e)^2 (D_S+D_N) R_Q \left(\frac{1}{r}+\frac{1}{R_1}+\frac{1}{R_2}\right).
\label{Tstar}
\ee 
This is the level spacing on the grain ($1/C_Q=D_S+D_N$) times a dimensionless resistance dependent factor, and it is also of the order of the inverse escape time from the grain.  

{\it High temperature limit.} When $T\gg T^*$ the denominator in (\ref{action1}) is dominated by
$|\omega|\gg T^*$, and the effective action for high temperatures
is
\be
\ba{c}
S\approx\frac{R_Q}{2\pi}\beta\sum\limits_{\omega_n}\left(\frac{1}{2}\l|\Delta_{1,\,(\omega_n)}\r|^2\l(\frac{|\omega_n|}{R_1}+C_1\omega_n^2/\hbar\r)+\r.\vspace{2mm}\\\l.
\frac{1}{2}\l|\Delta_{2,\,(\omega_n)}\r|^2\l(\frac{|\omega_n|}{R_2}+C_2\omega_n^2/\hbar\r)+\r.\vspace{2mm}\\\l.
\Delta_{1,\,(\omega_n)}\Delta_{2,\,(-\omega_n)}\r.
\vspace{2mm}\\\l.
\frac{|\omega|}{R_1R_2}\left(\frac{\hbar/C_Q(1+ |\omega_n| \eta_g C_1 R_1/\hbar) 
(1+ |\omega_n| \eta_g C_2 R_2/\hbar)}
{|\omega|}\right)\right)\vspace{2mm}\\
+S_J. 
\label{highTaction}
\ea
\ee
In this limit we see that the interaction term between the two junctions (which is $T$-independent to leading order in $C_1/C_Q$, $C_2/C_Q$) is
negligible compared to the other resistive and capacitative parts of the action; 
%($|\omega|\gg T^*$) 
the two junctions are thus effectively decoupled for $T\gg T^*$.
The dissipations for the two junctions in this limit are set simply by the individual shunt resistances
$R_1$ and $R_2$. This is the limit that has been discussed
in the literature; its validity at low temperatures relies on the basic
assumption of {\it macroscopic} grains, for which $T^*=0$.

{\it Low temperature limit.}
At temperatures $T$ below $T^*$  
(we assume that $T^*<\hbar/(R_1C_1)$ and
$T^*<\hbar/(R_2 C_2)$) a qualitatively 
different picture emerges in which coupling between the two junctions becomes important. The low-energy effective theory is  
\be
\ba{rcl}
Z &\approx& \int {\cal D} \Delta_1 {\cal D} \Delta_2 
\,\,e^{- S_d - S_C - \tilde{S}_J}\vspace{2mm}\\
\tilde{S}_J &=& \int_0^\beta d \tau 
(-J_1 \cos(\Delta_1)- J_2 \cos(\Delta_2)\vspace{2mm}\\
&&-J_+ \cos(\Delta_1+\Delta_2))\vspace{2mm}\\
S_C &=& \beta \sum\limits_{\omega_n}
\left(\frac{C_Q}{2(2e)^2}\left|\frac{rR_2 \Delta_{1}+rR_1
\Delta_{2}}{rR_1+rR_2+R_1R_2}\right|^2\omega_n^2\right)\vspace{2mm}\\
S_d &=& \beta \sum\limits_{ \omega_n }~ \frac{|\omega_n|}{2}~
\vec{\Delta}^{\dagger}~
\hat{G}~
\vec{\Delta}
\label{lowTaction}
\ea
\ee
with 
\be
\vec{\Delta}\equiv(\Delta_1,\Delta_2)
\ee
and the matrix
\be
\hat{G} = \frac{R_Q}{2\pi Y}\,\,
\left( \begin{array}{cc}
r+R_2& r \vspace{2mm}\\
r & r+R_1
\end{array} \right)
\label{Gmatrix}
\ee
where 
\be
Y\equiv rR_1+rR_2+R_1R_2.
\ee
In the equations above we have added, for future purposes, a lead--to--lead Josephson coupling term representing  cotunneling processes via the grain. 
This term describes a Cooper pair tunneling (pair-tunnel event) from the left electrode
to the right electrode (see Fig. \ref{Figure8}) via a virtual intermediate state with an additional pair on the grain.
Such processes appear perturbatively at  
second order in $J_1$ and $J_2$ and
will be generated in the RG flows for the action (\ref{lowTaction})
(see discussion below Eq. (\ref{Jflowsecond})).  

It is important to note that level spacing
$D_g = D_{sg} + D_{ng}$ only appears in $S_C$
via the quantum capacitance $C_Q=D_g^{-1}$, 
whose precise form will not matter except to yield a high frequency cut-off.
By the same token,  a different form of the capacitative energy of the leads and the grains would only modify $S_C$ and not change any of the
analysis presented in this paper.

Action (\ref{lowTaction}) is one of the main results of this paper, and
in the following sections we will mostly be concerned with
studying its properties.

\begin{figure}
\includegraphics[width=8.5cm]{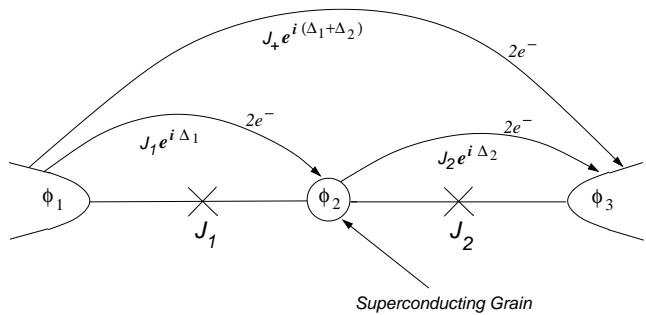}
\caption{Physical interpretation of expanding the $J_i\cos\Delta_i$ terms in the action (\ref{lowTaction}). 
The weak Josephson coupling action can be mapped to a theory of interacting Cooper-pair tunneling
 events (Coulomb-gas representation) with each pair-tunnel ``charge" corresponding to a Cooper-pair transferred through one of the junctions. The co-tunneling events which transfer  
Cooper-pairs from lead to lead are also shown; in the Coulomb-gas representation these correspond to pair-tunnel dipoles.\label{Figure3}}
\end{figure}

\section{Weak coupling analysis}

In this section we analyze the low energy properties of the system in the  weak Josephson coupling limit.

\subsection {Renormalization Group equations \label{ptc}}

In the limit of weak Josephson couplings $\{J_i\}$, the 
quantum action (\ref{lowTaction}) can be analyzed directly in the generalized
sine-Gordon representation. 
In Appendix \ref{22RG}, RG flow equations are derived to second order in the Josephson couplings:
\be
\ba{c}
\frac{dJ_1}{dl}=J_1\left(1-\frac{R_1+r}{R_Q}\right)+\frac{R_2}{R_Q}J_2 J_+\vspace{2mm}\\
\frac{dJ_2}{dl}=J_2\left(1-\frac{R_2+r}{R_Q}\right)+\frac{R_1}{R_Q}J_1 J_+\vspace{2mm}\\
\frac{dJ_+}{dl}=J_+\left(1-\frac{R_1+R_2}{R_Q}\right)+\frac{r}{R_Q}J_1 J_2
\label{Jflowsecond}
\ea 
\ee
In writing these we have set a combination of the short-time cut-offs to be
equal to one. In physical units, the energy cut-off is of order 
the charge relaxation rate of the junctions in units of which we are here measuring the $\{J_i\}$.  
The first order terms in the RG flows
arise, as usual, from integrating out fast modes in the quadratic part of action
(\ref{lowTaction}).  The second order terms are obtained from recombinant terms in the expansion in powers of $J$'s of
(\ref{lowTaction}). These can be understood physically; pair-tunnel 
events on junctions one and two can combine to form a
cotunneling event between the two leads, while a
cotunneling event plus a pair-tunnel in the opposite direction across one of the junctions
is equivalent to pair-tunneling across the other junction (for details see
Appendix \ref{22RG}).  From (\ref{Jflowsecond}) we see that, as claimed in the previous section,  $J_+$ gets
generated at low energies even if we start with a
 model in which $J_+=0$.

\subsection{Weak coupling phase diagram \label{weakpd}}

Surprisingly, the simple flow equations (\ref{Jflowsecond}) give rise to five different regimes. When all $J$'s are irrelevant about the uncoupled fixed line so that they flow to
zero, the system is in the normal state with no supercurrents between the leads or between either lead and the grain. This {\it normal} (NOR) phase occurs if $R_1$ and $R_2$ are both sufficiently large. When all $J$'s are
relevant and grow under the RG flows, the systems is in a fully  superconducting phase that
we denote FSC. This occurs if all the resistances are sufficiently small. For intermediate ranges of the resistances, the situation is more complicated.

When only one out of the three
$J$s is relevant while the other two flow  to zero at low energies,
the system is in a  ``mixed phase''; as we shall see,  there are three such phases. 
When the only relevant coupling is $J_1$, junction one is
superconducting and junction two is normal, we call this phase
S1-N2. With respect to lead-to-lead transport this is like the normal
phase. Analogously we will have an N1-S2 phase when $J_2$ is relevant
but $J_1$ and $J_+$ are not. Rather surprisingly, there can also be a
situation in which $J_+$ is relevant but $J_1$ and $J_2$ are not. This
is a phase in which individual junctions are normal, but
the circuit as a whole is superconducting and Cooper pairs can flow freely
between the leads. We denote this phase $SC^{\star}$.  Physically, it corresponds to Cooper pairs being localized on the grain, so that the 
individual junctions are normal, however the cotunneling processes, via virtual Cooper pair excitations on the grain,
induce superconducting coherence  between the leads.
A similar phase was discussed by Korshunov \cite{korshunov}
and Bobbert et. al. \cite{schoen} in the context of one dimensional
Josephson junction arrays. 

Inspection of the flow equations shows that as long as
$R_1,R_2,r>0$
there {\it cannot} be phases in which two of the three $J$s grow while the third flows to zero: the coupling terms in (\ref{Jflowsecond}) from the two growing ones will drive the third $J$ to grow as well. The system will then be in the {\it fully superconducting}  (FSC) phase.

\begin{figure*}
\includegraphics[width=16.5cm]{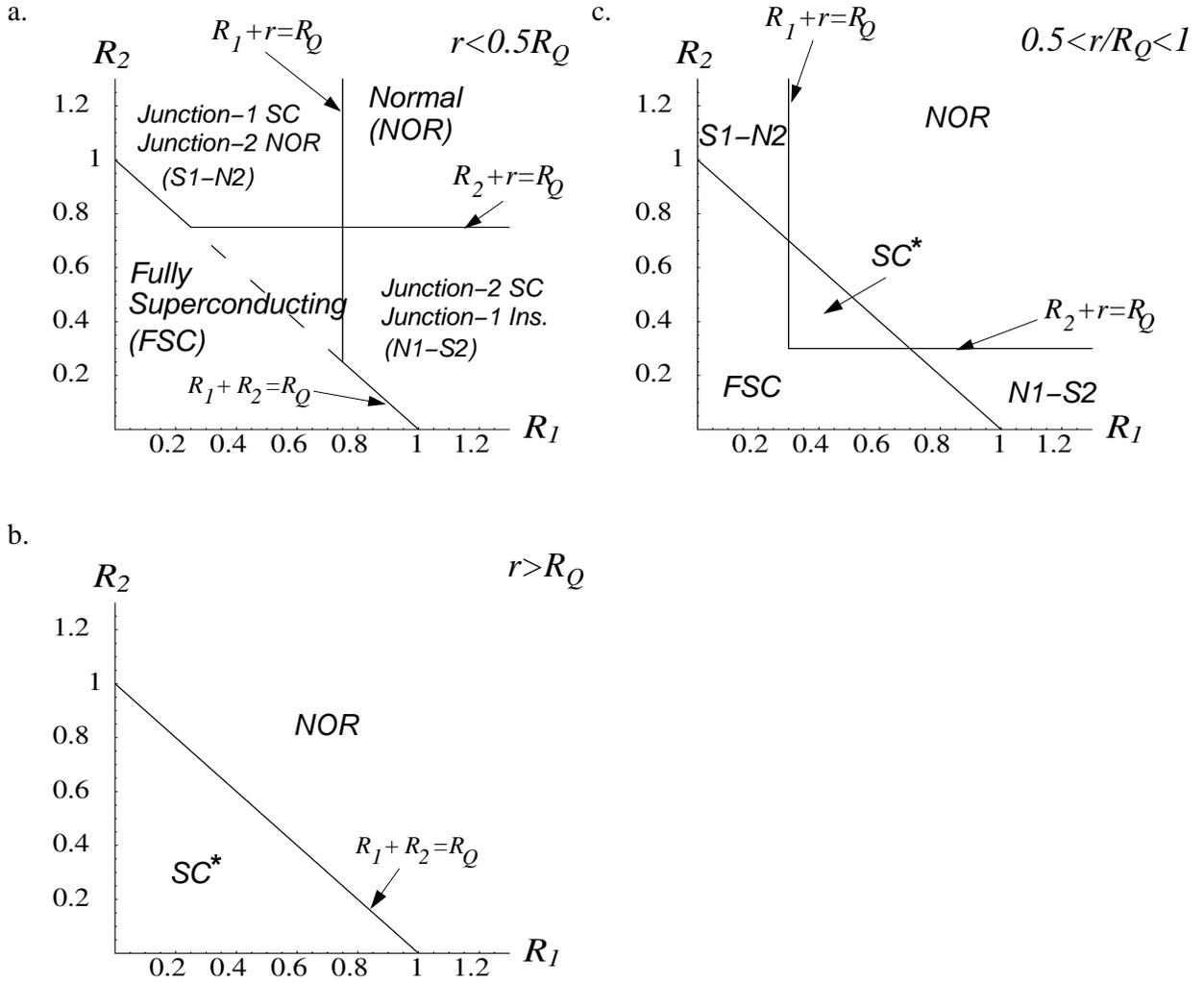}
\caption{Naive weak Josephson coupling phase diagram from first order RG. (a) For $r<0.5R_Q$ four phases are present. In the superconducting (FSC) phase both junctions are superconducting, in the normal phase (NOR) both 
junctions are normal. In the two remaining phases one junction is conducting and the other
normal (N1-S2 and S1-N2). (b) When $0.5<r/R_Q<1$ there is a new phase, $SC^{\star}$, 
in which charges are localized on the grain. Both junctions are thus normal, nevertheless 
Cooper-pairs can coherently tunnel between the leads (co-tunneling) so that this phase is 
superconducting with respect to lead-to-lead transport.
(c) For $r>R_Q$ the FSC phase and the N1-S2 and S1-N2 
phases disappear. The system is either in the $SC^{\star}$ phase, or in the
NOR phase. %The superconducting-insulating transition here is
%determined solely by the {\it total} resistance, $R_1+R_2$.
\label{Figure4} }
\end{figure*}

To lowest order for small $J$s, the phase boundaries between these phases are set by the relevance of $J_1$, $J_2$, and $J_+$ about the decoupled (normal) fixed line; these are determined
by the combinations $R_1+r$, $R_2+r$, and $R_1+R_2$ respectively,
 leading to the weak-coupling {\it approximate phase diagram} shown in Fig. \ref{Figure4}. 

This simple analysis, however, is not sufficient to obtain the correct phase diagram. For example, by looking at
Fig. \ref{Figure4}(a) we see that it has a superconductor to normal transition of junction one {\it inside} the superconducting phase of
junction 2 (the N1-S2 to FSC transition). In this situation the RG equations
derived in the vicinity of the uncoupled (normal) fixed line are no longer valid. 

 A better approximation for the N1-S2 to FSC transition can be obtained by noting that the fluctuations in phase difference across junction two, $\Delta_2$, will be small in the N1-S2 phase.  Thus in this regime we can approximately set $\Delta_2=0$ 
in (\ref{lowTaction}). This modifies the RG flow for $J_1$
to 
\be
\frac{dJ_1}{dl}=J_1\l(1-\frac{R_1+ \frac{r R_2}{r+R_2}}
{R_Q}\r)
\label{j1screened}
\ee
[We will see later that in the Coulomb gas language, equation (\ref{j1screened})
corresponds to including the screening effects of unbound
type two charges when considering the unbinding transition
for charges of type one (see Appendix \ref{22RG} for details).]
From (\ref{j1screened}) we find that the N1-S2 to FSC boundary gets
shifted to
\be
R_1+ \frac{r R_2}{r+R_2} = R_Q.
\label{j2rel0}
\ee
Similar modifications of the phase boundaries appear
for all transitions that involve ordering
of one field in the presence of order in another:
S1-N2 to FSC (ordering of $\Delta_2$ when $\Delta_1$ is ordered); and
$SC^{\star}$ to FSC (ordering of $\Delta_1$ and $\Delta_2$ when
$\Delta_1+\Delta_2$ is ordered).

\begin{figure*}
\includegraphics[width=16.5cm]{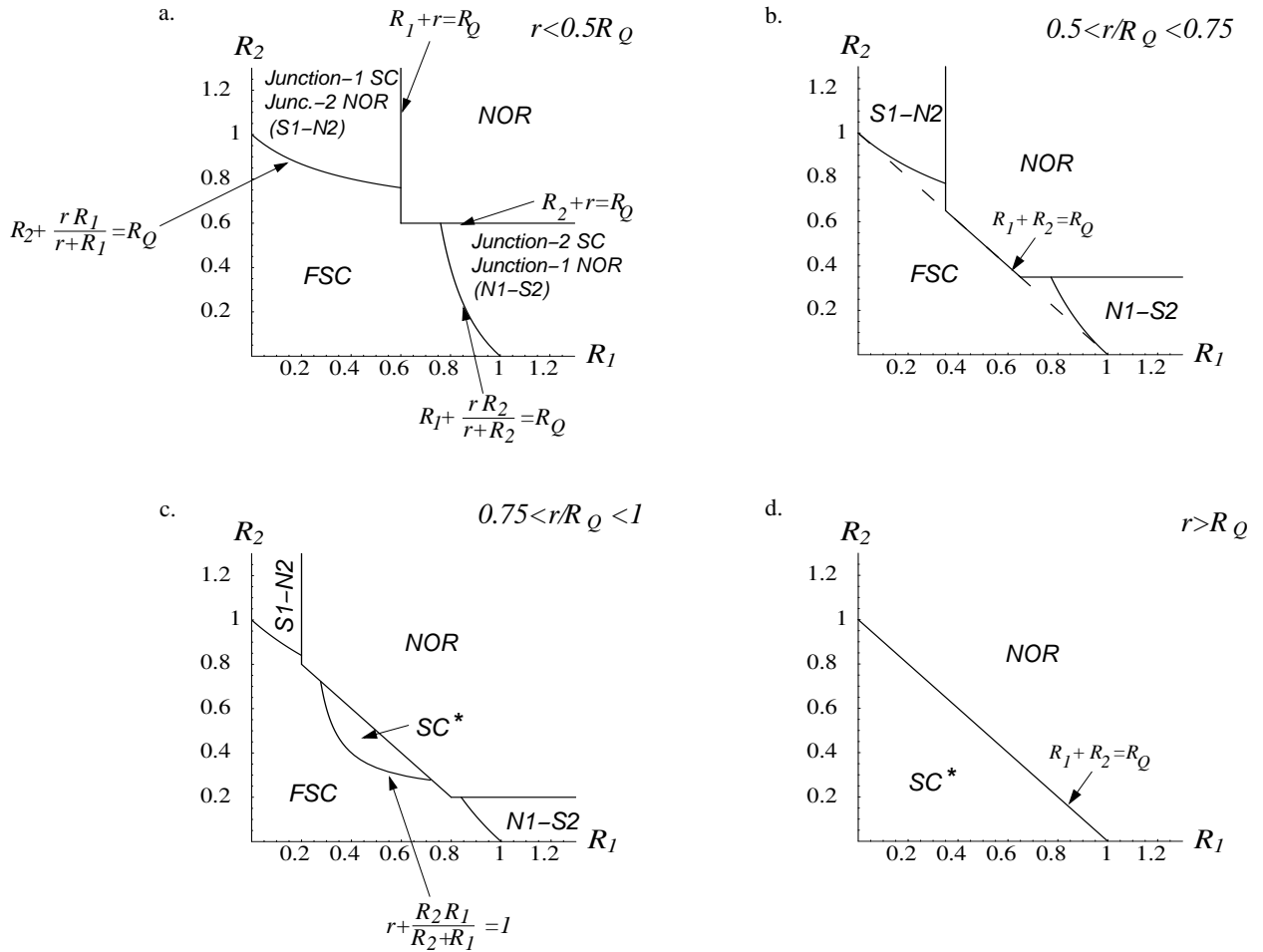}
\caption{Complete weak Josephson coupling phase diagram. Phase boundary formulas apply 
everywhere, although they are each given in only one graph. (a) When $r<0.5R_Q$ four of the five 
phases are present; each junction is either normal or superconducting. (b) For larger $r$, the shape of the phase boundary between the FSC and NOR phases changes. 
(c) When $0.75<r/R_Q<1$ the $SC^{\star}$ appears and all five phases are present. 
(d) When $r>R_Q$ only $SC^{\star}$ survives of the mixed phases, and FSC disappears. 
%The circuit behaves like a single junction with shunt resistance being the total 
%resistance $R_1+R_2$.  
\label{Figure5} }
\end{figure*}

The corrected --- and rather complicated --- weak-coupling phase diagram is shown in Fig. \ref{Figure5}. 
A particularly interesting regime occurs for $r>R_Q$. In this regime the two-junctions 
cease to behave as such, instead, they behave much like a single junction shunted by the total resistance, $R_1+R_2$, which therefore determines the location of the superconducting--to--normal transition between the two leads. 
This  will be discussed further in Section VII. 

\subsection{Circuit theory for weak coupling \label{ctcircuits}}

In this subsection we show how the phase diagram of Fig. \ref{Figure5}
can be obtained by simple physical arguments.  Before proceeding
it is useful to recall such an argument
for a single junction. 

We want to investigate the stability of the
superconducting state of a single Josephson junction with an ohmic shunt. In the
superconducting phase, Cooper pairs are delocalized between the
leads. Each Cooper pair tunneling event changes the charge on the
junction by $2e$. This charge needs to be screened by the normal
electrons in the shunt thereby causing a voltage drop to appear across the
junction. By the Josephson-relation, this voltage drop induces a change in the phase
difference across the junction. The superconducting phase
with delocalized Cooper pairs will survive only when the phase change due to one Cooper
pair tunneling event is less than $2\pi$ (otherwise the phase becomes delocalized). 
From circuit equations and the Josephson relation we find
\be
2e= \int I_N dt = \int \frac{\Delta V}{R_S} dt
= \frac{\hbar}{2e R_S} \int \frac{d\phi}{dt} dt = 
 \frac{\hbar}{2e R_S} \Delta \phi
\ee
where $I_N$ is the normal screening current, $\Delta V$
is the voltage difference across the junction,
and $\Delta \phi$ is the phase change due to a Cooper
tunneling. Rewriting the last relation as
\be
\frac{\Delta \phi}{2\pi} = \frac{R_S}{R_Q}
\ee
we obtain the usual condition; the
shunted Josephson junction is superconducting when $R_S < R_Q$.
We can summarize this argument by saying that
a Cooper pair tunneling event provides a current source with a magnitude that
depends on the shunting resistance. By the Josephson relation, this leads
to a phase fluctuation across the junction, and the superconducting
phase is only stable when this phase fluctuation is less than $2\pi$.[Note that this argument does not really yield the exact condition:  a multiplicative factor of order unity could have arisen. A fuller analysis, as from the RG flows,  is needed to obtain the correct coefficient.]

Applying this approach to the two junction system of Fig. \ref{Figure2}
effectively reduces the problem to determining the effective
shunting resistance associated with Cooper pair tunneling events in various
situations. As in the single junction case, a Cooper-pair tunneling can be simply modeled as a {\it current source}.

\begin{figure}
\includegraphics[width=8.5cm]{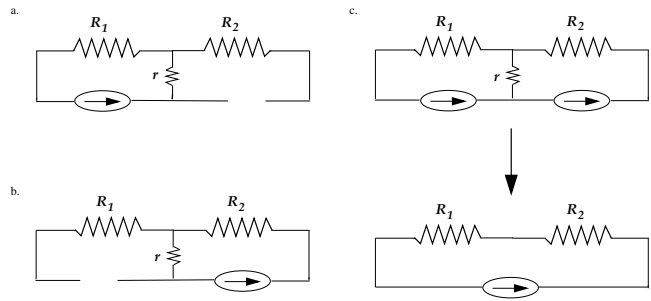}
\caption{Effective circuits for pair tunneling events. A pair-tunnel corresponds to a 
current source, whereas a junction without pair tunneling acts as an open-circuit. 
(a) Effective circuit for a pair tunneling through junction 1. (b)  Effective circuit for a 
pair tunneling through junction 2. (c) Effective circuit for a coherent lead-to-lead pair 
tunneling event. Since the current through junction 1 and 2 is the same, the resistance $r$ 
is effectively disconnected in this case.  
\label{Figure6} }
\end{figure}

\begin{figure}
\includegraphics[width=8.5cm]{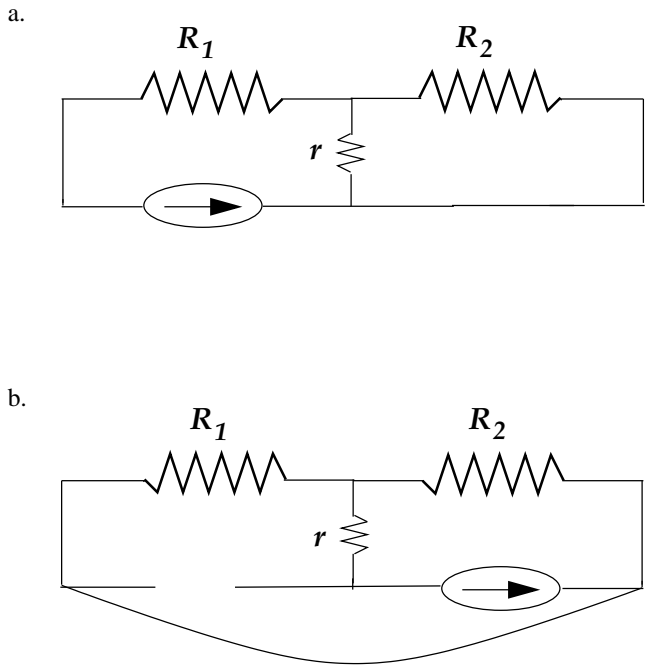}
\caption{(a) Effective circuit for a pair tunneling through junction-1 when junction-2 is 
superconducting ($J_2$ is relevant about the weak coupling limit). (b)  Effective circuit for a pair tunneling event 
through junction-1 (or 2) when $J_+$ is relevant and coherent lead-to-lead pair tunneling 
events proliferate.  
\label{Figure7} }
\end{figure}

\begin{itemize}
\item To find the transition between S1-N2 and the NOR phase, consider
 a Cooper pair tunneling across junction one
with junction two insulating and acting
as a circuit disconnect. The effective resistance 
that makes a circuit with the current source is then $R_1+r$ (see Fig. \ref{Figure6}(a)), and the phase boundary
is at $R_1+r=R_Q$. Analogously for the N1-S2 to NOR
transition we have the circuit shown in Fig. \ref{Figure6}(b) and a phase boundary at
$R_2+r=R_Q$.

\item The $SC^{\star}$ to NOR transition is marked by the proliferation of
 cotunneling processes in which a Cooper pair moves between
the leads, but with both junctions individually insulating. The circuit describing this case is 
depicted in Fig. \ref{Figure6}(c), with the
cotunneling process described as two current sources forcing the same current through both 
Josephson junctions.
The cotunneling process leaves no charge 
on the grain, and is hence screened only by normal
currents flowing in the resistors $R_1$ and $R_2$. 
The effective shunting resistance in this case is $R_1+R_2$ and the phase boundary
is at $R_1+R_2=R_Q$. 

\item The transition between N1-S2 and FSC occurs  while junction two is already superconducting
and can hence be replaced by a short in the circuit (see Fig. \ref{Figure7}(a)).
The effective shunting resistance across junction one then involves $r$ and $R_2$ in {\it  parallel}, as well as $R_1$; therefore 
the phase boundary for this transition occurs at
$R_1+rR_2/(r+R_2)=R_Q$.
By the same token the transition between S1-N2 and FSC takes place
when $R_2+r R_1/(r+R_1)=R_Q$.

\item To understand the FSC to $SC^{\star}$ transition we need to consider the
regime in which cotunneling maintains coherence between the leads; therefore
these are effectively connected by a short in the circuit as shown in Fig. \ref{Figure7}(b).
Now consider a Cooper pair tunneling from the grain to one of the
leads, say two. The effective resistance seen by a tunneling
Cooper pair is $r+R_1 R_2/(R_1+R_2)$ and the phase boundary is hence at
$r+R_1 R_2/(R_1+R_2)=R_Q$. The effective shunt resistance for tunneling from the grain to lead one is the same.  The nature of the $SC^{\star}$ phase is as follows: Cooper pair tunneling events
scramble phases across junctions one and two too much for the junctions
to be coherent, so Cooper pairs become localized on the grain.
Nevertheless, cotunneling events allow  Cooper pairs to move
between the leads, so there is a well defined phase difference
between them that acts as a short between the two macroscopic leads as far as dissipation across the individual junctions. 

\item The FSC to NOR transition line is, naively,  a continuation of the S1-N2 to NOR and
N1-S2 to NOR lines. This suggests that when considering fluctuations of the phase difference across
junction one, we assume junction two to be insulating, and vice versa. The consistency of such an approximation is highly questionable and reflects the limit of small $J$'s as our starting point: by a weak-coupling analysis: for a weak-coupling limit to be valid, we should only approach  
phase boundaries from normal phases of the junction under consideration.

\end{itemize}

It is worth pointing out that in all cases described above,
the effective dissipation is {\it decreased} relative to that in the
high temperature action (\ref{highTaction}). The most
extreme case happens for the  $SC^{\star}$ to NOR transition which
is determined by the {\it total} shunting resistance at low temperatures
rather than individual resistances $R_1$ and $R_2$, which would determine the transitions between macroscopic grains.
In the $SC^{\star}$ phase the whole system behaves as a single junction,
and the dissipation is determined by the resistance across the whole of the system.

\section{Strong coupling analysis \label{sta}}

We have seen that much can be concluded from the weak Josephson
 coupling analysis, in particular the nature of the five possible
 phases and some of the transitions between them. Yet some of the
 transitions could only be understood via a hybrid analysis involving
 some large and some small couplings,  and, as pointed out above,  the
 FSC to NOR  transition cannot be analysed in a controlled manner from
 a   weak-coupling analysis.  Even to  solidify the 
 identification of all of the superconducting phases, we really need to
go beyond weak coupling: as soon as one or more of the $J$s
 grows without bound, the system flows out of the regime of validity of
 the RG flow equations used thus far and we must ask where it flows to. 

 In this Section we turn to the  limit of large Josephson coupling and
attempt to analyse the phases, phase diagram, and transitions in that
limit.

\subsection{Sine-Gordon action for quantum phase slips}

When the Josephson couplings are large, the system is usually
in the vicinity of one of the classical minima of the Josephson potentials so that
$\Delta_1 \approx  2 \pi n_1$, $\Delta_2 \approx  2 \pi n_2$
with $n_1$ and $n_2$ integers. Only rarely does the system
undergo a tunneling event  in which one or both of the phases winds by $2\pi$.   
Such phase tunneling processes between minima of the classical
potential
are {\it quantum phase slips} (QPS). \cite{schoen-zaikin}  
When QPS across it are suppressed at low
temperatures  a Josephson junction is superconducting, but when
they proliferate the junction is incapable of supporting supercurrents  and becomes normal.

  In weak coupling we analyzed the low energy action in terms of Cooper
pair tunneling events. As discussed in Appendix B, this is equivalent
to a  classical  Coulomb gas with two types of charges corresponding to
pair tunneling events through the two junctions; dissipation gives rise
to effective logarithmic interactions among these.  Various of the superconductor
to normal transitions can be described as binding-unbinding transitions
of this two component plasma.

In the strong coupling case we can write a Coulomb gas representation
for the quantum phase slips instead of  the Cooper pair tunneling
events.   The phase slips also  behave as a two-component gas ---
phase-slips on the two junctions ---  with logarithmic interactions
between them. When the phase-slips across a junction proliferate, it
becomes  normal; if instead their fugacity tends to zero at low-energy-scales,
the junction is superconducting. Mathematically, the strong coupling
case can be analyzed by performing a Villain transformation to
represent the partition function (\ref{lowTaction}) in terms of two
types of interacting phase slips. This classical Coulomb gas can then
be transformed into a new sine-Gordon model that is {\it dual} to
(\ref{lowTaction}).  Appendix C describes the details of such
transformations. We find 
\be  
\ba{c} 
Z= \int D[\theta_1]\int D[\theta_2]\exp(-S)\vspace{2mm}\\ 
{\rm with} \\  S=  \beta \sum\limits_{\omega_n}
|\omega_n|  \vec{\theta}_{-\omega_n}^{~T}
\hat{M}\vec{\theta}_{\omega_n}\vspace{2mm}\\ 
-\int_0^\beta
d\tau\left(\zeta_1\cos(\theta_1)+\zeta_2\cos(\theta_2)+\zeta_-\cos\left(\theta_1-\theta_2\right)\right),
\ea \label{psaction} 
\ee 
where $\vec{\theta}=(\theta_1,\theta_2)$ and 
\be \hat{M}= \hat{G}^{-1} = \frac{1}{2\pi R_Q}\left( \begin{array}{cc}
r+R_1 & -r\vspace{2mm}\\ -r & r+R_2 \end{array} \right)  \ee  is the
scaled resistance matrix. The variables $\zeta_1,\,\zeta_2,\,\zeta_-$
are the fugacities corresponding to the three types of phase-slips:
$\zeta_1$ across junction 1; $\zeta_2$ across junction 2; and $\zeta_-$ a
combination of these that corresponds to a phase slip across
1 and a simultaneous  anti-phase slip across 2, thereby slipping the
phase on the grain with respect to {\it both} of the superconducting
leads.  

\subsection{Phase diagram \label{spd}}

Following the steps leading to Eq. (\ref{Jflowsecond}) we readily obtain the flow equations for the phase slip fugacities $\zeta_1,\,\zeta_2$ 
and $\zeta_-$:
\be
\ba{c}
\frac{d\zeta_1}{dl}=\zeta_1\left(1-\frac{R_Q}{R_1+\frac{R_2 r}{R_2+r}}\right)+\frac{R_1}{Y}\zeta_2\zeta_-\vspace{2mm}\\
\frac{d\zeta_2}{dl}=\zeta_2\left(1-\frac{R_Q}{R_2+\frac{R_1 r}{R_1+r}}\right)+\frac{R_2}{Y}\zeta_1\zeta_-\vspace{2mm}\\
\frac{d\zeta_-}{dl}=\zeta_-\left(1-\frac{R_Q}{r+\frac{R_1R_2}{R_1+R_2}}\right)+\frac{r}{Y}\zeta_1\zeta_2,\vspace{2mm}\\
\label{flowsecond}
\ea
\ee
where we use $Y\equiv R_1R_2+rR_1+rR_2$.
These flow equations are correct
to second order in the $\zeta$s, being simply 
the analog of Eqs. (\ref{Jflowsecond}) for the weak coupling
limit. We again work in units in which the short time cut-off --- here related to the ``transit time" for a least-action phase slip --- is unity.

Growth under renormalization of
a fugacity  $\zeta_i$ corresponds to proliferation of the 
corresponding QPS and
hence destruction of superconductivity across the respective junction in
the case of $\zeta_1$ or $\zeta_2$, or between the grain and the rest of the
system in the case of $\zeta_-$.

\begin{figure*}
\includegraphics[width=16.5cm]{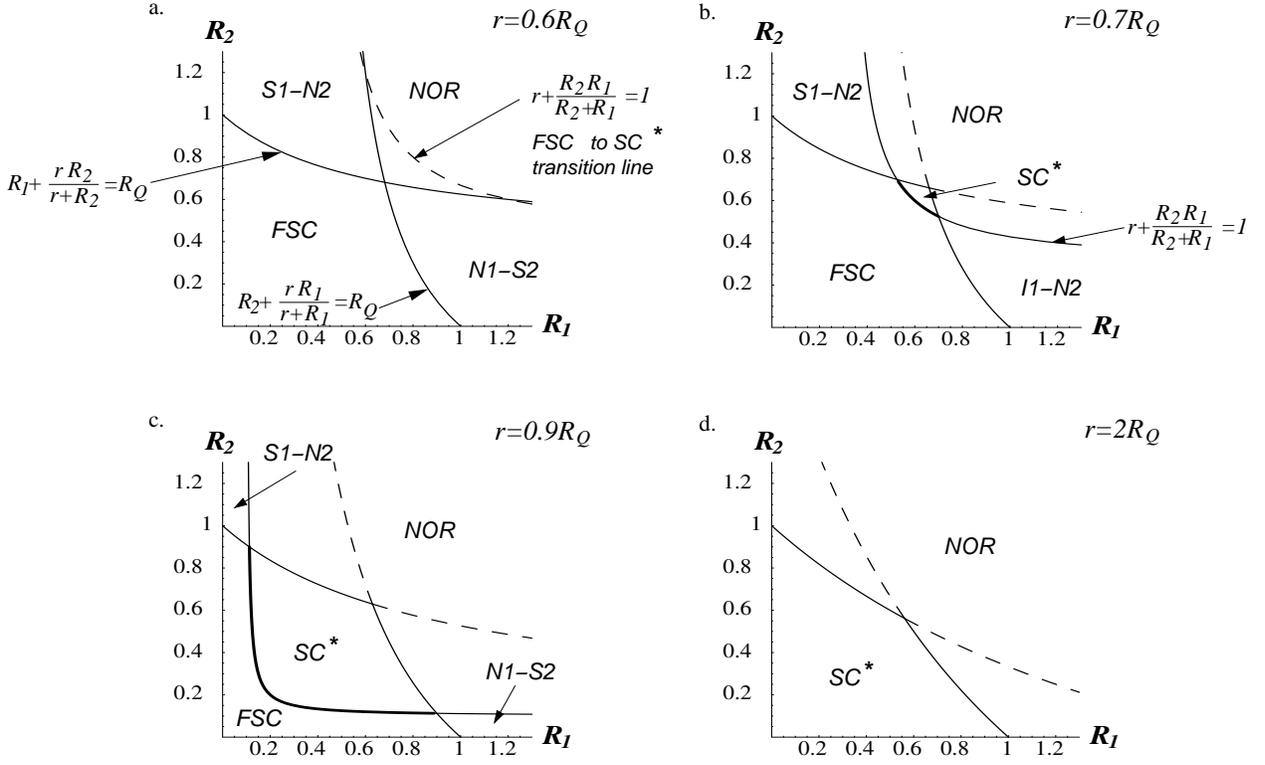}
\caption{Naive strong Josephson coupling phase diagram from first-order RG flows. Phase boundary formulas apply everywhere 
although they are each given in one graph only. (a) In the range $0<r/R_Q<2/3$, 
every junction is either normal or superconducting; (b) and (c) show the range 
$2/3<r/R_Q<1$ for which the $SC^{\star}$ phase takes up some of the FSC parameter space, and 
also pushes the NOR phase towards the axis at the expense of N1-S2 and S1-N2. 
The dashed line is where the boundary of the insulating phase would have been 
if phase-slip dipoles had not been taken into account; (d) obtains in the range $r>R_Q$ for which  
the SC, N1-S2 and S1-N2 phases no longer occur, so that only $SC^{\star}$ and NOR phases survive.  
\label{Figure8}}
\end{figure*}

Equation (\ref{flowsecond}) gives rise, as did the weak coupling
analysis, to five phases.  When all $\zeta$'s are irrelevant and flow
to zero, the system is in the fully superconducting state (FSC) since 
isolated phase slips all cost infinite action. Conversely, if all
$\zeta$'s are relevant,  we expect the normal state (NOR) to obtain. 
As in the weak coupling case, three mixed phases appear  when only
{\it one} of the fugacities is relevant. When $\zeta_1$ is relevant and
$\zeta_2$ and $\zeta_-$ are not,  the system is in the N1-S2 phase;
analogously a relevant $\zeta_2$ and irrelevant $\zeta_1$ and $\zeta_-$
signal the S1-N2 phase. 

If $\zeta_-$ is relevant but $\zeta_1$ and  $\zeta_2$ are not, the
special $SC^{\star}$ phase occurs. In this phase only QPS {\it dipoles}
proliferate; these consist of a phase slip across one junction and an
{\it anti}-phase slip across the other.   Isolated phase slips across
individual junctions will not occur in the $SC^{\star}$ phase.
Superconducting phase coherence between the two leads is thus
{\it maintained}, since the phase difference between them is the sum of the
phase differences for the two junctions, and a phase-slip on junction 
one gets canceled by its accompanying anti-phase-slip on junction two.
But the phase difference between the leads and the grain  is  ill
defined in $SC^\star$ as a result of the proliferated QPS dipoles. We
thus see that proliferation of the QPS dipoles
induces charge localization on the grain.  

The transition between the two superconducting phases,
$SC^{\star}$ and FSC, is, from the point of view of phase slips on the individual junctions,  a transition between  a dipole-free state, FSC, in which all the phase slips will be bound in
quadrapoles, and a phase, $SC^\star$, in which dipoles proliferate but single quantum phase slips still do not occur.  Because of the free dipoles in the $SC^\star$ phase, a single quantum phase slip between the two leads can consist of any combination of phase slips across the two junctions that add up to a total phase difference between the leads of $2\pi$.

As in the weak coupling limit (Sec. \ref{weakpd}), we could attempt to construct 
a naive phase diagram showing all five phases from the {\it first order} strong coupling flow equations. 
This approach would give phase  
boundaries that depend on $R_1+\frac{R_2 r}{R_2+r},\,R_2+\frac{R_1 r}{R_1+r}$ 
and $r+\frac{R_1 R_2}{R_1+R_2}$ (see Fig. \ref{Figure8}). 
But such an analysis, as in the weak coupling limit,  is not sufficient: when one type of phase slip proliferates it will partially screen the interactions between the other types of phase slips. 

\begin{figure*}
\includegraphics[width=16.5cm]{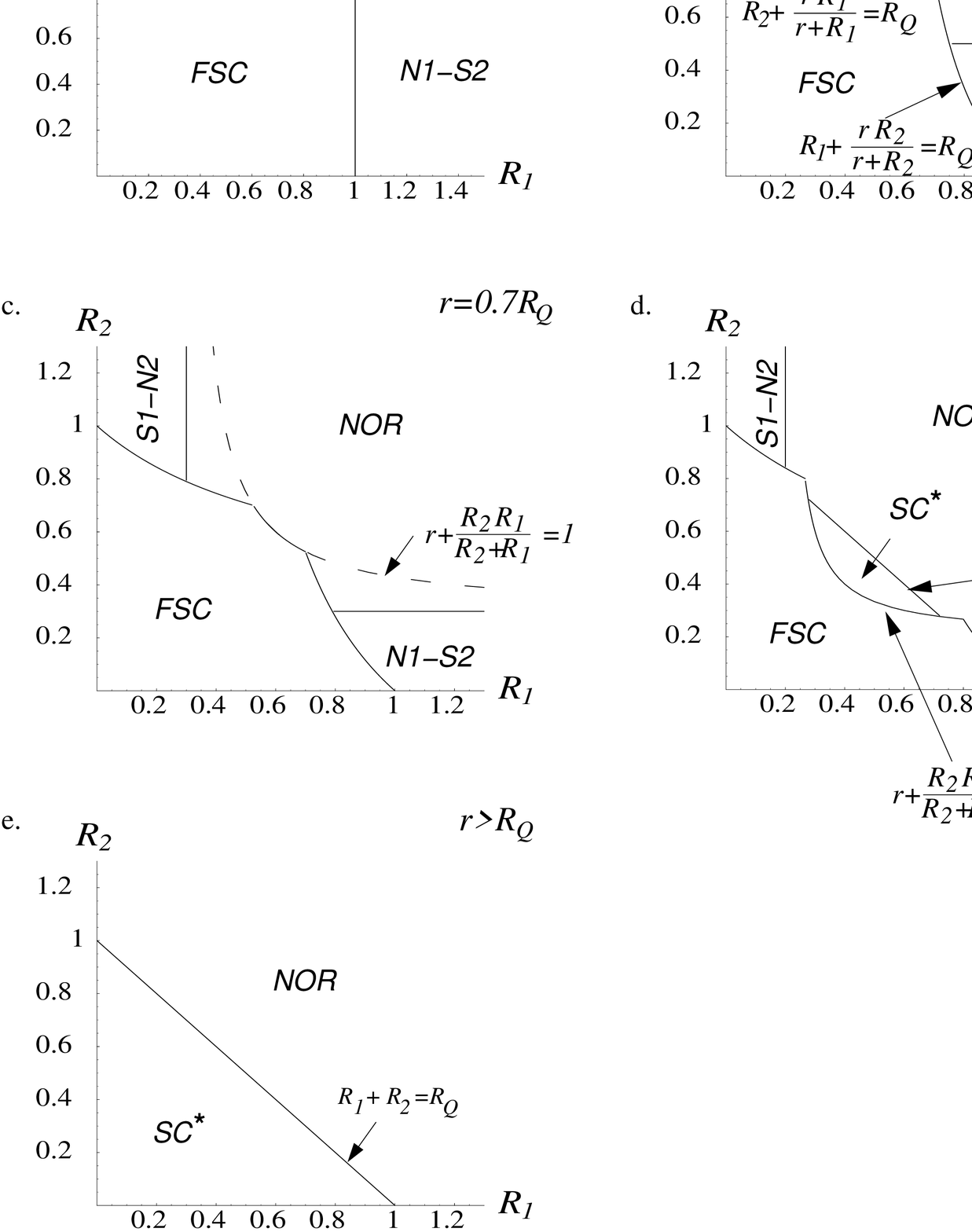}
\caption{Complete strong Josephson coupling phase diagram. Phase boundary formulas
apply everywhere, although they are each given in only one graph. 
(a) $r=0$ for which  the two junctions are effectively independent; 
(b) range $0<r/R_Q<2/3$; (c) range $2/3<r/R_Q<3/4$; 
(d) range $3/4<r/R_Q<1$ where all five phases are present 
(e) range $r>R_Q$ for which the two junctions act  like a 
single junction with a shunt resistor $R_1+R_2$.  
\label{Figure9} }
\end{figure*}

To do better we must consider the effects of the relevance of a $\zeta \cos\theta$ term: this will cause the dual phase, $\theta$, associated with the proliferating phase slips to become localized at an integer multiple of $2\pi$. As the $\theta$ will then not fluctuate appreciably about this at low energies, we can set it to zero.
As for weak coupling, this suppression of some of the fluctuations will change the flows of the remaining fugacities and thereby modify the phase diagram. The complete  phase diagram from such a strong coupling analysis is shown in 
Fig. \ref{Figure9}.

\subsection{Circuit theory for strong coupling}

The strong coupling phase diagram of Fig. \ref{Figure9} can be  simply interpreted
in terms of the effective electronic circuits.  Although
these arguments are dual to the ones used for weak coupling, we present them here
for completeness. 

Again it is useful to start by considering the case of a single junction, now starting from the superconducting regime.
The normal state occurs when quantum phase slips proliferate.
When a QPS occurs, the phase difference across 
the junction changes by $2\pi$.  By the Josephson relation,
this generates both a voltage drop and charge flow through the normal shunt.
In the normal state the Cooper pairs should be localized, therefore such a state
can only be stable if the charge fluctuation caused by an individual QPS
is less than $2e$ (again, the justification of the factor being exactly two really needing the fuller analysis). From Kirkhoff's laws and the Josephson relation
we have
\be
2\pi = \int \frac{d\phi}{dt} dt = \frac{2e}{\hbar} \int V dt =
\frac{2e}{\hbar} R_s \int I dt = \frac{2e}{\hbar} \Delta q
\ee
Here $\Delta q$ is the amount of charge that passes through the  shunt resistor 
as a result of the QPS.  In units of the charge of a Cooper pair, $2e$, this is
\be
\frac{\Delta q}{2e} = \frac{R_Q}{R_S}.
\ee
We thus guess that the normal state is stable when $R_S > R_Q$.  The basic physics is that fluctuating QPS's act as  voltage noise
that gives rise to charge fluctuations on the junction. The
insulating state is only stable when these charge fluctuations are sufficiently small: less
than $2e$. 

The generalization of the single-junction argument to the system in Fig.
\ref{Figure2} requires analysis of the effective shunting resistances for the 
various QPS configurations. The quantum phase slips are effectively  
voltage sources.  The phase slip dipole corresponding to $\zeta_-$ is thus equivalent to two
equal but opposite voltage sources across the two junctions so that there is no voltage between the two leads, but the grain is at a different voltage than the leads.  

\begin{figure}
\includegraphics[width=8.5cm]{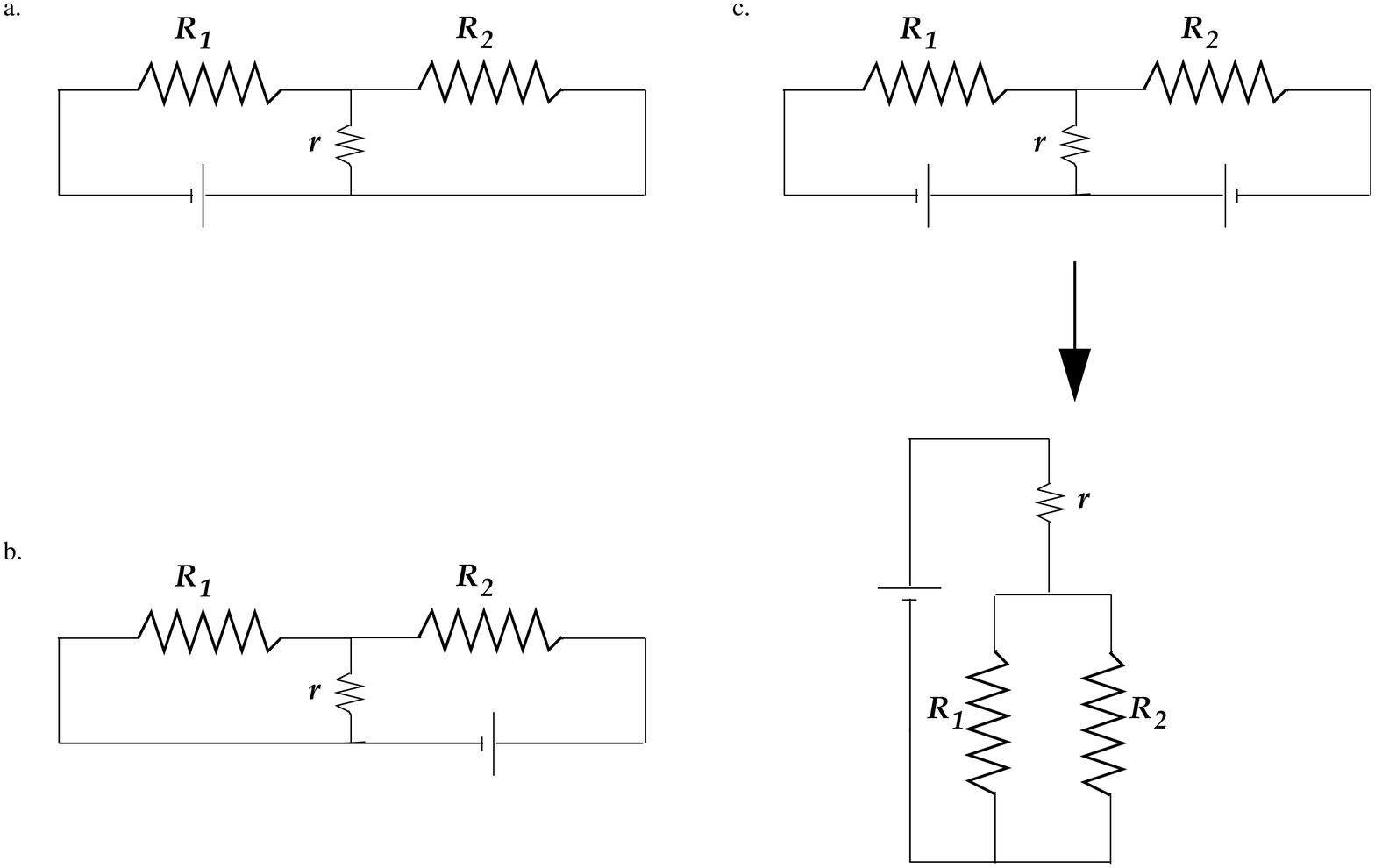}
\caption{Effective circuits for transitions to the FSC phase in the phase slip picture. Phase slips correspond to a voltage source across the corresponding junction.
(a) phase slip on junction 1; (b)  
phase slip on junction 2; (c) slip-anti-slip pair which corresponds to slipping the phase of the grain relative to both the leads.  
\label{Figure10} }
\end{figure}

\begin{figure}
\includegraphics[width=8.5cm]{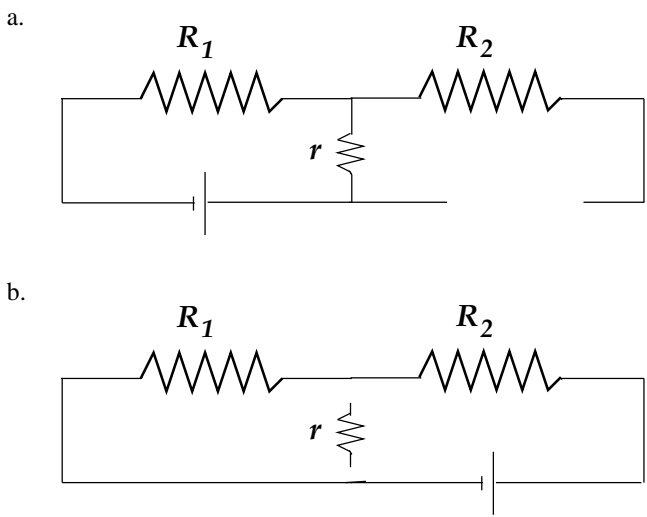}
\caption{Effective circuits for transitions to the NOR phase in the phase slip picture. 
(a) phase slip on junction 1 when junction 2 is insulating 
($\zeta_2$ is relevant); (b)  phase slip on junction 1 (or 
2) when $\zeta_-$ is relevant and slip-anti-slip pairs proliferate.  
\label{Figure11} }
\end{figure}

\begin{itemize}

\item The FSC to N1-S2 transition is determined by the circuit in Fig. \ref{Figure10}(a).  
In this case junction two can be replaced by a short as
it is  superconducting on both sides of the transition. This gives an effective
shunting resistance $R_1+r R_2/(R_2+r)$ for the phase slip, so the transition occurs at
$R_1+r R_2/(R_2+r)=R_Q$. Similarly, the transition between  S1-N2 and FSC
is determined by the circuit in Fig. \ref{Figure10}(b), with the effective shunting 
resistance at the transition being $R_2+r R_1/(R_1+r)=R_Q$.

\item To understand the FSC to $SC^{\star}$ transition we need to consider a dipole consisting of a QPS on junction one and a simultaneous anti-QPS on junction two,
corresponding to a $2\pi$ phase twist on the intervening grain. In particular, we need to know
how much charge flows from the super electrons on the grain to the normal electrons on the grain 
during such a phase twist. An equivalent circuit is shown in Fig. \ref{Figure10}(c), and we conclude that 
the phase boundary should occur at $r+R_1R_2/(R_1+R_2)=R_Q$, as the charge must flow through 
$r$ and either $R_1$ or $R_2$.

\item The transition between S1-N2 and the NOR phase is determined by the
relevance of the QPS on junction one when junction two is
insulating. The corresponding circuit is shown in Fig. \ref{Figure11}(a); since the
effective shunting resistance is $R_1+r$, we find a phase boundary at $R_1+r=R_Q$. 
Similarly, the N1-S2 to NOR transition is at $R_2+r=R_Q$.

\item The transition between $SC^{\star}$ and NOR is determined by the effective 
circuit in Fig. \ref{Figure11}(b). 
In the $SC^{\star}$ phase the component that is incoherent with the rest of the system is the grain. 
Since phase coherence between the leads is maintained, 
charge can flow freely from lead to lead via virtual super-conducting electrons on the grain unhindered by the phase fluctuations on the grain. But if 
some charge flows through $r$ to the normal electrons on the grain, this current will couple to the 
phase-slip dipoles and induce a large voltage drop; hence $r$  
becomes effectively a disconnect in the $SC^{\star}$ phase.  
The destruction of lead-to-lead superconductivity that characterizes the $SC^\star$ to NOR transition thus occurs at $R_1+R_2 = R_Q$.

\item The FSC to NOR line is naively a continuation of the S1-N2  and N1-S2 
lines.  This suggests that to approach the transition line from the
superconducting side, when we consider a QPS in junction one we
assume junction two to be superconducting and vice versa. This highly questionable approximation reflects the limitations of
our strong coupling analysis for the FSC to NOR transition; we will analyze it more carefully below.

\end{itemize}

\section{Intermediate coupling fixed point \label{nfp} }

In the previous two sections we have analysed the zero temperature states and transitions between them in both the weak and the strong Josephson coupling limits.  In both cases, we found that there were some regimes that could not be adequately analysed.  In this section analyze the intermediate coupling behavior, finding that transitions occur whose locations and properties are not given 
correctly by either the weak or strong coupling approaches.

A comparison of Fig. \ref{Figure5} and Fig. \ref{Figure9} 
reveals that there is a difference between
weak and strong coupling phase diagrams for $r<R_Q$.  In particular, the inferred phase boundaries
between the FSC and NOR phases differ in these two limits. This transition is special in that {\it both} junctions go from superconducting to normal,
but the transition is driven by the dynamics of just one of them (note that
there was no direct FSC to NOR phase boundary in the naive weak-coupling phase
diagram in Fig. \ref{Figure4}). In the weak coupling limit, when we analyzed 
the superconductor to
normal transition of junction one, our underlying assumption
was that junction two was normal. By contrast, for the same transition in the strong coupling case,
junction two was assumed to be effectively superconducting. This distinction between the approximate descriptions
accounts for the difference in inferred phase diagrams.  What is the actual behavior in this regime?  Does it, in contrast to the other regimes, depend on the magnitudes of the Josephson couplings as well as the resistances?

In Fig. \ref{Figure12} we indicate parts of the phase diagram for which weak
and strong coupling analyses suggest different natures of the ground
state. These regimes of the resistances would be fully superconducting (FSC) in the strong coupling  approximation and
normal (NOR) in the weak coupling approximation: the FSC fixed manifold is stable to small fugacities of the phase slips, and the NOR fixed manifold is stable to small Josephson couplings.  This suggests that in  such regimes,
there should be a transition from NOR to FSC as the $J$s are varied at a finite non-zero value of the Josephson couplings.  Specifically, if an appropriate combination of the Josephson couplings is
greater than some (resistance dependent) critical value  then the system will be in the FSC state, while if this combination is
 less than the critical value, the system will be in the normal state. As such a transition is presumably controlled by an intermediate coupling fixed point, it will have very different character than the other transitions; from now on we will refer to regimes in which such critical
fixed points  occur as  simply {\it intermediate } regimes.

It is useful to
remember that the original microscopic model had $J_+=0$, so for fixed resistances in the intermediate regime, on the
$J_1,\,J_2$ plane there will be a  manifold below which the
system flows to the normal fixed-point, and above which it flows to
the FSC fixed-point; this is the critical manifold
of the FSC to NOR transition. 
Alternatively, the microscopic model could be  defined in terms of the phase slip fugacities, 
$\zeta_1,\,\zeta_2$ with $\zeta_-=0$. For fixed resistances in the intermediate regime, the critical manifold would
show up here too, separating the FSC and the
normal phase in, for fixed resistances, the $\zeta_1,\ \zeta_2$ plane.

In general, an analysis of the critical behavior in the intermediate regime is beyond the methods of this paper, but we can make use of the weak and strong coupling limits to analyze {\it parts} of this regime: specifically, when the critical values of either the Josephson couplings or the QPS fugacities, respectively, are small.

\subsection{Weak Coupling Limit}

We first study the weak coupling limit. In order to find the critical values of $J_1,\,J_2,\,J_+$ in the
intermediate regime, we need to analyze the effects of the non-linear terms in  the RG flow equations
(\ref{Jflowsecond}), and, if there is indeed a perturbatively accessible critical fixed point, find it and the corresponding critical manifold.  
Truncating at second order, we indeed find a fixed point:
\be \ba{c}
(J^*_1)^2=\frac{(R_2+r-R_Q)(R_1+R_2-R_Q)}{rR_1} \vspace{2mm}\\
(J^*_2)^2=\frac{(R_1+r-R_Q)(R_1+R_2-R_Q)}{rR_2} \vspace{2mm}\\
(J^*_+)^2=\frac{(R_2+r-R_Q)(R_1+r-r_Q)}{R_2R_1},
\label{fpJ}
\ea 
\ee
with an overall cutoff-dependent proportionality coefficient having been set equal to unity when the RG equations were first derived.   
As we see below, this fixed point can be shown to be critical provided each of the three resistance combinations in parentheses are positive. These factors, which we will call 
\be
u\equiv R_2+r-R_Q \qquad v\equiv R_1+r-R_Q \qquad w\equiv R_1+R_2-R_Q
\ee
are the negatives of the eigenvalues of the three couplings along the normal fixed manifold so that the normal phase is stable to small $J$s in this regime as indicated by the weak coupling phase diagram. 
Naively, one might have expected the non-linear perturbative analysis to be valid only when all three of these eigenvalues are small, but we see that in fact all that is needed is  {\it two of the three eigenvalues small and negative} with the third being arbitrarily negative. Correspondingly, we require that all three of $u,\ v,\ w$ are positive with two of them being small.
 
By rescaling the $J$s appropriately, the RG flows can be put in a simple symmetric form in terms of $u,\ v,\ w$, and the fixed point values written as 
\be
J_1^\ast= \sqrt{\frac{uw}{rR_1}} \qquad J_2^\ast= \sqrt{\frac{vw}{rR_2}} \qquad J_+^\ast=\sqrt{\frac{uv}{R_1R_2}}.
\ee

The linearized flows around this intermediate coupling fixed point yield the eigenvalues which are given by
\be 
\lambda_i\approx\Lambda_i(u+v+w)
\label{ev1}
\ee
with the $\{\Lambda_i\}$ being the three roots of 
\be
\Lambda^3+\Lambda^2=m
\ee
in terms of  the dimensionless combination of the resistances
\be
m\equiv \frac{4uvw}{(u+v+w)^3} \ .
\label{ev2}
\ee
We see immediately that for $m$ positive, as it must be, there is always a unique positive eigenvalue, $\lambda_+$, which controls the growth of deviations from the critical manifold; the two others have negative real parts and are hence irrelevant at the intermediate coupling critical fixed point. Note that if only two of $u,\ v,\ w$ are small, with, say $w$ being much larger than the other two, then $\Lambda_+\approx \sqrt{\frac{4uv}{w^2}}\ll 1$ so that $\lambda_+\approx 2\sqrt{uv}$. If all three are small and comparable, $\lambda_+$ will be of the same order but depend in a somewhat complicated way on their ratios.  

\vspace{3mm}

 \subsection{Strong Coupling Limit}
 
It is clear by examining the limits of validity of the weak coupling expansion above that we cannot extract the critical behavior throughout the intermediate regime from this analysis.  Fortunately, we can access another part of this regime from the strong coupling direction.

Using the second order RG flows in terms of the fugacities of phase slips, we find a critical fixed point at  
\begin{widetext}
\be \ba{c}
(\zeta^*_1)^2=\l(\frac{1}{R_Q}-\frac{R_2+R_1}{Y}\r)\l(\frac{1}{R_Q}-\frac{r+R_1}{Y}\r)
\frac{Q^2}{rR_2} \vspace{2mm}\\
(\zeta^*_2)^2=\l(\frac{1}{R_Q}-\frac{R_2+R_1}{Y}\r)\l(\frac{1}{R_Q}-\frac{r+R_2}{Y}\r)
\frac{Q^2}{rR_1} \vspace{2mm}\\
(\zeta^*_-)^2=\l(\frac{1}{R_Q}-\frac{R_2+r}{Y}\r)\l(\frac{1}{R_Q}-\frac{r+R_1}{Y}\r)
\frac{Q^2}{R_1 R_2} \vspace{2mm}\\
\label{fpz} \ea 
\ee 
\end{widetext}
with $Y=r(R_1+R_2)+R_1R_2 $. As for weak coupling, it is convenient to work in terms of the negatives of the eigenvalues of the three fugacities about the FSC fixed manifold, defining
\be
\ba{c} 
\overline{u}\equiv \frac{R_2+r}{Y}-R_Q \vspace{2mm}\\ 
\overline{v}\equiv\frac{R_1+r}{Y}-R_Q  \vspace{2mm}\\
\overline{w}\equiv\frac{R_1+R_2}{Y}-R_Q
\ea
\ee
with the condition for the validity of the expansion being that all these must be positive with at least two of them small.   The expansion is carried out in exactly the same manner as for the weak coupling limit, and the eigenvalues about the intermediate coupling critical fixed point determined by exactly the same conditions as in Eqs. (\ref{ev1}-\ref{ev2}), with simply $u,\ v,\ w$ replaced by their (overbared) strong coupling equivalents.

\begin{figure*}
\includegraphics[width=16.5cm]{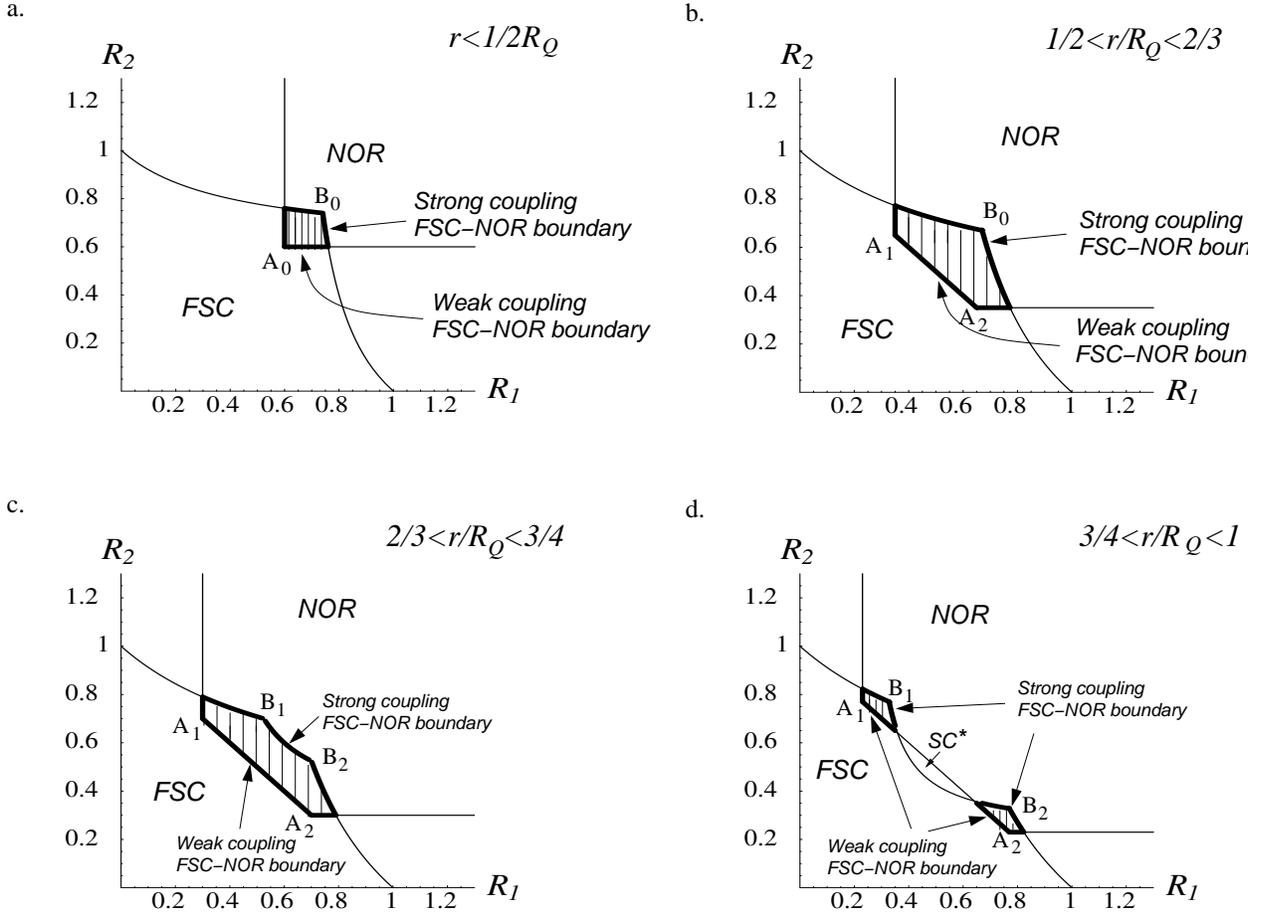}
\caption{ Intermediate coupling fixed point regions of the phase diagram. The shaded regions surrounded by a bold line lie in a superconducting phase for strong-coupling 
and in an insulating phase for weak-coupling. For $r>R_Q$ 
there are no such regions. For intermediate $J$ the 
 phase boundaries will be in the shaded regions. 
The points $A_0,\,A_1,\,A_2$ mark where the critical fixed point, $J^*$ goes to zero, and the points $B_0,\,B_1,\,B_2$ 
mark where $\zeta^*$ goes to zero corresponding to $J^\ast\to \infty$. Near these multicritical points the RG analyses in the text becomes exact. 
\label{Figure12} }
\end{figure*}

\begin{figure*}
\includegraphics[width=16.5cm]{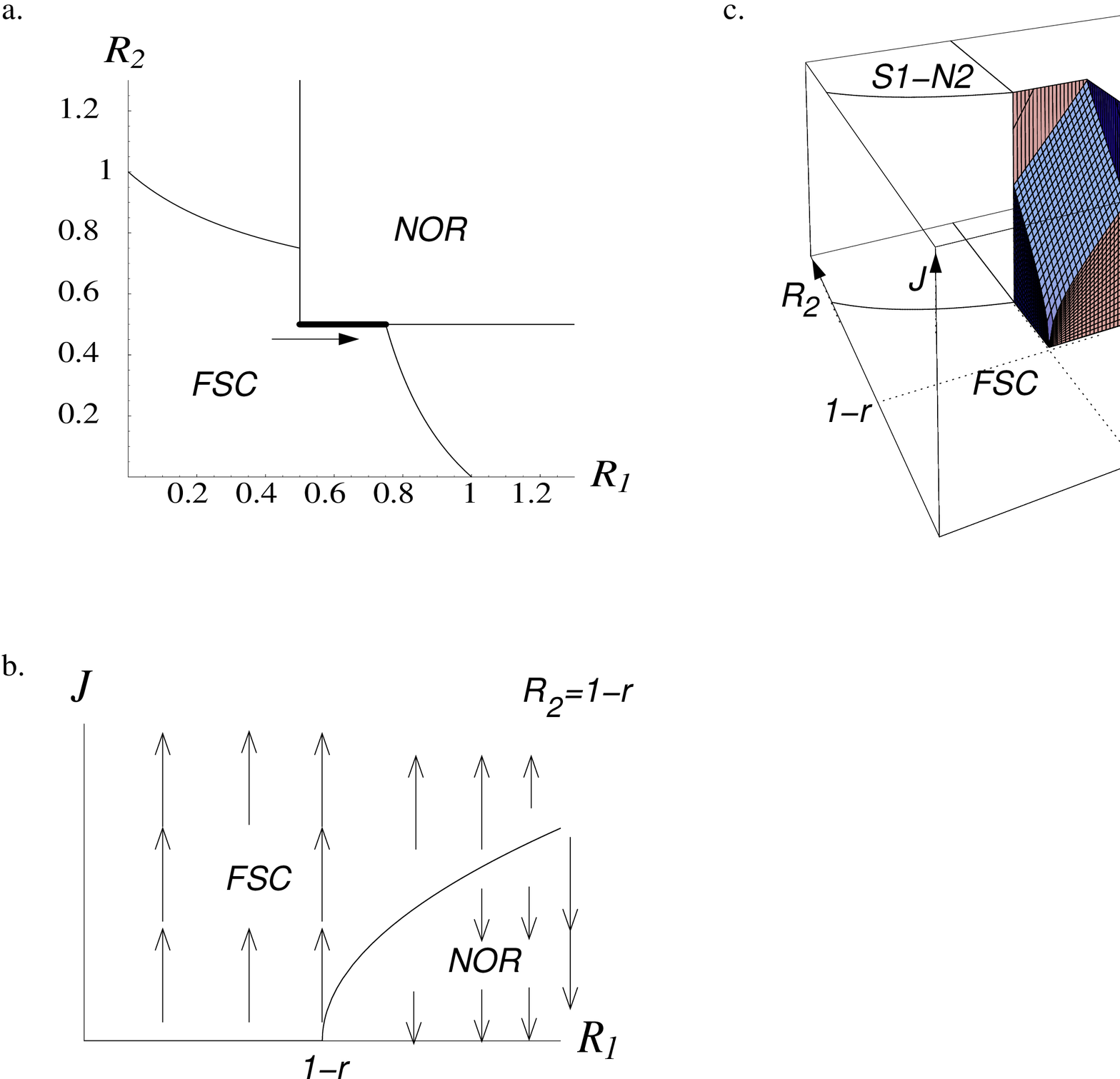}
\caption{Example of phase diagram in vicinity of a transition between the 
FSC and NOR phases for fixed $r<1/2$. The critical manifold in the intermediate regime depends on 
the Josephson-coupling strengths. (a) Small $J$ phase diagram in $R_1,\ R_2$ plane showing FSC - NOR $J=0$
transition line (bold) at $R_2=1-r$. (b) Schematic cross-section of phase diagram along line with $R_2=1-r$ showing the jump in $J_c$ suggested by  the truncated second order RG analysis for crossing the  phase boundary 
from $R_2<1-r$ to $R_2>1-r$. The arrows indicate the RG flow 
of the Josephson couplings. Higher order terms in RG flows are likely to drive the critical $J_c$ to zero on the line $R_2=1-r$ (c) Three-dimensional view of the phase diagram, 
focusing on the FSC - NOR transition. The solid lines in the x-y plane 
mark the phase boundary between the mixed phases and the insulating and the FSC 
phases. These phase boundaries are independent of $J$.      
\label{Figure13} }
\end{figure*}

\subsection{Superconducting-normal critical manifold}
From the above discussion we see that direct transitions between the FSC and NOR phases
will always be controlled by intermediate coupling fixed points. Although we thus cannot find the full phase boundary exactly in the intermediate region of the resistance space, we can use the weak and strong coupling analysis to find it in some regimes of the intermediate region. 
Eqs. (\ref{fpJ}, \ref{fpz})  apply   in the weak and strong Josephson coupling limits, respectively so that  we can  locate the phase boundaries accurately in the intermediate region  from the flow equations   
provided that both the bare and the fixed point values of the Josephson couplings are either all large or all small.  In particular, we have found that the $J^*$ go to zero along certain lines in the $r,\  R_1,\ R_2$ space which intersect the constant $r$ surfaces shown in Fig. \ref{Figure12} at the  points $A_0$, $A_1$, $A_2$; our
weak coupling analysis is controlled in their
vicinity providing the bare $J$s are small.  Analogously, the fixed point values $\zeta_i^*$ vanish at points $B_0$, $B_1$, $B_2$ of the constant $r$ surfaces as 
shown in Fig. \ref{Figure12} and the strong coupling analysis is controlled in their vicinity provided the bare $J$s are large . 

The finite values of the  $J^*$'s at the fixed point on 
the critical lines has interesting implications for
the phase boundaries in the full $R$s and $J$s parameter space as sketched in Fig. \ref{Figure13}. If we cross from 
FSC to NOR phase by changing resistances and keeping
$J$s fixed, the exact location of the transition will
generally depend on the values of the $J$s.  However, there is 
a whole range of small $J$s (which we can schematically
denote as $0< J < {J}^*$) for which, in the second order RG approximation, this transition
occurs exactly at the FSC to intermediate region boundary; if we consider higher order terms in the RG, the location of the  transition in this range 
will be modified slightly.
Analogously
there is a range of large $J$'s for which the FSC to NOR
transition happens very close to the intermediate region to NOR line
(in strong coupling this occurs for $0< \zeta < {\zeta}^*$).

\begin{figure}
\includegraphics[width=8.5cm]{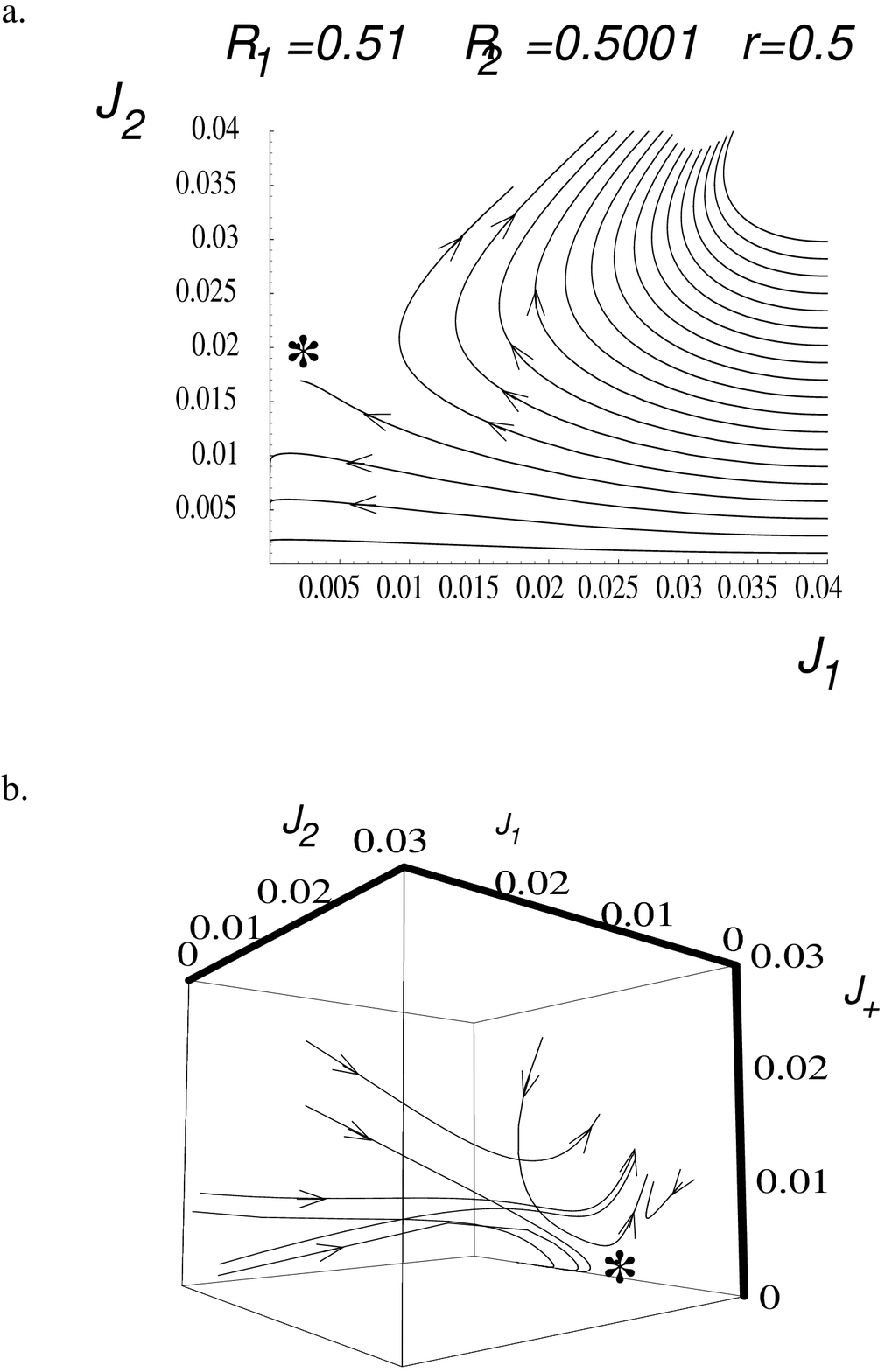}
\caption{RG flows in the intermediate region with  $R_1=0.51,\,R_2=0.5001,\,r=0.5$. Note the 
typical flow pattern in the vicinity of the unstable critical fixed point 
marked by an asterix.  (a) Projection of the RG flow trajectories on the $J_1,\,J_2$ plane. (b) A 3D flow diagram for near-critical trajectories. 
\label{Figure14} }
\end{figure}

For illustrative purposes we calculate explicitly
the phase boundary as a function of weak $J_{1,2}$ in the part of the intermediate regime of resistances in which the fixed point is at small but non-zero coupling.  In particular, we consider the  FSC to NOR transition  for: 
\be
\ba{ccc}
r<0.5, 	& 	R_2=1-r+u,		&	R_1=1-r+v
\ea
\label{iline}
\ee
with $u$ and $v$ small, and for convenience, we set $R_Q=1$ for this section. The third parameter, 
\be
w=R_1+R_2-1=1-2r+u+v\approx 1-2r
\ee
is generally {\it not} small.  It is convenient to define rescaled couplings by
\be
K_1\equiv \sqrt{\frac{r(1-r)}{1-2r}}J_1 \qquad K_2\equiv \sqrt{\frac{r(1-r)}{1-2r}}J_2
\ee
which have fixed point values $K_1^\ast\approx \sqrt{u}$ and $K_2^\ast\approx \sqrt{v}$.  From the RG flow equations, it can be seen that $J_+$ rapidly approaches its nullcline value, $J_+^n(K_1,K_2)$, and then evolves slowly with the other variables. Substituting $J^n_+$ for $J_+$ in the flow equations for $K_{1,2}$, we can find the invariant manifold on which  the critical fixed point lies. This is parametrized by
\be
K_1^2-u[1+\ln(K_1^2/u)]\approx K_2^2-v[1+\ln(K_2^2/v)] 
\ee
which has two branches of solutions; the branch with one of $K_1$ or $K_2$ larger than its fixed point value and the other smaller is the desired critical manifold. Note that as $J_1$ increases above its fixed point value, the critical value of $J_2$ decreases exponentially, and visa versa.  Although we have taken the bare $J_+=0$, even a $J_+$ of order the fixed point values of the other $J$s will not appreciably change their critical values in this regime with $w\gg u,v$. 

Similar analysis can be done with either of the other pairs, $u,\ w$ or $v,\ w$ both small and the third of order unity. In these cases, however, the smallness of the bare $J_+$ means that the early stages of the renormalization will give rise to a  non-zero value of $J_+$ at  intermediate scales whose value is needed to estimate the critical condition that relates the other $J$s. 

{\it Symmetric case.}  Although unrealistic for the physical model of two junctions, it is instructive to consider the case in which there is a symmetry between the three superconducting components and the Josephson couplings linking them.  In this case we take
\be
r=R_1=R_2=R \qquad {\rm and} \qquad J_+=J_1=J_2=J
\ee
and the RG flow equations become simply
\be
\frac{dJ}{d\ell}\approx J(1-2R)+RJ^2
\ee
with 
\be
w=u=v=2R-1
\ee
so that the weak coupling part of the intermediate region occurs for $R$ slightly bigger than $\frac{1}{2}R_Q$.  The critical value of $J$ is then simply
\be
J_c \approx J^\ast\approx 4(R-\frac{1}{2})
\ee
and the RG eigenvalue controlling flows away from this is
\be
\lambda \approx 2(R-\frac{1}{2}) \ .
\ee
In the strong coupling limit, we can similarly use a single QPS fugacity $\zeta$ and write
\be
\frac{d\zeta}{d\ell}\approx \zeta(1-\frac{2}{3R})+\frac{1}{R}\zeta^2
\ee
so that the intermediate region occurs for
\be
\frac{1}{2}<R<\frac{2}{3} \ .
\ee
Near the upper end of this range, $R$ slightly less than $\frac{2}{3}$, the critical value of the fugacity is small, and the RG eigenvalue for deviations from criticality becomes
\be
\lambda\approx \frac{3}{2}(\frac{2}{3}-R) \ .
\ee
 Comparing the two limiting expressions for 
$\lambda$, we see that, for the symmetric case, it is unlikely to get above a small value of order $0.2$ anywhere in the intermediate region.

\section{Symmetries of the two-junction system \label{dualities} }

From the microscopic model of Fig. \ref{Figure2},
the only obvious symmetry --- more properly a simple duality --- is the exchange of the two
junctions, $R_1\leftrightarrow R_2$ and $J_1\leftrightarrow J_2$.
Analysis presented in this section uncovers
additional symmetries in the phase diagram of the system at zero temperature; indeed, in the  analysis of the previous section we have already seen evidence of these. 
Here we will show more generally that the junction interchange is only
one part of a larger permutation symmetry, or triality, that involves the interchange
of all resistors $r$, $R_1$, and $R_2$ and the corresponding Josephson couplings. We also
show how the familiar weak to strong coupling duality
of a single shunted Josephson junction \cite{fisher-zwerger, schoen-zaikin} can be generalized
to the two-junction system. These symmetries allow one to relate in a non-trivial
way many of the phase boundaries shown in Fig. \ref{Figure12}.

\subsection{Permutation triality}

The two-junction system exhibits three normal phases and two
superconducting ones. The simplest insulating phase involves proliferations of all three kinds of phase slips.  Conversely, the simplest superconducting
phase, the fully-superconducting one, FSC, has none of the phase slips proliferating.
Of the three remaining phases, two are normal as far as inter-lead properties are concerned, because of phase slips
that proliferate in {\it one} of the two junctions. The last phase is the
$SC^{\star}$ phase, which is superconducting because it exhibits
dissipationless lead to lead transport due to Cooper pair cotunneling
processes.  This phase, however, also has signatures of normal phases, in particular
localized charges on the middle grain and
the proliferation of QPS - anti-QPS pairs that decouple the phase of this grain from the linked superconductivity of the two leads. 

An alternative way to group the five phases is thus as
one purely normal phase; one
purely superconducting phase; and three mixed phases, in which part of
the system is normal and part is superconducting. In terms of phase-slip fugacities, these correspond respectively to one
phase in which all fugacities grow under the RG transformation, one
phase in which all fugacities renormalize to zero, and three phases in
which only one of the fugacities, $\zeta_1$, $\zeta_2$, or $\zeta_-$, grows
under RG, while the remaining two renormalize to zero.
Such grouping is very suggestive of a permutation symmetry of the full phase diagram in  which
the phases N1-S2, S1-N2, and $SC^{\star}$ are transformed into each other, and phases
FSC and NOR are invariant. In this section we show that such triality
is indeed present in the low energy properties of the microscopic models
(\ref{lowTaction}) and (\ref{psaction}) describing the system. Note that other  systems possessing
triality have been discussed earlier by R. Shankar in Ref. \cite{shankar}; as in our case, these are non-trivial in some representations but easy to see in others. 

To demonstrate the triality in the original quantum action
we consider the strong coupling representation of equation (\ref{psaction}),
although equivalent arguments can be made for the weak coupling representation described 
of equation (\ref{lowTaction}).
Let us begin with the mathematical formulation of this symmetry.

The action in (\ref{psaction}) reads:
\be
\ba{c}
Z=
\int D[\theta_1]\int D[\theta_2]\exp\left(- \beta
\sum\limits_{\omega_n} |\omega_n| 
\vec{\theta}_{-\omega_n}^T \hat{R}\vec{\theta}_{\omega_n}+\r.\vspace{2mm}\\\l.
\int_0^\beta d\tau\left(\zeta_1\cos(\theta_1)+\zeta_2\cos(\theta_2)+\zeta_-\cos\left(\theta_1-\theta_2\right)\right)\right)
\ea
\label{psaction1}
\ee
where  the resistance matrix is
\be
\hat{R}= \left( \begin{array}{cc}
r+R_1 & -r\vspace{2mm}\\
-r & r+R_2
\end{array} \right)
\label{symaction}
\ee
and the vector $\vec{\theta}$ has components $\theta_{1,2}$. An interchange of the two junctions, $R_1\leftrightarrow R_2$ and $\zeta_1
\leftrightarrow \zeta_2$, will leave the phase diagram invariant, 
exchanging the two mixed states in which one junction is
superconducting and the other is normal (N1-S2 and S1-N2). In
(\ref{symaction}) this interchange of
junctions corresponds to transforming the fields, $\theta_1\leftrightarrow
\theta_2$ or: \be \left( \ba{c} \theta_1\vspace{2mm}\\ \theta_2 \ea \right)= \l(
\ba{cc} 0 & 1\vspace{2mm}\\ 1 & 0 \ea \r) \left( \ba{c} \theta'_1\vspace{2mm}\\ \theta'_2 \ea
\r)
\label{lin1}
\ee											
In terms of the new variables $\vec{\theta}'=\hat{S}^{-1}\vec{\theta}$:
\be
\ba{c}
Z=
\int D[\theta_1']\int D[\theta_2']\exp\left(- \beta
\sum\limits_{\omega_n} |\omega_n| 
\vec{\theta}_{-\omega_n}^{~'T} \hat{R}'\vec{\theta}'_{\omega_n}+\r.\vspace{2mm}\\\l.
\int_0^\beta d\tau\left(\zeta_1\cos(\theta'_1)+\zeta_2\cos(\theta'_2)+\zeta_-\cos\left(\theta'_1-\theta'_2\right)\right)\right)
\ea
\label{psactionprime}
\ee
where  
\be
\ba{rcl}
\hat{R}' & = & \hat{S}^T \hat{R} \hat{S} = 
\l( \ba{cc}
							r+R_2 	&	-r\vspace{2mm}\\
							-r	&	r+R_1  
								\ea			\r)
\vspace{2mm}\\
\zeta'_1 &=&\zeta_2 \hspace{1cm} \zeta'_2 =\zeta_1
\hspace{1cm} \zeta'_- =\zeta_-
\label{prime1}
\ea
\ee
This new action (\ref{psactionprime}), (\ref{prime1}),  has $R_1\leftrightarrow R_2$
and $\zeta_1\leftrightarrow \zeta_2$ but otherwise
exactly the same physics with simply relabeling the fields $\theta_i$.

A less trivial symmetry involves the transformation 
\be \left(
\ba{c} \theta_1\vspace{2mm}\\ \theta_2 \ea \right)= \l( \ba{cc} 1 & 0\vspace{2mm}\\ 1 & -1 \ea
\r) \left( \ba{c} \theta'_1\vspace{2mm}\\ \theta'_2 \ea \r)
\label{lin2}
\ee									
leading to the action (\ref{psactionprime}) with
\be
\ba{rcl}
\hat{R}'&=& \l( \ba{cc} R_1+R_2 & -R_2\vspace{2mm}\\ -R_2 & r+R_2 \ea \r)
\nonumber\vspace{2mm}\\ \zeta'_1 &=&\zeta_1 \hspace{1cm} \zeta'_2 =\zeta_-
\hspace{1cm} \zeta'_- =\zeta_2
\label{prime2}
\ea
\ee
This new symmetry is surprising as it swaps $R_2$ with $r$. 
One way of understanding this  is 
as a change of basis for the quantum phase slips.
Earlier we took QPS on junctions one and two as a basis (schematically,
we can label them as $(1,0)$ and $(0,1)$) and considered a QPS
dipole as their composite: $(1,-1)=(1,0)+(0,-1)$. An equivalent
basis set, however, can be obtained by taking one of the QPS and the dipole as the basic objects,
and viewing the other QPS as their composite
e.g., $(0,1)=(1,0)+(-1,1)$. The corresponding transformation (\ref{lin2}) maps
phases S1-N2 and $SC^{\star}$ into each other, while leaving the other
ones intact.
\newline

Using  transformations (\ref{lin1}) and (\ref{lin2}) one can construct 
transformations that permute {\it any} of the three resistances
and connect any of the phases N1-S2, S1-N2, and $SC^{\star}$.
The physical basis of this  symmetry
follows from the observation  that the circuits corresponding to the
three kinds of phase-slips are similar; one resistor is connected in
series to the two other resistors, which are connected in parallel. (The strong 
coupling representation we are using here implies starting from the FSC phase as in 
Section IVc).  From the circuit diagrams
in Fig. \ref{Figure10} we see the origin of the permutation symmetry:
circuits associated with all three kinds of phase slips
differ only in the exchange of resistors. The strong coupling permutation
triality thus generally corresponds to
\be
\ba{cc}
\ba{c}
\zeta'_i=\zeta_{\pi_{(i)}}\vspace{2mm}\\
R_i=R_{\pi_{(i)}}
\ea
&	{\rm with} \qquad i=1,2,-
\ea
\ee
where we have paired the fugacities with the corresponding resistance, so that $R_+=r$, and $\pi$ is a permutation of the three indices.

In the weak coupling regime, the nature of the triality  is
the same: the  circuits corresponding to the three Cooper pair
tunneling
events are similar with two resistors in series and a third taken out of
the circuit. (Use of the weak coupling representation implies starting from the NOR phase, 
see Sec. III(c).) If we now pair the
Josephson couplings with the corresponding missing resistor in the equivalent circuits,   $r_1=R_2,\,r_2=R_1$, and $r_+=r$, the permutation symmetry in the weak-coupling limit becomes 
\be
\ba{cc} \ba{c} J'_i=J_{\pi_{(i)}}\vspace{2mm}\\ r'_i=r_{\pi_{(i)}} \ea & {\rm with} \qquad
i=1,2,+ \ea \ee
where $\pi$ is again a permutation.

%----------------------------------weak-strong duality

\subsection{Weak to strong coupling duality}

The similar form of the strong-coupling and weak-coupling representations of the quantum actions (\ref{lowTaction}) and
(\ref{psaction}) suggests that there is a duality between the two regimes.

The duality we find is a generalization of that  of a single resistively shunted Josephson
junction (see {\it e.g.} \cite{weiss-grabert}).  
For the single
junction the duality is equivalent to the observation
that quantum phase slips in a junction with shunt resistance $R$ behave
similarly,  as far as their quantum statistical-mechanics, to 
Cooper pair tunneling events in a junction with shunt resistance $\tilde{R}=R_Q^2/R$.  
In
the two junction problem discussed in this paper we expect  that
Cooper pair tunneling events across any of the junctions in weak
coupling should be dual to quantum phase slips on the same junction in strong
coupling; and Cooper pair cotunneling processes across the two junctions should be dual to QPS
dipoles on the two junctions. But a complication is that the effective
resistance for a Cooper pair tunneling event (or a QPS) in one of the
junctions depends on the state of the other junction (see Secs. III(c)
and IV(c)).  

The duality transformation maps Cooper pairs into QPS and
superconducting phases into normal ones. Hence, when we discuss the
duality between Cooper tunneling events and QPS on any given
junction, we need the duality transformation to
change the  {\it state} of the other
junction. For example, consider a Cooper pair tunneling through
junction one with junction two normal. The dual
of this will be a QPS on junction one, with junction
two superconducting. Comparison of the effective shunting resistances
in the two cases immediately gives the duality relation
\be
\tilde{R}_1+\tilde{r}=\frac{R_Q^2}{R_1+\frac{rR_2}{r+R_2}}
\label{ws_duality1}
\ee
Analogous arguments give
\be
\ba{c}
\tilde{R}_2+\tilde{r}=\frac{R_Q^2}{R_2+\frac{rR_1}{r+R_1}}\vspace{2mm}\\
\tilde{R}_1+R_2'=\frac{R_Q^2}{r+\frac{R_1 R_2}{R_1+R_2}}.
\label{ws_duality2}
\ea
\ee

An alternative way of seeing the duality is to
take $\theta_1 \rightarrow \Delta_1'$ and 
$\theta_2 \rightarrow - \Delta_2'$
in action (\ref{psaction}). The cosine terms
of the resulting action in terms of $(\Delta_1',\Delta_2')$ 
and those of the weak coupling action (\ref{lowTaction}) 
then have the same form. If we compare the quadratic terms in these actions,
we find the same duality relations (\ref{ws_duality1})
and (\ref{ws_duality2}).

\begin{figure}
\includegraphics[width=8.5cm]{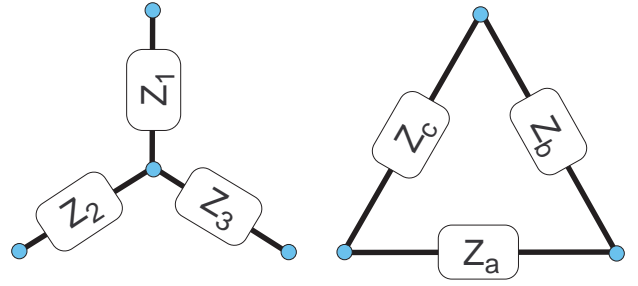} 
\caption{ \label{fg:DY} $Y \leftrightarrow \Delta$ transformation. 
 ``Y'' resistor network on left is mapped to $\Delta$ network on  right via $Z_1 Z_a=Z_2 Z_b=Z_3 Z_c =Z_1 Z_2 +Z_2 Z_3 + Z_3 Z_1$. The inverse transformation ($\Delta\rightarrow Y$) is  ${Z_b Z_c}/{Z_1}={Z_c Z_a}/{Z_2}={Z_a Z_b}/{Z_3}=Z_a+Z_b+Z_c $.}
\end{figure}

\begin{figure}
\includegraphics[width=8.5cm]{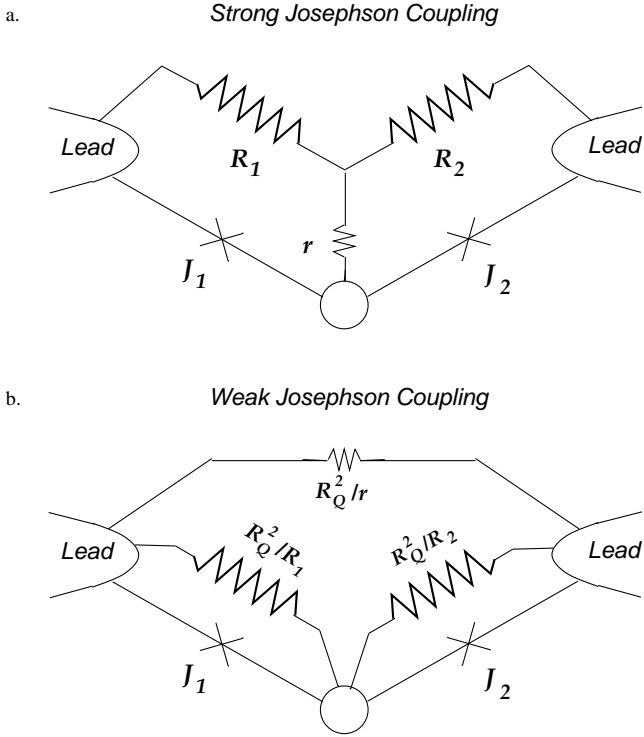} 
\caption{(a) Original circuit of Fig. \ref{Figure2} in  weak coupling limit showing a ``Y'' resistor network . 
(b) Strong coupling dual of the circuit showing a ``$\Delta$'' shaped network. The network in (b) captures the duality Eqs. (\ref{wsdual}). \label{Ydelta2}}
\end{figure}

We can solve the duality relations (\ref{ws_duality1}) and (\ref{ws_duality2}) 
for $\tilde{r},\,\tilde{R}_1,\,\tilde{R}_2$:
\be
\ba{c}
\tilde{r}=R_Q^2\frac{r}{Y}\vspace{2mm}\\
\tilde{R}_1=R_Q^2\frac{R_2}{Y}\vspace{2mm}\\
\tilde{R}_2=R_Q^2\frac{R_1}{Y}
\label{wsdual}
\ea
\ee
with $Y\equiv rR_1+rR_2+R_1R_2$. This mapping of the resistors to dual resistors may seem rather unintuitive, however 
Eqs. (\ref{wsdual}) coincides with the well known ``Y-$\Delta$" transformation of resistor networks. 
The Y-$\Delta$ transformation is depicted in Fig. \ref{fg:DY}. By comparing the Y-$\Delta$ transformation equations
in Fig. \ref{fg:DY} we see that the duality transforms the system in Fig. \ref{Ydelta2}(a) to the 
system in Fig. \ref{Ydelta2}(b). In  Fig. \ref{Ydelta2}(a) the resistors $R_1,\,R_2$ and $r$ are connected in a 
``Y'' pattern; the transformed system has the resistances $R_Q^2/R_2,\,R_Q^2/R_1$ and $R_Q^2/r$ connected a $\Delta$
pattern. 

This statement of the duality is simple; pair-tunneling events (current sources) with a Y resistance network and 
resistors $r,\,R_1,\,R_2$ (Fig. \ref{Ydelta2}(a)) are dual to quantum phase-slips (voltage sources) with a $\Delta$ network 
of resistances $R_Q^2/r,\,R_Q^2/R_1,\,R_Q^2/R_2$ (Fig. \ref{Ydelta2}(b)). This is a simple generalization 
of the single junction duality. From Fig. \ref{Ydelta2} we see that as $r\rightarrow 0$ the duality 
reduces to:
\[
\ba{c}
\tilde{r}=0\vspace{2mm}\\
\tilde{R}_1=\frac{R_Q^2}{R_1}\vspace{2mm}\\
\tilde{R}_2=\frac{R_Q^2}{R_2}
\ea
\]
which is simply the duality of a single junction applied to the two uncoupled junctions, 
as should be 
expected in this limit in which the middle grain is macroscopic.  
\begin{figure}
\includegraphics[width=8.5cm]{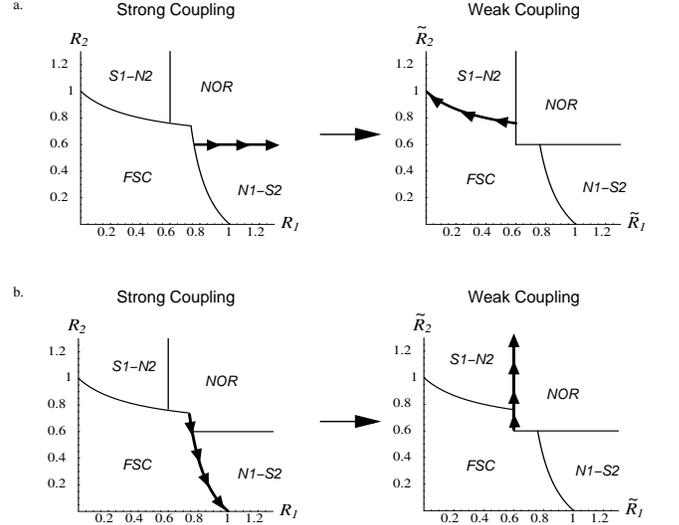}
\caption{Weak --- strong duality (a) Mapping of the strong-coupling critical 
line $R_2+r=1$ to the weak-coupling regime. (b)  Mapping of the strong-coupling 
critical line $R_1+\frac{R_2 r}{R_2+r}=1$ to the weak-coupling regime.  
\label{Figure15}}
\end{figure}

\subsection{Phase boundaries controlled by weak or strong coupling}

The weak-to-strong coupling duality yields a mapping between various of the phase boundaries
in Fig \ref{Figure12}. The nature of this mapping
is such that weak coupling transitions
will be mapped to strong coupling ones, e.g.
the NOR to N1-S2 boundary gets mapped into the FSC to S1-N2
boundary. Here NOR to N1-S2 corresponds to a weak
coupling transition, since it involves ordering of $\Delta_2$
with $\Delta_1$ remaining disordered on both sides of 
the transition, i.e., $J_2$ becomes
relevant, while $J_1$ and $J_+$ stay irrelevant.
By contrast S1-N2 to FSC is really a strong coupling
transition because it involves
$J_2$ becoming relevant
with $J_1$ already relevant.
This latter transition is simple 
in terms of the QPS fugacities, corresponding to $\zeta_2$ becoming relevant about the FSC manifold
with  $\zeta_1$ and $\zeta_-$ irrelevant on both
sides of the phase boundary.

First, we map the phase boundary $R_2+r=R_Q$  via 
(\ref{wsdual}). After substituting $r=R_Q-R_2$ this yields:
\be
\ba{c}
\tilde{r}=\frac{R_Q^2}{R_1+R_2+\frac{R_1 R_2}{R_Q-R_2}}=R_Q\frac{R_Q-R_2}{R_1+R_2-R_2^2/R_Q}\vspace{2mm}\\
\tilde{R}_1=\frac{R_Q^2}{R_Q+R_1-R_2+\frac{R_1 (R_Q-R_2)}{R_2}}=R_Q\frac{R_2}{R_1+R_2-R_2^2/R_Q}\vspace{2mm}\\
\tilde{R}_2=\frac{R_Q^2}{R_Q+\frac{R_2(R_Q-R_2)}{R_1}}=R_Q\frac{R_1}{R_1+R_2-R_2^2/R_Q}\vspace{2mm}\\
\ea
\ee
These apparently complicated expressions are simply the boundary of the FSC phase, since:
\be
\ba{c}
\frac{1+\frac{\tilde{R}_1}{\tilde{r}}}{\tilde{R}_1+\tilde{R}_2+\frac{\tilde{R}_1\tilde{R}_2}{\tilde{r}}}=\frac{1}{\tilde{R}_2+\frac{\tilde{r}\tilde{R}_1}{\tilde{r}+\tilde{R}_1}}=\vspace{2mm}\\
\frac{1}{R_Q}\frac{R_1+R_2-R_2^2/R_Q}{R_1+(R_Q-R_2)R_2/R_Q}=\frac{1}{R_Q},
\ea
\ee
as shown in Fig. \ref{Figure15}(a).
As a second example, consider the critical line $R_1+\frac{R_2 r}{R_2+r}=1$, which 
separates the FSC phase from the mixed phase in which junction one 
is normal. The duality equations yield:
\be
\ba{c}
\tilde{r}=\frac{R_Q^2}{R_1+R_2+\frac{R_1R_2(R_1+R_2-R_Q)}{R_2(R_Q-R_1)}}=R_Q\frac{R_Q-R_1}{R_2}\vspace{2mm}\\
\tilde{R}_1=\frac{R_Q^2}{R_1+\frac{R_2(R_Q-R_1)}{R_1+R_2-R_Q}+\frac{R_1R_2(R_Q-R_1)}{(R_2+R_1-R_Q)R_2}}=R_Q\frac{R_1+R_2-R_Q}{R_2}\vspace{2mm}\\
\tilde{R}_2=\frac{R_Q^2}{R_2+\frac{R_2(R_Q-R_1)}{R_1+R_2-R_Q}+\frac{R_2^2(R_Q-R_1)}{(R_2+R_1-R_Q)R_1}}=R_Q\frac{R_1(R_1+R_2-R_Q)}{R_2^2}\vspace{2mm}\\
\ea
\ee
so that
\be
\tilde{R}_1+\tilde{r}=R_Q
\ee
which is the condition for the phase boundary between the normal phase 
and the mixed phase in which junction one is superconducting, as in
Fig. \ref{Figure15}(b).

\subsection{Duality in the intermediate region}
 
In the intermediate regime of  the resistance parameter space, the behavior under duality is more complicated. Since the controlling critical fixed point that determines the fully normal to fully superconducting phase boundary is at non-zero Josephson coupling in this regime, the early stages of the renormalization will affect the {\it location} of the critical manifold in the full parameter space.  Thus duality cannot be used to locate the phase boundaries.  Nevertheless, duality is still useful in this intermediate region.   

The low energy properties of the system will be given by the effective actions that {\it do} exhibit duality. Thus universal properties near the transitions at pairs of points in resistance space should be dual even when the location of the transitions as functions of the Josephson couplings are not.  In particular, as we have seen in the explicit perturbative calculations of the critical behavior in the intermediate region in the regimes in which the critical fixed point is  at either very strong or very weak coupling, the critical exponents, such as  the RG eigenvalue $\lambda$ that controls deviations from criticality, will be universal functions of the resistances with values on twelve-member sets of points being the same by the duality and the three fold permutation symmetry.  

For the highly symmetric case, $R_1=R_2=r=R$, the duality is simply 
\be
\tilde{R}=\frac{R_Q^2}{3R}
\ee
so that there is a self-dual point at $R=\frac{1}{\sqrt{3}}$ at which we expect the eigenvalue $\lambda$ to attain its maximum and the associated correlation time exponent that controls the scaling of the temperature at which crossover will occur from critical to non-critical to be minimum

More generally, the fact that the duality of (\ref{wsdual}) involves the combination $Y$ in a simple way, enables us to immediately find a self dual condition: 
\be
Y=rR_1+rR_2+R_1R_2=R_Q^2 \ .
\ee
When this condition is satisfied,  the system will be on the self-dual surface.  In the intermediate region, we thus expect the exponent $\lambda$ to be maximal on this surface and decrease in both directions away from it.  On this surface, it will presumably vary.
 
%This is depicted in Fig. \ref{Figure16}. 

\section{Discussion}
%\section{Experimental Implications and Broader Perspective}

\subsection{Relation to experiments \label{exp}}

We now consider the consequences of the  results obtained in this paper for the two junction system shown in Fig. \ref{Figure2}.

{\it Existence of the $SC^{\star}$ phase}. One new  prediction is the $SC^{\star}$ phase
that is superconducting for  lead to lead transport but has localized
Cooper pairs on the middle grain. 
A similar phase has been discussed previously in the context of one dimensional Josephson junction arrays
\cite{korshunov,schoen}.

\begin{figure}
\includegraphics[width=8.5cm]{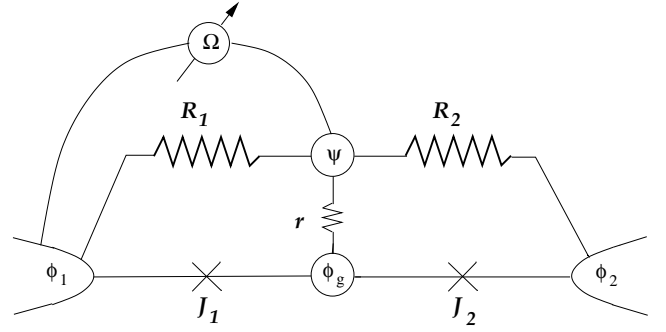}
\caption{Detection of the 
FSC to $SC^{\star}$ phase transition. 
The transition between FSC and $SC^{\star}$ will induce a jump in the effective resistance between the leads and the grain.
This can be observed by measuring the resistance between the lead 1 and the normal part of the grain. The resistance measured by $\Omega$ will increase from 
$\l(\frac{1}{R_1}+\frac{1}{R_2}+\frac{1}{r}\r)^{-1}$ in the FSC phase to 
$\frac{R_1 R_2}{R_1+R_2}$ in the $SC^{\star}$ phase. A similar discontinuity in the resistance will also occur at other phase boundaries; see the discussion in Sec. \ref{exp}.
 \label{Figure16}} 
\end{figure}

To observe the difference between the $SC^{\star}$ and the fully superconducting phase in the transport between the two leads (labeled by $\phi_1$ and $\phi_2$ in Fig. (\ref{Figure16})),  one must consider the non-linear behavior, as in both phases there is no inter-lead resistance  at
zero current. But the transition $SC^{\star}$ to FSC will be characterized by a
discontinuous jump in the {\it exponent} of the non-linear current-voltage
characteristics, reflecting a change in the nature of
the quantum phase slips in the two phases. In the $SC^{\star}$ phase
the  system behaves  essentially as one junction, and current is
carried by lead to lead Cooper pair cotunneling processes that are shunted
by the effective resistance $R_1+R_2$.  At $T=0$, for small currents we thus
expect 
\be
V \propto I^{\alpha_1}\ ,
\ee
 where 
 \be
 \alpha_1=2(R_Q/(R_1+R_2)-1)\ .
 \ee
  This form will also obtain at low temperatures and fixed current as long as 
$k_B T<hI/e$. But at low currents for positive temperature we
expect 
\be
V \propto T^{\alpha_1}
\ee
 (see \cite{schoen-zaikin}).  
 
 In the FSC
phase both junctions are superconducting and quantum phase slips can
appear in each of the junctions.  The shunting resistances for QPS in
junctions one and two are $R_1+r R_2/(r+R_2)$ and $R_2+r R_1/(r+R_1)$
respectively, so we expect at $T=0$ and small currents 
\be
V \propto
[max(T,I/e)]^{\alpha_2}
\ee
 with 
 \be
 \alpha_2=2(R_Q/(R_{max}^{eff}-1)
 \ee
  in terms of  
  \be
R_{max}^{eff}=max(R_1+r R_2/(r+R_2), R_2+r R_1/(r+R_1))\ .
\ee

Another way to distinguish the FSC and $SC^{\star}$ phases 
is to measure resistances directly between the 
leads and the grain, as shown in Fig. \ref{Figure16}.
The effective resistances between the grain and the leads should jump at the transition between the 
FSC and $SC^{\star}$ phases. In order to measure this jump, consider adding to the circuit 
an ohm-meter, $\Omega$,  measuring the resistance between the (normal) grain and   lead one.
%  We could first measure $R_1$ and $R_2$ either by making 
%  a resistance measurement above $T_c$ 
%  or by measuring the resistance using currents (imagine an ohm-meter as a current-source 
%  with a volt-meter connected in parallel to the current-source, measuring the voltage drop on it) 
%  that exceed the 
%  critical currents of the two junctions $J_1,\,J_2$.  
The transition between  FSC and $SC^{\star}$ will be characterized by the measured
resistance
increasing from $\l(\frac{1}{R_1}+\frac{1}{R_2}+\frac{1}{r}\r)^{-1}$ 
to $\frac{R_1 R_2}{R_1+R_2}$ which is a large change if $r$ is small. This occurs because in the $SC^{\star}$ phase the superconductivity on the grain 
is effectively decoupled to that current cannot flow through $r$.  

The ohm-meter could also probe other phase transitions. For instance, in the NOR and N1-S2 phases, the measured resistance would be $R_1$, while in the S1-N2 phase, it would be $\frac{R_1 r}{R_1+r}$. 

{\it Observation of $T^*$}. Another result of our analysis 
is the existence of a new temperature scale $T^*$ set by the grain level-spacing like parameter $\delta$. At high
temperatures, $ T>>T^*$, the Josephson junctions are effectively decoupled
with the dissipation set by individual shunt resistances $R_1$ and $R_2$
(see discussion below Eq. (\ref{highTaction})).
At temperatures below $T^\ast$, in contrast, we have a system of strongly coupled Josephson
junctions with the dissipation determined by the whole circuit. For example,
in the case $r>R_Q$, the effective dissipation is the total shunting
resistance $ R_1+R_2$. One possible way to observe the crossover at $T^\ast$
is to choose parameters so  that $r>R_Q$, $R_{1,2}<R_Q$ but
$ R_1+R_2 > R_Q$. For $T>T^*$ dissipation is then strong enough
to stabilize superconductivity on the individual junctions and we expect
that the measured resistance of the system will decrease with decreasing temperature.
But below $T^*$ the dissipation is no longer sufficient
to stabilize phase coherence between the leads as $(R_1+R_2)/R_Q>1$. At this point the phase slip fugacities
become relevant, and we expect an upturn in the linear
resistance as the temperature is lowered further. The basic reason for this is that
at lower temperatures, the superconductivity
is determined by longer length-scale fluctuations that involve less dissipation; the superconductivity is more vulnerable to these than the higher temperature more dissipative fluctuations. 

{\it Universal vs. non-universal behavior of the resistance at the
transition}.  An interesting feature of the zero temperature phase diagram, which contrasts with that of a single junction, is the occurrence of 
some of the normal to superconductor transitions  at non-universal values of the {\it total resistance }. Other transitions will occur at universal values of the appropriate resistance.

\begin{enumerate}
\item
In the mixed phase S1-N2, the linear  resistance of the
whole circuit is $R_2+ r R_1/(r+R_1)$ (junction one is superconducting,
and junction two is insulating, see Fig. \ref{Figure10}(b)). When this resistance
becomes equal to the quantum of resistance $R_Q$ there is a transition into the
superconducting state FSC. That this transition occurs at a universal value of the
total resistance  is not surprising:
it is due to the ordering of the  ``last''
non-superconducting junction in the otherwise superconducting circuit.
\item
In the fully normal phase,  the system has resistance $R_1+R_2$.  At
the transition point into the superconducting $SC^{\star}$ phase $R_1+R_2=R_Q$,
so we again have a universal total resistance. This transition into the $SC^{\star}$
phase is like a global or ``long wavelength'' one: it
involves superconducting fluctuations of the longest lengthscale
available: lead to lead cotunneling of Cooper pairs .
\item
At the direct transition from NOR to FSC, $R_1+R_2$ does {\it not} assume
a universal value. For example, in the limit of
small $r$ the transition takes place when both resistances are close
to $R_Q$ (see Figs. \ref{Figure5} and \ref{Figure9}), so  the total resistance will be around $2 R_Q$ at the transition. When
$r \rightarrow 0$ the two junctions are decoupled
even at zero temperatures (see Eq. (\ref{Gmatrix})). This limit is an
example of a ``local'' superconductor to normal
transition in which the   {\it resistance
 per junction } is equal to $R_Q$ at the transition point.
This is the limit that has been extensively considered in the literature.
\cite{chakravarty,fisher}
\end{enumerate}

{\it Tuning the superconductor to normal transition by changing
the Josephson couplings}.  We have shown that the superconductor to
normal transition in a two junction system may be tuned by changing the
Josephson couplings, $J_1$ or $J_2$, as well as by changing the shunting resistances
$R_{1,2}$. The former may be easier to control in experiments as
demonstrated recently in Refs. \cite{haviland}. 

{\it Non-Universality of the critical exponents.}  In
Section \ref{dualities} we showed that
the transition between the fully superconducting
and fully normal phases  is controlled by a fixed point at
intermediate values of the Josephson couplings. The critical exponents of this
transition are non-universal and vary continuously as  the three resistances in
the system change. Non-universality of the critical exponents at superconductor-normal
transitions in the presence of dissipation has also been discussed in
Refs. \cite{wagenblast,voelker,ashvin}.

{\it Symmetries of the two junction system}.  In Section
\ref{dualities} we discussed the rich symmetries of the two-junction
system. In addition to the usual weak-strong coupling duality
\cite{schoen-zaikin, fisher-zwerger} it exhibits a
permutation-triality. Exchanging the three resistances
$R_1,\,R_2,\,r$  leaves the action and the phase
diagram essentially unchanged. These symmetries provide a powerful
tool for studying the two-junction system; one need only investigate
one corner of the phase diagram to be able to construct it in its
entirety.  The boundaries of the
 region in which there is an intermediate coupling fixed point (see Fig. \ref{Figure15}), can be found from the triality and
weak-strong duality transformations.

\subsection{Broader relevance and open questions}

The results obtained in this paper should provide hints that may help understand
other superconductor to normal transitions, such as in thin wires
\cite{tinkham, bezryadin} and in films \cite{dynes}. It is often conjectured that such
transitions can be described in terms of models of resistively shunted
Josephson junctions. For example, in wires one might  perhaps think of segments
of wire of length $\xi_0$ (i.e. the superconducting coherence length or phase slip core size)
as individual grains. Then to estimate the crossover temperature analogous to our $T^*$ one 
%<<< WHY? IN PARTICULAR, WHY TAKE THIS FOR $r$?   COULD MAKE THE ARGUMENT THAT EXPECT THE SUPER TO NORMAL ELECTRON RELAXATION RATE ($\propto r/\delta$) TO BE OR ORDER $T_c$ WHICH WOULD YIELD $r\sim R$>>>  
could take
both $R$ and $r$ of order the normal state resistance of a single segment. This would yield a 
superfluid-to-normal relaxation rate that is of the order of $T_c$. The crossover temperature, $T^*$, is related to the energy
level separation parameter $\delta$ in such a segment of wire of length $\xi_0$. Using dirty limit exressions
$T_c=1.8 \hbar D/\xi_0^2$, $R=\xi_0/(e^2N_0 D A)$ and $\delta=(N_0 A
\xi_0)^{-1}$, in terms of  $D$, the diffusion coefficient, $N_0$, the density of
states per unit volume, and $A$, the wire's cross-section, we find
$T^* \approx T_c$. So at all temperatures one should consider the effects
of  interactions between the effective ``Josephson junctions" that link the ``grains"; i.e., effects analogous to those discussed in this paper.  

One
possibility is that for wires much longer than $\xi_0$, the
superconductor to normal transition will be determined not by
the resistance per coherence length, but by the {\it total} normal state
resistance. Such behavior has been observed recently in experiments of
Bezryadin {\it et.al.} \cite{bezryadin} where wires as long as fifteen times $\xi_0$ had
a normal to superconductor transition when their total normal state resistance was close to
$R_Q$ (see however [\onlinecite{tinkham, yuval1,yuval2}]). 

There is, however, another effect that must be considered in the long-wire regime. When normal metallic wires are long enough that their resistance is of order $\frac{\hbar}{e^2}=4R_Q$, localization effects start to be important at low temperatures, specifically below the temperature at which the inelastic mean-free path of the normal electrons is of order the length over which the wire has resistance of order $4R_Q$.  It is thus not clear that there is a regime in which the  dissipative effects discussed here can affect the superconductivity without localization effects also becoming important. At least naively, however, sections of length $\xi_0$ can not have resistance $R_Q$ for $T < T_c$, and the inelastic scattering length is smaller than the coherence length near $T_c$. Thus there may well be temperature regimes in which these collective effects are important but localization effects not. This clearly requires substantial further thought. 
Alternate geometries, such as configurations with a metal layer underlying the superconducting wire, may be the best candidates for avoiding some of these complications.

In the previous subsection we discussed the possibility of a surprising phenomenon in the two-junction system: a minimum  of the resistance at a crossover
temperature $T^*$ with an upturn at lower temperatures.  Qualitatively similar behavior has already been observed in
experiments on Josephson junction arrays and superconducting films. It is likely that the disorder plays an important role in such systems --- especially in
 granular films such as
$InO$.\cite{dynes} Close to  superconductor to normal
transitions in disordered materials, the  behavior may be dominated by 
weak links  that involve  connections via mesoscopic size grains. As the
temperature is lowered below the local $T^*$, 
the effective dissipation shunting these links will change in a manner analogous to that of the  pair of junctions in series through a  small grain discussed in this paper. This could potentially  account for the observed
saturation of the resistance at low temperatures in systems that would appear to be becoming superconducting  on the basis of their  behavior at higher
temperatures.  Understanding of such systems would benefit from  generalizing the analysis of the two junction system
presented here to arrays of superconducting grains and
Josephson junctions in both one and two dimensions.

An important issue that we have not addressed is the
microscopic nature of the charge relaxation between normal and
superconducting fluids that we have introduced phenomenologically.  We have assumed that at low frequencies this is 
ohmic even in the limit of zero temperature, but even if this is indeed the case, $r$ should certainly depend on details
of the experimental system. If, in fact, the relaxation is subohmic or superohmic  in the low
temperature limit, this will be roughly equivalent  to the $r \rightarrow \infty$ of $r \rightarrow 0$
cases discussed here.  However taking into account charge quantization effects
on the super-to-normal fluid relaxation and the role of quasiparticles and their non-conservation may lead to
qualitatively new effects.   One question that must be considered is whether there will be enough low energy excitations on scales below $T^\ast$ to give rise to the dissipative  effects that are crucial for the logarithmic dependence of the effective action of quantum phase slips on temperature.  We leave these issues for future research.

\section{Summary}

In this paper we have analyzed Cooper pair tunneling between two macroscopic leads via a
mesoscopic  superconducting grain in the presence of ohmic
dissipation. We treated this system in terms of a two fluid description to the grain
 by effectively
splitting it into normal and superconducting parts with
capacitative and galvanic couplings between the Cooper pairs and normal
electrons. A phenomenological ohmic resistance, $r$, was introduced to
describe the charge relaxation between the superconducting and normal
parts of the grain.  The corresponding microscopic Hamiltonian was
used to derive the quantum action in terms of which the analysis was carried out.   We showed that there is a new temperature scale $T^*$
that separates two very different regimes.  For macroscopic grains,
$T^*=0$, so that  the system is always in the high temperature
regime in which the two junctions are decoupled. In contrast, for small grains at temperatures below
$T^*$ there is strong coupling between the junctions and the system
can be described by a two component sine-Gordon model. We analyzed
this model in the limit of weak Josephson coupling and
showed that it leads to a rich quantum phase diagram with two
superconducting and three non-superconducting phases. The most surprising result is
the appearance of a novel superconducting phase, $SC^{\star}$ that has localized Cooper pairs on
the grain but phase coherence between the leads due to Cooper pair
cotunneling processes.  

The limit of strong
Josephson coupling was studied using a dual two component sine-Gordon model. Simple circuit theory for the two-junction system enabled us  
 to derive the phase 
diagram for both the weak and strong Josephson coupling limits.
In contrast to the single-junction case,  we demonstrated that 
the strong and weak coupling analysis predict different
locations of the transition between the fully superconducting and fully
normal phases  implying the
existence of an intermediate coupling fixed point controlling this transition. We analyzed the renormalization group flows in this intermediate regime and found non-universal critical behavior with the exponents depending continuously on the resistances involved, The rich symmetries of the two component sine-Gordon
model  include weak to strong coupling duality and permutation
triality of the shunting resistors $R_{1,2}$ and relaxation resistance
$r$.  

Experimental
implications of our model, including  the
crossover temperature $T^*$, the identification of the novel
superconducting phase $SC^{\star}$, and the lack of universality of the
measured resistance at the superconductor to normal transition were discussed briefly.  Finally, we noted that our results may be useful for understanding some
of the puzzling properties of superconductor to normal transitions
in thin wires and films.

Acknowledgments:

We would like to thank A. Amir, A. Bezryadin, S. Chakravarty, M. Dykman, E. Fradkin, L. Glazman, B. Halperin, W. Hofstetter, 
Y. Imry, R. Kapon, S. Kivelson, 
  N. Markovic, D. Podolsky, L. Pryadko, M. Tinkham, and G. Zarand for helpful discussions.  This
reseach was supported by the National Science Foundation via grants DMR-0132874 (E.D.), DMR-9976621
(G.R., Y.O. and D.S.F.), by Harvard's Materials Research Science and Engineering Center, by the Sloan Foundation (E.D.), and by the Israeli Science Foundation via grant 160/01-1 (Y.O.).

\appendix
%In the Appendices, we give various details of the analysis.  

\section{Microscopic model}

\subsection{Microscopic model for a two-fluid network \label{AppendixA1}}

In this Appendix we provide the derivation of several
important results used in Sec. \ref{ita}. For generality,
the first part of our analysis is not restricted
to the system shown in Fig. \ref{Figure2}, but applies to any
two-fluid network. The network  consists of superconducting
islands (which may be electrodes or grains). Each
island $i$ in this network is assumed to have part of its
charge in the form of superconducting Cooper pairs,
$Q_{Si}$, and part of the charge, $Q_{Ni}$, in the form of normal
fluid. The Hamiltonian of the system consists of three pieces:
\be
{\cal H}(Q_{Ni}, Q_{Si}, \phi_{i}, \psi_{i}) 
= {\cal H}_Q + {\cal H}_J + {\cal H}_{dis}.
\label{NetworkHamiltonian}
\ee
The charging part ${\cal H}_Q$ is given by Eq.
(\ref{ChargingHamiltonian}), with $\kappa_{ij}$ defined
as in equation (\ref{NetworkKappa}).
The Josephson energy of the Cooper pair tunneling between the grains
is
\be
\ba{cc}
{\cal H}_J= - \frac{1}{2} \sum_{ij}
J_{ij} \cos(\phi_i-\phi_j).
\ea
\ee
Dissipation between the islands, as well as charge relaxation
between the Cooper pairs and normal fluid inside the islands,
is described using the Caldeira-Leggett heat bath model
(see discussion in Secs. \ref{TwoJSHamiltonian}, 
\ref{ActionSection}) with resistances $R_{ij}$ and $r_i$
respectively.
\be 
\ba{cc} 
{\cal H}_{dis} = \frac{1}{2}
\sum_{ij} {\cal H}_{bath}(R_{ij}, 2 \psi_i - 2 \psi_j) \vspace{2mm}\\
 + \sum_i {\cal H}_{bath}(r_i, \phi_i - 2 \psi_i) 
\ea 
\ee
The commutation relations between charges and phases are
given by equation (\ref{QPhiCommutation_relations}).
Note that the Heisenberg equations of motion on $\phi_i$
and $\psi_i$ correctly reproduce Josephson relations
as in Eqs. (\ref{NetworkJosphson_relations}).

We use the Hamiltonian (\ref{NetworkHamiltonian}) and the commutation relations (\ref{QPhiCommutation_relations})
to construct the imaginary time quantum action
\be
\ba{c}
Z= \int {\cal D} Q_{Ni} {\cal D} Q_{Si} {\cal D} \phi_i 
{\cal D} \psi_i \vspace{2mm}\\
\exp \left( 
2ie\sum_i\int_0^\beta d\tau\, Q_{Si}\, \dot{\phi}_i 
+ ie \sum_i \int_0^\beta d\tau\, Q_{Ni}\, \dot{\psi}_i \r.\vspace{2mm}\\\l.
- \int_0^\beta d\tau {\cal H}(Q_{Ni}, Q_{Si}, \phi_{i}, \psi_{i})  \right)
\ea\label{NetworkZwithQ}
\ee
We remind the reader that in the presence of ohmic dissipation
the phase variables, $\phi_i$ and $\psi_i$, should be
periodic at $\tau=0$ and $\tau=\beta$ (no phase twists by multiples of
$2\pi$ are allowed).

After integrating out $Q_{Ni}$ and $Q_{Si}$ in (\ref{NetworkZwithQ}) we find
\begin{widetext}
\be
\ba{rcl}
Z &=& \int 
{\cal D} \phi_i {\cal D} \psi_i 
\exp \left( -  S_Q - S_J - S_{dis} \right)
\vspace{2mm}\\
S_Q &=& \int_0^\beta d\tau
\left( \frac{1}{2(2e)^2} \sum_{ij} \dot{\phi}_i M_{Sij}
\dot{\phi}_j
+ \frac{1}{2 e^2} \sum_{ij} \dot{\psi}_i M_{Nij}
\dot{\psi}_j
+ \frac{1}{ (2e^2)} \sum_{ij} \dot{\phi}_i M_{SNij}
\dot{\psi}_j \right)
\vspace{2mm}\\
S_J &=& - \frac{1}{2} \sum_{ij} \int_0^\beta d\tau
J_{ij} \cos(\phi_i-\phi_j)
\vspace{2mm}\\
S_{dis} &=& \beta  \sum_{\omega_n}  \left( \frac{1}{2} \sum_{ij} 
\frac{R_Q|\omega_n|}{2\pi R_{ij}} \; 
| 2\psi_{i,(\omega_n)}-2 \psi_{j,\,(\omega_n)}|^2
+ \sum_i 
\frac{R_Q |\omega_n|}{2\pi r_i}\;
| 2\psi_{i,(\omega_n)}-\phi_{i,\,(\omega_n)} |^2
 \right),
\label{NetworkAction}
\ea
\ee
\end{widetext}
where the matrices $M$ satisfy the equation
\begin{eqnarray}
\left( \begin{array}{cc} \hat{\kappa}^{-1}_S & \hat{C}^{-1} 
\\ \hat{C}^{-1} & \hat{\kappa}^{-1}_N \end{array} \right)
\left( \begin{array}{cc} \hat{M}_S & \hat{M}_{SN} \\ 
\hat{M}^T_{SN} & \hat{M}_N \end{array}
\right) =  \left( \begin{array}{cc} \hat{1} & 0 \\ 0 & \hat{1} \end{array}
\right),
\label{NetwokMmatrices}
\end{eqnarray}
where we defined (Eq. \ref{NetworkKappa})
\be
\ba{rcl}
\kappa^{-1}_{Sij} &=& C^{-1}_{ij} +D_{Si} \delta_{ij},\vspace{2mm}\\
\kappa^{-1}_{Nij} &=& C^{-1}_{ij} +D_{Ni} \delta_{ij}.
\ea
\ee
$D_{Si},\, D_{Ni}$ are the level spacings of the island i, and $C_{ij}$ is the capacitance of the island network.

In mesoscopic grains, level spacings are already much smaller than the
electrostatic capacitances, and this condition is even better
satisfied in macroscopic electrodes. Hence, we can expand
(\ref{NetwokMmatrices}) in $D_{S,N}$.
It is useful to point out that this approximation
does not require that every $D_{S,Ni}$ is smaller
than any island of the $C_{ij}^{-1}$ matrix,
but only that $D_{S,Ni}$ is smaller
than $C_{ii}^{-1}$. Hence, this expansion can be applied even when
we have a combination of macroscopic electrodes
and mesoscopic grains. We obtain
\begin{eqnarray}
M_{Sij} &=& \frac{\delta_{ij}}{D_{Si}+D_{Ni}}
+ s_i\; C_{ij}\; s_j 
\nonumber\\
M_{Nij} &=& \frac{\delta_{ij}}{D_{Si}+D_{Ni}}
+ \eta_i\; C_{ij}\; \eta_j 
\nonumber\\
M_{NSij} &=& - \frac{\delta_{ij}}{D_{Si}+D_{Ni}}
+ \eta_i\; C_{ij}\; s_j,
\end{eqnarray}
where
\begin{eqnarray}
s_i &=& \frac{D_{Ni}}{D_{Si}+D_{Ni}}
\nonumber\\
\eta_i &=& \frac{D_{Si}}{D_{Si}+D_{Ni}}.
\end{eqnarray}
Therefore, we can use the following simple expression
\be
\ba{c}
S_Q = \int_0^\beta d\tau
\left( \frac{1}{2(2e)^2} \sum_{i} 
\frac{( \dot{\phi}_i - 2 \dot{\psi}_i )^2}{(D_{Si}+D_{Ni})}\r.\vspace{2mm}\\\l.
+ \frac{1}{ 2(2e^2)} \sum_{ij} (s_i \dot{\phi}_i 
+\eta_i 2 \dot{\psi}_i)\; C_{ij}\; (s_j \dot{\phi}_j 
+\eta_j 2 \dot{\psi}_j)  \right)
\label{NetworkSq}
\ea
\ee
The first term in (\ref{NetworkSq}) tends to equilibrate
the normal and the superconducting fluids, by
introducing an energetic penalty for having different
chemical potentials. For macroscopic
grains level spacings are zero, so this term
requires $\dot{\phi} = 2 \dot{\psi}$, which is the case considered
in the literature previously. The second term in (\ref{NetworkSq})
describes the usual Coulomb interaction between the islands, but the potential on each island
is now give by the weighted average of the potentials of the two fluids:
\be
\overline{V}_i=(s_i V_{Si} 
+\eta_i V_{Ni}).
\ee

\subsection{Equations of motion \label{AppendixA2}}

As a consistency check on the quantum action (\ref{NetworkAction}), it
is useful to show that its equations of motion reproduce the familiar
equations of electrodynamics. After taking functional
derivatives of (\ref{NetworkAction})
with respect to $\phi_i$ and $\psi_i$ and analytically continuing 
into real time, we have
\begin{widetext}
\be
\ba{c}
\frac{1}{(2e)^2} \sum_j M_{Sij} \ddot{\phi}_j
+ \frac{1}{2e^2} \sum_j M_{SNij} \ddot{\psi}_j
- \sum_j J_{ij} sin (\phi_i - \phi_j)
+ \frac{r_i}{(2e)^2} ( \dot{\phi}_i - 2 \dot{\psi}_i )=0\vspace{2mm}\\
\frac{1}{2e^2} \sum_j M^T_{SNij} \ddot{\phi}_j
+ \frac{1}{e^2} \sum_j M_{Nij} \ddot{\psi}_j
- \sum_j \frac{1}{e^2 R_{ij}} ( \dot{\psi}_i -  \dot{\psi}_i )
+ \frac{1}{(2e)^2 r_i} ( \dot{\phi}_i - 2 \dot{\psi}_i )=0.
\label{NetworkEOM1}
\ea
\ee
\end{widetext}
From (\ref{NetworkKappa}), (\ref{NetworkJosphson_relations}),
and (\ref{NetwokMmatrices}) we have
\begin{eqnarray}
Q_{Si} &=& \frac{1}{2e} \sum_j M_{Sij} \dot{\phi}_j
+ \frac{1}{e} \sum_j M_{SNij} \dot{\psi}_j
\nonumber\\
Q_{Ni} &=& \frac{1}{e} \sum_j M^T_{SNij} \dot{\phi}_j
+ \frac{1}{e} \sum_j M_{Nij} \dot{\psi}_j.
\end{eqnarray}
Eqs. (\ref{NetworkEOM1}) may be written then as
\begin{eqnarray}
\frac{d Q_{Si}}{dt} &-&
\frac{1}{2e} \sum_j J_{ij} sin (\phi_i - \phi_j)
+\frac{V_{Si}-V_{Ni}}{r_i} =0
\label{NetworkSEOM}
\nonumber\\
\frac{d Q_{Ni}}{dt} &+& \sum_j \frac{V_{Ni}-V_{Nj}}{R_{ij}}
-\frac{V_{Si}-V_{Ni}}{r_i} =0.
\label{NetworkNEOM}
\end{eqnarray}
These are the usual charge conservation equations; the Josephson
form of the Cooper pair tunneling current, and Ohm's laws for the
normal currents and the ``conversion currents'' between the Cooper
pairs and the normal fluid.

\subsection{Two leads Josephson coupled via a 
mesoscopic superconducting grain \label{AppendixA3}}

We now apply our general discussion from Sec. \ref{AppendixA2} to
the system shown in Fig. \ref{Figure1}, a single mesoscopic grain between two
superconducting electrodes. We assume that the electrodes are
sufficiently large, so the superconducting and normal fluids are
perfectly coupled in them, $\phi_1=2\psi_1$ and $\phi_2=2\psi_2$.
From equation (\ref{NetworkSq}) the charging part of our system can be
written as
\begin{eqnarray}
S_Q = \frac{1}{2(2e)^2} \int_0^\beta d\tau \sum_{ij}
\; \dot{\chi}_i\; C^0_{ij}\; \dot{\chi}_j, 
\end{eqnarray}
where $\chi^T=(\phi_1,\phi_2,\phi_g,2\psi_g)$, and
\begin{widetext}
\be
\hat{C}^0=\left( \begin{array}{llll}
C_{11} & C_{12} & C_{1g} s_g & C_{1g} \eta_g \\
C_{12} & C_{22} & C_{2g} s_g & C_{2g} \eta_g \\
C_{1g}s_g & C_{2g} s_g & C_{gg} s^2_g + C_Q & 
C_{gg} s_g \eta_g - C_Q \\
C_{1g} \eta_g & C_{2g} \eta_g & C_{gg} s_g \eta_g - C_Q & 
C_{gg} \eta^2_g + C_Q
\end{array} \right),
\ee
where $C_Q^{-1} = D_{Sg} + D_{Ng}$,
$s_g=D_{Ng}/(D_{Sg} + D_{Ng})$,
and $\eta_g=D_{Sg}/(D_{Sg} + D_{Ng})$.
It is convenient to change variables to the phase differences
and the center of mass phase $\Phi$, defined as 
\begin{eqnarray}
\Delta_1 &=& \phi_g  -\phi_1 
\nonumber\\
\Delta_2 &=& \phi_2 - \phi_g 
\nonumber\\
\Delta_g &=& \phi_g - 2 \psi_g
\nonumber\\
\Phi &=& \frac{C_{11}+C_{12}+C_{1g}}{C_{tot}} \phi_1
+  \frac{C_{22}+C_{12}+C_{2g}}{C_{tot}} \phi_2
+  \frac{C_{1g}+C_{2g}+C_{gg}}{C_{tot}} \; s_g \; \phi_g
+  \frac{C_{1g}+C_{2g}+C_{gg}}{C_{tot}} \: \eta_g\; 2 \psi_g,
\end{eqnarray}
\end{widetext}
with $C_{tot}=C_{11}+2C_{12}+C_{22}+2C_{1g}+2C_{2g}+C_{gg}$.
We have 
\begin{eqnarray}
S_Q = \frac{1}{2(2e)^2} \int_0^\beta d\tau \left(
\; \sum_{\alpha\beta} \dot{\Delta}_\alpha \; \tilde{C}_{\alpha\beta}\; 
\dot{\Delta}_\beta  + C_{tot} \dot{\Phi}^2 \right),
\label{Sqrotated}
\end{eqnarray}
where the indices $\alpha$ and $\beta$ are summed over $1$,$2$, and $g$.
It is useful to observe that the center of mass phase, $\Phi$,
is decoupled from the phase differences in (\ref{Sqrotated})
and can be integrated out in the partition function. 

We do not discuss the most general case of the capacitance
matrix $C_{ij}$, but concentrate on the situation when
the dominant capacitances are the mutual capacitances
between the electrode one and the grain, $C_1$, and 
the electrode two and the grain, $C_2$. This corresponds to taking
$C_{11}=C_1+\Delta C_1$, $C_{12}=0$, $C_{1g}=-C_1$,
$C_{22}=C_2 +\Delta C_2$, $C_{2g}=-C_2$,
and $C_{gg}=C_1+C_2+ \Delta C_g$. After some straightforward
manipulations, we get
\be
\ba{c}
S_Q = \frac{1}{2(2e)^2} \int_0^\beta d\tau \left(
C_1 ( - \dot{\Delta}_1 + \eta_g \dot{\Delta}_g )^2\r.\vspace{2mm}\\\l.
+ C_2 ( \dot{\Delta}_2 + \eta_g \dot{\Delta}_g )^2
+ C_Q \dot{\Delta}^2_g +C_{tot}\dot{\Phi}^2 \right).
\ea
\ee

\subsection{Circuit-theory approach to the two-fluid model}

\begin{figure}
\includegraphics[width=7cm]{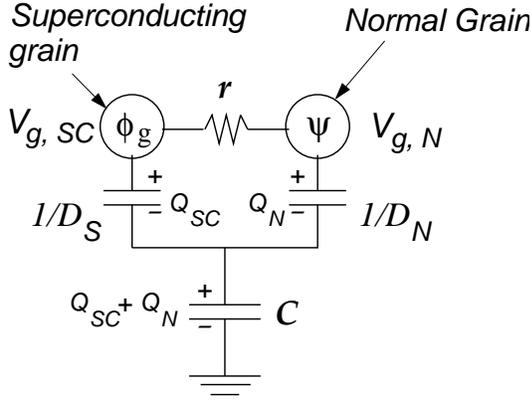}
\caption{The two fluid model description of a free-standing, mesoscopic, superconducting grain. The grain is split into two grains, a superconducting-fluid grain, which contains Cooper pairs, and a normal-fluid grain, which contains the normal electrons. Normal electrons can become superconducting by flowing through $r$. The potential on the grains is given by a sum of the electrical potential, $(Q_N+Q_{SC})/C$, and a chemical contribution, $D_N Q_N$, and $D_{S} Q_{SC}$. The finite level spacings are modeled as capacitors with capacitances $1/D_{N},\,1/D_{S}$. \label{Figx}}
\end{figure}

We can gain more intuition about the analysis presented in Section IIA
by considering effective circuits for the island network.
As a first example, let us take a free standing grain.
The electrochemical potentials for the normal and superconducting electrons on the grain can be written in the form
\be
\ba{c}
V_{g,\,N}=\frac{Q_N+Q_{SC}}{C}+D_N Q_N\vspace{2mm}\\
V_{g,\,SC}=\frac{Q_N+Q_{SC}}{C}+D_{S} Q_{SC}.
\label{potentials1}
\ea \ee
Here $C$ is the capacitance of the grain relative to the ground, and
the $D_{i}$'s are the inverses of the corresponding compressibilities.
Eq. (\ref{potentials1}) describes the electrical system in Fig. \ref{Figx}.
In addition to $C$, there are two more ``effective'' capacitors,
$1/D_{SC},\,1/D_{N}$, which describe the extra potential
drop produced by the level spacings in each part of the grain. As can be seen in
Fig. \ref{Figx}, the charge on the capacitor $C$ has to be equal to the total
charge on the grain, $Q_N+Q_{SC}$.

The  electro-chemical potentials in (\ref{potentials1}) yield
the charging part of the Hamiltonian:
\be
\ba{c}
{\cal H}_Q=\frac{1}{2C}\left(Q_N+Q_{SC}\right)^2\vspace{2mm}\\
+\frac{1}{2}D_N Q_N^2+\frac{1}{2}D_{S} Q_{SC}^2.
\label{energy11}
\ea
\ee
From here on we could proceed along the lines of Appendix \ref{AppendixA1} 
to obtain the action for this circuit. 

The general principal behind Eqs. (\ref{potentials1}) is that the potential 
on each island consists of a sum of the  
electrical contribution, $V_E$, due to Coulomb interactions, and the level 
spacing contribution:
\be
\ba{c}
V_{g,\,N}=V_E+D_N Q_N\vspace{2mm}\\
V_{g,\,SC}=V_E+D_{SC} Q_{SC}.
\label{potentials2}
\ea 
\ee
If we construct a circuit for an island network, Eqs. (\ref{potentials2}) 
indicate that we need to put the extra effective capacitors, $1/D_{Ni}$, $1/D_{Si}$, between the point at which a macroscopic 
island would be, and the normal and superconducting grains, respectively. Let us demonstrate this by constructing 
the effective circuit of the two-junction system. 

\begin{figure}
\includegraphics[width=8.5cm]{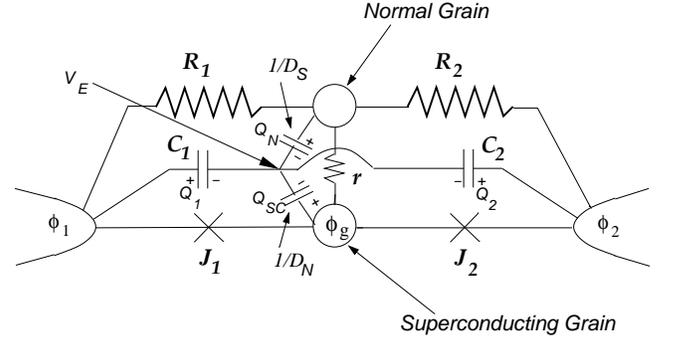}
\caption{The effective circuit of the two Josephson junction system. The mesoscopic grain is connected to the leads through Josephson junctions and resistors. It also interacts capacitatively with the leads. This interaction is modeled by the capacitors $C_1,\,C_2$ which connect to the two parts of the grain through additional capacitors, $1/D_N,\,1/D_{S}$. The additional capacitors account for the finite level spacings in the grain. The ``bare'' electrical potential on the grain (the electro-chemical potential without the level-spacing contribution) is given by $V_0$, as noted in this figure. \label{Figxx}}
\end{figure}

The two junction system consists of a mesoscopic superconducting grain situated 
between two macroscopic superconducting leads (Fig. \ref{Figxx}). 
The capacitors $C_1$ and $C_2$ describe the ``bare'' interaction between the
leads and the grain. In addition to them, there are also the effective capacitors $1/D_{N}$ and $1/D_{S}$,
which describe, respectively, the level spacing of the normal part and the superconducting part of the mesoscopic grain (Fig. \ref{Figxx}). These capacitors connect the point $V_0$, at which a macroscopic grain would have been, to the normal and superconducting parts of the mesoscopic grain. 

The electrostatic part of the Hamiltonian of the two-junction system as shown in Fig. \ref{Figxx} is given by
\be
\ba{c}
{\cal H}_Q=\frac{1}{2C_1}Q_1^2+\frac{1}{2C_2}Q_2^2\vspace{2mm}\\
+\frac{1}{2}D_N Q_N^2+\frac{1}{2}D_{S} Q_{SC}^2,
\label{energy1} 
\ea
\ee
with the constraint
\be
Q_1+Q_2+Q_N+Q_{SC}=0.
\label{cc}
\ee
This constraint merely reflects the fact that the capacitors $1/D_{SC}$ and $1/D_N$ are 
not real capacitors, but an electrical analogy to the effects of the level spacings in the mesoscopic grain. The 
charge on the grain is $-Q_1-Q_2$ (where the minus sign is due to the convention in Fig. \ref{Figxx}), 
and it is split into a superconducting part, $Q_{SC}$, and a normal part, $Q_N$. In turn, $Q_{SC}$ and $Q_N$ 
increase the electro-chemical potential on the grain, which is taken into account using the fictitious 
capacitors, $1/D_{SC},\,1/D_N$. 

We can use the constraint (\ref{cc}) to eliminate the charge of 
the normal grain:
\be
\ba{c}
{\cal H}_Q=\frac{1}{2C_1}Q_1^2+\frac{1}{2C_2}Q_2^2\vspace{2mm}\\
+\frac{1}{2}\delta_N \l(Q_1+Q_2+Q_{SC}\r)^2+\frac{1}{2}\delta_{SC} Q_{SC}^2.
\label{energy3} 
\ea
\ee
One can now  proceed by defining the phases $\phi_1,\,\phi_2$ and $\phi_g$, which 
obey the commutation relations
\be
\ba{c}
\l[Q_{1},\phi_1\r]=-2ie\vspace{2mm}\\
\l[Q_{2},\phi_2\r]=-2ie\vspace{2mm}\\
\l[Q_{SC},\phi_g\r]=-2ie.
\ea
\ee
and following steps presented in Appendix \ref{AppendixA1}.

%--------------------------------------------Yuval's dissipation

\section{Low temperature dissipation \label{altdiss}}

\begin{figure*}
\includegraphics[width=15cm]{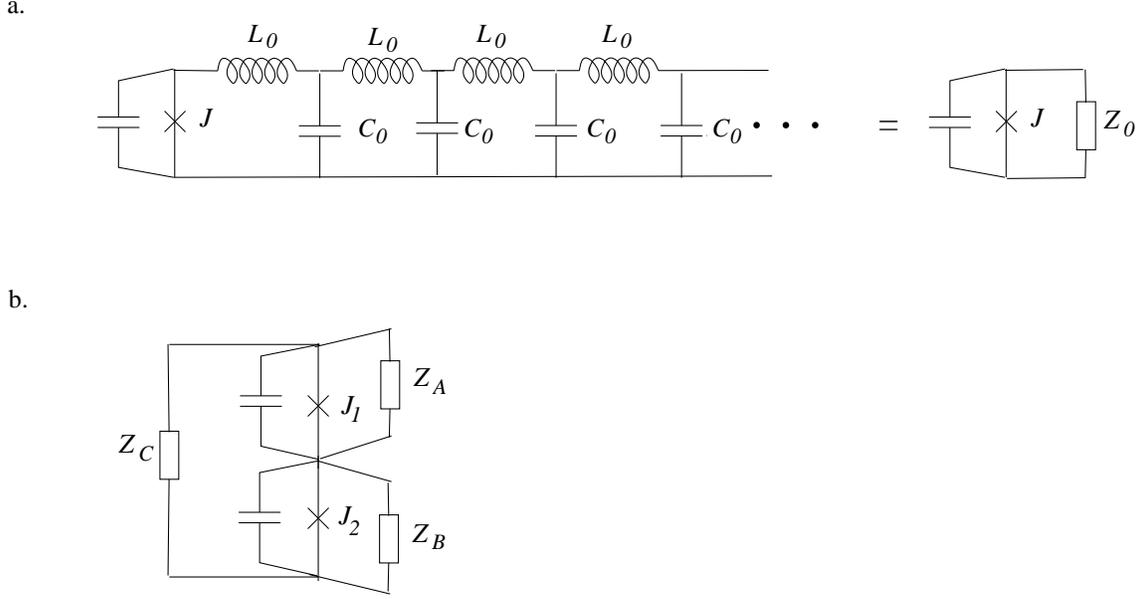}
\caption{An infinite transmission line as a source of dissipation. (a) A line of coils and capacitors, $L_0,\, C_0$, has an effective real impedance $Z=\sqrt{L_0/C_0}$ at low frequencies. This line could describe electromagnetic modes that are excited by tunneling of Cooper-pairs across junction $J$. (b) The two-junction system may have more than one transmission line. These lines are reduced in the figure to the effective impedances $Z_A,\,Z_B,\,Z_C$. The schematic circuit shown has three effective shunt resistors, as in the model in Eq. (\ref{lowTaction}). This configuration of effective impedances is the same as the ``$\Delta$'' resistors network shown in Fig. \ref{fg:DY}. This will translate (using the Y-$\Delta$ transformation, Fig. \ref{fg:DY}) to the model in Eq. (\ref{lowTaction}), with $R_1=\frac{Z_AZ_C}{Z_A+Z_B+Z_C},\, R_2=\frac{Z_BZ_C}{Z_A+Z_B+Z_C},\, r=\frac{Z_AZ_B}{Z_A+Z_B+Z_C}$. 
\label{Fig_LCline}}
\end{figure*}

In the discussion in Sec. II we introduced
the normal fluid of gapless quasiparticles
as the origin of the dissipation for
the junctions. This is not, however,
a unique way of getting
dissipation, including its ohmic variety. From the various
 possibilities, let us mention
exciting electromagnetic waves in the
environment by fluctuations of
the voltage and charge on the junctions.
 A well studied
example is a junction connected
to an LC line (see Fig. \ref{Fig_LCline}(a)).
Sudden changes of the voltage in the junction excites
plasmons, which carry the energy
off to infinity (away from the junction) leading
to dissipation. It can be described using effective
impedance formalism:  \cite{Ingold92}
\begin{eqnarray}
S_{dis} = \beta \sum_{\omega_n}  
{\rm Re}\left[ \frac{R_Q}{Z[\omega]} \right] |\omega_n|\;
| \Delta \phi_{\omega_n} |^2,
\end{eqnarray}
where $\Delta \phi$ is the phase difference across
the junction and $Z(\omega)$ is the impedance
of the environment seen by the junction.
In the case of an infinite LC line $Z=(L_0/C_0)^{1/2}$,
where $L_0$ and $C_0$ are inductance and capacitance per unit
length respectively, so we arrive at the Caldeira-Leggett
type ohmic dissipation given in equation (\ref{Sdis1}).
For the system considered in this paper (see Fig. \ref{Figure1})
such LC line (or its analogues) may come from
the edges of the electrodes or the connecting wires.
The crucial observation is that different Cooper pair tunneling
processes (between the two electrodes and the grain, and
the cotunneling process) should in general excite
different electromagnetic waves. This can be seen from the schematic 
circuit shown in Fig. \ref{Fig_LCline}(b). The effective transmission lines in the figure 
give rise to three different
resistors, which are related to $R_1$, $R_2$, and $r$ from the model in Eq. \ref{lowTaction}, as discussed in the caption of Fig. \ref{Fig_LCline}.
It is useful to point out that in this model we can relax the assumption of the small size of the grain, since the electromagnetic 
interactions discussed here, as well as the Josephson couplings, are present at all temperatures. In that case
wires can be connected to each of the superconductors
separately, allowing direct measurement of the rich phase diagram
discussed in the bulk of the text, and effects related to charge
discreetness are expected to be less significant.
Dissipation due to other low energy degrees of freedom
in the system \cite{Webb} is also possible.

It is worth emphasizing that the precise form of the 
quantum model for dissipation depends crucially on its
nature. A common choice of the Caldeira-Legget ohmic heat bath
model comes from the fact that it is the simplest quantum model
consistent with the classical equations of motion.
One expects that many effects of the dissipation would be at least qualitatively 
independent of its nature, \cite{Leggett84}
although considerable differences may also be present.

%------------------------------------------------------ RG

\section{Frequency Shell RG \label{arg}}

In order to find the phase diagram of a resistively shunted Josephson 
junction in the weak or strong coupling regime, it is best to 
employ a frequency shell RG. Generally we start with a 
sine-Gordon partition function such as:
\be
Z=\int D[\theta]\exp\left(-\int\frac{d\omega}{2\pi}\theta^2\frac{R|\omega|}{2\pi R_Q}+\int d\tau\frac{\zeta}{a}\cos(\theta)\right)
\label{simpleS-Gappend}
\ee
where we have taken the $T\rightarrow 0$ limit and changed the $\omega$ sums 
into integrals. It is useful to redefine
the amplitude of the anharmonic term using the short time cut-off
$a \sim \frac{1}{\omega_p}$, where $\omega_p$ is the plasma frequency of the junction. The sharp high frequency cut-off we use 
is
\be
\Lambda\equiv\frac{\pi}{a}
\label{Lambda}  
\ee

As is well known \cite{Minnhagen}, the partition function (\ref{simpleS-Gappend}) is also the partition function of an 
interacting Coulomb gas in one dimension, with fugacity $\zeta$ and interaction energy:
\be
E_{ij\,(\tau)}=-2\sigma_i \sigma_j\frac{R_Q}{R} \log\l|\frac{\tau}{a}\r|
\ee 
where $\sigma_i$ is the charge of the i'th particle. In the weak coupling limit (\ref{ptc}) the ``particles" are pair-tunnel events, and $\zeta=J$. In the strong coupling limit (\ref{sta}) the ``particles" are quantum phase-slips. 

Presently we would like to integrate out the "fast" degrees of 
freedom associated with the field $\theta$ in (\ref{simpleS-Gappend}). 
This has the physical meaning of reducing the frequency cutoff
$\Lambda$, and
can be thought of as {\it increasing} the effective size of a  particle, 
and eliminating all particle - anti-particle pairs whose separation 
is lower than this size.  
We write the action (\ref{simpleS-Gappend}) as:
\be
\ba{c}
Z=\int D[\theta_<]\exp\left(-\int_{|\omega|<\Lambda-d\Lambda}\frac{d\omega}{2\pi}\theta^2\frac{R|\omega|}{2\pi R_Q}\right)\vspace{2mm}\\
\int D[\theta_>]\exp\left(-\int_{\Lambda-d\Lambda<|\omega|<\Lambda}\frac{d\omega}{2\pi}\theta^2\frac{R|\omega|}{2\pi R_Q}\right)\vspace{2mm}\\
\left(1+\frac{\zeta}{2a}\int d\tau\left(e^{i\theta<}e^{i\theta_>}+e^{-i\theta_<}e^{-i\theta_>}\right)+\ldots\right)\vspace{2mm}\\
=Z_<C\left(1+\int d\tau\frac{\zeta e^{-\int_{\Lambda-d\Lambda}^{\Lambda}d\omega\frac{1}{R|\omega|}}}{2a}\left(e^{i\theta_<}+e^{-i\theta_<}\right)+\ldots\right)\vspace{2mm}\\
=C\int D[\theta_<]\exp\left(-\int_{|\omega|<\Lambda-d\Lambda}\frac{d\omega}{2\pi}\theta^2\frac{R|\omega|}{2\pi R_Q}\r.\vspace{2mm}\\\l.
+\frac{\zeta e^{-d\Lambda\frac{1}{R\Lambda}}}{a}\int d\tau\cos(\theta)\right)
\label{fastmodeseg}
\ea
\ee

From this we obtain the RG flow:
\be
\frac{\zeta}{a}\rightarrow\frac{\zeta}{a}\left(1-\frac{R_Q}{R}\frac{d\Lambda}{\Lambda}\right)
\ee
We still need to restore the variables to their original scale so that Eq. 
(\ref{Lambda}) is fulfilled. Since $\Lambda\rightarrow\Lambda-d\Lambda$, $a\rightarrow a+da$. This leads to:
\be
\ba{c}
\frac{\zeta}{a}\rightarrow\frac{\zeta}{a+da}\frac{a+da}{a}\left(1-\frac{R_Q}{R}\frac{d\Lambda}{\Lambda}\right)\vspace{2mm}\\
\rightarrow\frac{\zeta}{a+da}\left(1-\frac{R_Q}{R}\frac{d\Lambda}{\Lambda}+\frac{d\Lambda}{\Lambda}\right)
\ea
\ee
which we can write as:
\be
\ba{c}
\zeta\rightarrow\zeta\left(1-\frac{R_Q}{R}\frac{d\Lambda}{\Lambda}+\frac{d\Lambda}{\Lambda}\right)\vspace{2mm}\\
\frac{d\zeta}{dl}=-\Lambda\frac{d\zeta}{d\Lambda}=\zeta\left(1-\frac{R_Q}{R}\right)
\label{zetaflow}
\ea
\ee
Where the minus sign on the  LHS of the middle equation denotes 
the fact that $\Lambda$ is decreasing, and $dl\equiv-d\log\Lambda$ with $l$ 
the differential logarithmic flow scale-parameter, $\Lambda=\Lambda_0 e^{-l}$.

From (\ref{zetaflow}) we see that when $R>R_Q$ $\zeta$ is relevant and 
the particles proliferate. When $R<1$ the opposite happens: all 
particles form dipoles that disappear when the scale increases.

%-------------------------------------------------------------------Coulomb gas weak coupling - appendix D
\section{Coulomb gas representation of the weak coupling limit}

In this Sec. \ref{22RG} of this appendix we will derive the RG equations to second order of the two-component sine-Gordon model of Eq. (\ref{lowTaction}). Before doing that, we will derive the Coulomb gas representation of this model (Sec. \ref{cb}). This representation shows that the model (\ref{lowTaction}) describes a gas of interacting pair-tunnel events. The Coulomb gas representation makes it conceptually easier to derive the second order RG equations. In Sec. \ref{scpt} we use the Coulomb gas description to demonstrate how proliferated pair-tunnel events (or their strong coupling counterparts, phase slips) screen other events. This adds to the discussion of the mixed phases in Sec. \ref{weakpd} and Sec. \ref{spd}.

\subsection{Coulomb gas representation \label{cb}}

To analyze the two junction system we use the mapping of the 
partition function (\ref{lowTaction}) to a partition function 
of a Coulomb gas.
The starting point for this investigation is the free 
energy that appears in Eq. (\ref{lowTaction}):
\be
\ba{c}
S\approx \int\frac{d\omega}{2\pi}\frac{R_Q}{2\pi}\left(\Delta_1^2\frac{|\omega|\left(1+\frac{R_2}{r}\right)}
{R_1+R_2+\frac{R_1R_2}{r}}+
 \Delta_2^2\frac{|\omega|\left(1+\frac{R_1}{r}\right)}
{R_1+R_2+\frac{R_1R_2}{r}}\right.\vspace{2mm}\\\left.+
2\Delta_1\Delta_2\frac{|\omega|}{R_1+R_2+\frac{R_1R_2}{r}}\right)\vspace{2mm}\\
+\int\frac{d\tau}{a}\l(J_1 \cos\Delta_1+J_2 \cos\Delta_2+J_+\cos(\Delta_1+\Delta_2)\r)
\ea
\label{lowTaction1}
\ee
where we have redefined the anharmonic terms by a factor of  $a\sim \omega_p^{-1}$. 

The first step is to make use of the weak coupling statement, 
$J_{1,\,2}\ll 1$ and expand the exponent in a power law in the $J$'s. 
Following that, an integration over the fields $\Delta_{1,\,2}$ 
reduces action (\ref{lowTaction1}) to a partition function of an 
interacting gas with two kinds of charges, $\mu=1,\,2$ 
corresponding to $\exp\l(i\Delta_\mu\r)$:
\be
\ba{c}
Z=\sum\limits_{n_1,\,n_2}\frac{J_1^{2n_1}}{(n_1!)^2}\frac{J_2^{2n_2}}{(n_2!)^2}\int \Pi_{i=1}^{2n_1} d\tau^{(1)}_i\Pi_{j=1}^{2n_2} d\tau^{(2)}_j\vspace{2mm}\\
 \exp\l(-\frac{1}{2}\int d\tau_1 d\tau_2 \sum\limits_{\mu_1,\,\mu_2=1}^2 \rho^{(\mu_1)}_{(\tau_1)}\rho^{(\mu_2)}_{(\tau_2)} E^{(\mu_1,\,\mu_2)}_{(\tau_1-\tau_2)}\r),
\label{2gas}
\ea
\ee
where $\rho^{(\mu)}_{(\tau)}=\sum\limits_{i=1}^{n_{\mu}}\sigma_i\delta_{(\tau-\tau_i)}$ is the density of the gas, and $\sigma_i$ is the charge of the $i$'th particle. $J_1$ and $J_2$ play the role of fugacities for the two types of gas particles.  

The interaction energies are:
\be
\ba{c}
E^{(11)}_{(\tau)}=-2 \frac{(R_1+r)}{R_Q}\log\l|\frac{\tau}{a}\r|\vspace{2mm}\\
E^{(22)}_{(\tau)}=-2 \frac{(R_2+r)}{R_Q}\log\l|\frac{\tau}{a}\r|\vspace{2mm}\\
E^{(12)}_{(\tau)}=2 \frac{r}{R_Q}\log\l|\frac{\tau}{a}\r|,
\ea
\label{intweak}
\ee
where $E^{ij}$ is the interaction energy between a type-i 
particles and type-j particles.
As we can see, the $\log$ divergent interactions impose 
the neutrality condition satisfied in (\ref{2gas}). 
Notice that gas particles of type-1 and type-2 of the 
same charge actually attract. 

The meaning of each of the particles is very simple; 
a particle represents a {\it cooper-pair tunneling} event 
through the corresponding junction (see Fig. \ref{Figure3}).\cite{schoen-zaikin} 
To see this recall that, for example, $\Delta_1=\phi_g-\phi_1$ and the $\phi$'s 
are conjugate to the number of cooper-pairs on the corresponding grain or 
leads, hence the expansion in powers of $J$'s leads to products of 
terms like $\exp(i(\phi_g-\phi_1))$, which are translation operators 
for the charge-difference between the middle grain and lead $1$.

\subsection{Two component gas RG \label{22RG}}

To find the phase diagram of the two-junction system, we need to use 
both the mapping to a Coulomb gas from \ref{cb}, and the angular-frequency 
RG as described in Appendix \ref{arg}, extended to include second order 
contributions. In order to make the discussion general, we will treat 
the following form of the action (\ref{lowTaction}):
\be
\ba{rcl}
Z&=&\int {\cal D} \Delta_1 {\cal D} \Delta_2 
\,\,e^{-S_d-\tilde{S}_J}\vspace{2mm}\\
\tilde{S}_J &=& \int_0^\beta \frac{d \tau}{a} 
(-J_1 \cos(\Delta_1)- J_2 \cos(\Delta_2)
-J_+ \cos(\Delta_1+\Delta_2))\vspace{2mm}\\
S_d &=& \int_{|\omega|<\Lambda}\frac{d\omega}{2\pi}~ |\omega|~
\vec{\Delta}^T_{(-\omega)}~
\hat{G}~
\vec{\Delta}_{(\omega)}
\label{lowTactionmod}
\ea
\ee
with the capacitative part omitted, the sums over $\omega_n$ 
approximated by integrals, and the upper frequency cutoff 
$\Lambda=\pi/a$ introduced.
The first order contributions to the RG flow equations come 
from the terms linear in the $J$s and exactly following 
Appendix \ref{arg}, but using
\be
<\Delta_i \Delta_j>_{\omega}=\frac{1}{2\pi}{\bf G}^{-1}_{ij}
\ee
we get the following first-order RG equations:
\be
\ba{c}
\frac{dJ_1}{dl}=J_1\left(1-\frac{1}{2\pi}{\bf G}^{-1}_{11}\right)\vspace{2mm}\\
\frac{dJ_2}{dl}=J_2\left(1-\frac{1}{2\pi}{\bf G}^{-1}_{22}\right)\vspace{2mm}\\
\frac{dJ_+}{dl}=J_+\left(1-\frac{1}{2\pi}\l({\bf G}^{-1}_{11}+{\bf G}^{-1}_{22}+2{\bf G}^{-1}_{12}\r)\right).
\label{flowsecondapp}
\ea
\ee

The interactions between the two gas components give rise 
to second-order contributions to the RG equations. 
In addition, the second order terms in the power-law 
expansion in $J$'s of (\ref{lowTactionmod}) produce 
corrections to the plasma frequency of the problem and 
other irrelevant operators. 

First we will demonstrate how to derive all second order 
contributions to the flow equations by deriving one such 
contribution. Let us consider an example for a second 
order term. Then we will proceed to derive the 
unimportant plasma-frequency corrections. 

Consider the term that results from the product of a $J_1$ 
and $J_2$ first order terms In a power law expansion of 
\ref{lowTactionmod}:
\be
\ba{c}
Z=\ldots+\int D[\vec{\Delta}_<]D[\vec{\Delta}_>]exp\left(-\int\frac{d\omega}{2\pi}\vec{\Delta}^\dagger{\bf G}\vec{\Delta}\right)\vspace{2mm}\\
\int d\tau_1 \int d\tau_2\frac{J_1J_2}{4a^2}(e^{i(\Delta_{1\,(\tau_1)}+\Delta_{2\,(\tau_2)})}+e^{i(\Delta_{1\,(\tau_1)}-\Delta_{2\,(\tau_2)})}\vspace{2mm}\\
+c.c.))+\ldots
\label{2ndorder}
\ea
\ee
At this point we would like to integrate out the fast modes 
of the fields $\Delta_1,\,\Delta_2$ as in (\ref{fastmodeseg}). 
But we need to be careful since if $\tau_1=\tau_2$ the 
suppression resulting from the contraction of the fast modes 
is {\it not} the product of the two factors obtained from the 
first order terms in $J_1$ and $J_2$:
\be
\ba{c}
\l<\cos\l(\Delta_1+\Delta_2\r)\r>_{\Delta_1^>,\,\Delta_2^>}=\vspace{2mm}\\
\exp\l(-\frac{1}{2}\int_{\omega>\Lambda}\frac{d\omega}{2\pi}<(\Delta_1^>+\Delta_2^>)^2>\r)\cos\l(\Delta_1+\Delta_2\r)\vspace{2mm}\\
=\exp\l(-\frac{1}{2}\int_{\omega>\Lambda}\frac{d\omega}{2\pi}\l({\bf G}^{-1}_{11}+{\bf G}^{-1}_{22}+2{\bf G}^{-1}_{12}\r)\r)\cos\l(\Delta_1+\Delta_2\r).
\ea
\ee
The renormalized second order term, as in (\ref{2ndorder}), will 
only contain the self interaction of a Cooper-pair tunneling 
event in junction 1 and 2, completely dropping the $\exp\l(2{\bf G}^{-1}_{12}\r)$. 
This difference produces a $J_+$ renormalization term. To calculate this term we 
first need to separate the contribution to the partition function that comes 
from the term (\ref{2ndorder}) to same time and different time contributions. 
Define $\tau=\frac{\tau_1+\tau_2}{2},\,x=\tau_2-\tau_1$ and write:
 
\be
\ba{c}
\int d\tau_1 \int d\tau_2=\int d\tau \int dx=\vspace{2mm}\\
\int d\tau \int_{|x|>a+da} dx+\int d\tau \int_{|x|<a} dx+\int d\tau \int_{a<|x|<a+da} dx
\label{intsum}
\ea
\ee

The first integral is unaltered in the RG step (except for the influence on 
$J_i$ of the integration of fast modes as in first order) and can be 
re-exponentiated. The second term represents two gas-particles of type 1 and 2 
occurring at the same time, with the resolution of this RG step. This event too 
should be re-exponentiated since it is also obtained as a second order event in 
the renormalized variables. However, as pointed out in the previous paragraph, 
there is a discrepancy in the RG suppression coming from the fast modes contraction. 
Hence we write:
\begin{widetext}
\be
\ba{c}
\int \frac{d\tau}{a}\int_{|x|<a+da} \frac{dx}{a}\frac{J_1J_2}{4}\l<\left(e^{i(\Delta_{1\,(\tau+x/2)}+\Delta_{2\,(\tau-x/2)})}
+e^{i(\Delta_{1\,(\tau+x/2)}-\Delta_{2\,(\tau-x/2)})}+c.c.\right)\r>_{\Delta_1^>,\,\Delta_2^>}\vspace{2mm}\\
\approx\int \frac{d\tau}{a}\frac{a+da}{a}J_1J_2\exp\l(-\frac{1}{2}\int_{\omega>\Lambda}\frac{d\omega}{2\pi}\l({\bf G^{-1}_{11}+G^{-1}_{22}}\r)\r)\vspace{2mm}\\
\l(\cos\left(\Delta_{1\,(\tau)}-\Delta_{2\,(\tau)}\right)
\l(1+\l(\exp\l(-\frac{1}{2}\int_{\omega>\Lambda}\frac{d\omega}{2\pi}(-2){\bf G}^{-1}_{12}\r)-1\r)\r)\r.\vspace{2mm}\\\l.
+\cos\left(\Delta_{1\,(\tau)}+\Delta_{2\,(\tau)}\right)
\l(1+\l(\exp\l(-\frac{1}{2}\int_{\omega>\Lambda}\frac{d\omega}{2\pi}(2){\bf G}^{-1}_{12}\r)-1\r)\r)\r)\vspace{2mm}\\
=\int \frac{d\tau}{a+da}\frac{J'_1J'_2}{2}\l(\l(1+\frac{2}{2\pi}{\bf G}^{-1}_{12}\frac{|d\Lambda|}{\Lambda}\r)\cos\left(\Delta_{1\,(\tau)}-\Delta_{2\,(\tau)}\right)
+\l(1-\frac{2}{2\pi}{\bf G}^{-1}_{12}\frac{|d\Lambda|}{\Lambda}\r)\cos\left(\Delta_{1\,(\tau)}+\Delta_{2\,(\tau)}\right)\r).
\label{extraterm}
\ea
\ee
\end{widetext}
The second term in each of the brackets multiplying the $\cos$ terms 
are the corrections that feed into $J_-\cos(\Delta_1-\Delta_2)$ 
and $J_+\cos(\Delta_1+\Delta_2)$.

Eq. (\ref{extraterm}) leads to an additional $J_1J_2$ in $\frac{dJ_{-,\,+}}{dl}$. 
The same could be done to second order terms that are products of $J_+$ and $J_{1,\,2}$. 
For instance, in the case of $J_+$ the complete flow equation to second order would be:
\be
\ba{c}
\frac{dJ_+}{dl}=J_+\left(1-\frac{1}{2\pi}\l({\bf G}^{-1}_{11}+{\bf G}^{-1}_{22}+2{\bf G}^{-1}_{12}\right)\r)\vspace{2mm}\\
+\frac{1}{2\pi}{\bf G}^{-1}_{12}J_1J_2
\label{flowsecondexp}
\ea
\ee
The same equation with the sign of ${\bf G}^{-1}_{12}$ reversed 
applies to $J_-$. Similarly, the flow equation for $J_1$ ($J_2$) 
would have a contribution proportional to $J_2J_+$ ($J_1J_+$) and $J_2J_+$ ($J_1J_+$).   
 
The third term in Eq. (\ref{intsum}) leads to terms in the action 
proportional to $\omega^2$ and hence are unimportant. To see this 
we will use two examples that exhaust all possibilities. As the 
first case, let us look at the term in the second order 
expansion (\ref{2ndorder}) proportional to $J_1^2$:
\begin{widetext}
\be
Z=\ldots+\int D[\Delta]exp\left(-\int\frac{d\omega}{2\pi}\vec{\Delta}^\dagger{\bf G}\vec{\Delta}\right)
\int d\tau_1 \int d\tau_2\frac{J_1^2}{8a^2}\left(e^{i(\Delta_{1\,(\tau_1)}+\Delta_{1\,(\tau_2)})}+e^{i(\Delta_{1\,(\tau_1)}-\Delta_{1\,(\tau_2)})}
+c.c.\right)+\ldots
\label{2ndorder2}
\ee
We ignore the first term in the brackets as it implies a very 
costly configuration of two particles very close to each other, 
and concentrate on the second:
\be
\int D[\Delta]exp\left(-\int\frac{d\omega}{2\pi}\vec{\Delta}^\dagger{\bf G}\vec{\Delta}\right)
\int d\tau \int_{a<|x|<a+da} dx\frac{J_1^2}{8a^2}\left(e^{i(\Delta_{1\,(\tau+x/2)}-\Delta_{1\,(\tau-x/2)})}+c.c.\r)
\ee
Now:
\be
\ba{c}
\int_{a<|x|<a+da}\l(e^{i(\Delta_{1\,(\tau+x/2)}-\Delta_{1\,(\tau-x/2)})}+c.c.\r)\vspace{2mm}\\
\approx 2\l(e^{i\dot{\Delta}_{1\,(\tau)}a}+e^{-i\dot{\Delta}_{1\,(\tau)}a}\r)\vspace{2mm}\\
\approx 2\l(2-\dot{\Delta}_{1\,(\tau)}^2 a^2\r)
\ea
\ee
but this can be re-exponentiated to give a correction piece 
for the action:
\be
\Delta S=\int d\tau da \frac{J_1^2}{4}\dot{\Delta}_{1\,(\tau)}^2=\int \frac{d\omega}{2\pi} da \frac{J_1^2}{4} \omega^2 \Delta_1^2 ,
\ee
which is just a $\omega^2$ contribution that renormalizes 
the plasma frequency of the model.

A more complicated case would be considering again the term 
that mixes the two components of the gas:
\be
\ba{c}
Z=\ldots+\int D[\Delta]exp\left(-\int\frac{d\omega}{2\pi}\vec{\Delta}^\dagger{\bf G}\vec{\Delta}\right)\vspace{2mm}\\
\int d\tau \int_{a<|x|<a+da} dx\frac{J_1J_2}{4a^2}\left(e^{i(\Delta_{1\,(\tau+x/2)}+\Delta_{2\,(\tau-x/2)})}
+e^{i(\Delta_{1\,(\tau+x/2)}-\Delta_{2\,(\tau-x/2)})}+c.c.\right)+\ldots
\label{2ndorder3}
\ea
\ee
Here we use the following derivation:
\be
\ba{c}
\int_{a<|x|<a+da}\l(e^{i(\Delta_{1\,(\tau+x/2)}+\Delta_{2\,(\tau-x/2)})}+c.c.\r)
\approx da\l(e^{i(\Delta_{1\,(\tau)}+\Delta_{2\,(\tau)})}\l(e^{i(\dot{\Delta}_{1,\,(\tau)}-\dot{\Delta}_{2,\,(\tau)})a}
+e^{-i(\dot{\Delta}_{1,\,(\tau)}-\dot{\Delta}_{2,\,(\tau)})a}\r)+c.c\r)\vspace{2mm}\\
\approx 2da\cos\l(\Delta_{1\,(\tau)}+\Delta_{2\,(\tau)}\r)\l(2-\frac{a^2}{4}\l(\dot{\Delta}_{1,\,(\tau)}-\dot{\Delta}_{2,\,(\tau)}\r)^2\r).
\ea
\ee
This results then in the introduction of a term:
\be
\Delta S=\int d\tau da \frac{J_1J_2}{2}\cos\l(\Delta_{1\,(\tau)}+\Delta_{2\,(\tau)}\r)\frac{\l(\dot{\Delta}_{1,\,(\tau)}-\dot{\Delta}_{2,\,(\tau)}\r)^2}{4}
\ee
once again, proportional to $\omega^2$, and hence, unimportant. Similarly:
\be
\ba{c}
\int_{a<|x|<a+da}\l(e^{i(\Delta_{1\,(\tau+x/2)}-\Delta_{2\,(\tau-x/2)})}+c.c.\r)
\approx da\l(e^{i(\Delta_{1\,(\tau)}-\Delta_{2\,(\tau)})}\l(e^{i(\dot{\Delta}_{1,\,(\tau)}+\dot{\Delta}_{2,\,(\tau)})a}+e^{-i(\dot{\Delta}_{1,\,(\tau)}+\dot{\Delta}_{2,\,(\tau)})a}\r)+c.c\r)\vspace{2mm}\\
\approx 2da\cos\l(\Delta_{1\,(\tau)}-\Delta_{2\,(\tau)}\r)\l(2-\frac{a^2}{4}\l(\dot{\Delta}_{1,\,(\tau)}+\dot{\Delta}_{2,\,(\tau)}\r)^2\r)
\ea
\ee
Giving:
\be
\Delta S=\int d\tau da \frac{J_1J_2}{2}\cos\l(\Delta_{1\,(\tau)}-\Delta_{2\,(\tau)}\r)\frac{\l(\dot{\Delta}_{1,\,(\tau)}+\dot{\Delta}_{2,\,(\tau)}\r)^2}{4}
\ee
This exhausts all second order contributions to the RG flow equations. 
\end{widetext}

%------------------------------------------------------screening 
\subsection{Screening of pair tunneling events \label{scpt}}

When discussing the phase diagram obtained from the RG flow Eqs. (\ref{Jflowsecond})
in Section III, we had to account to parts of the phase diagram 
in which one of the three Josephson couplings is relevant;
then we used the procedure of setting the respective phase-difference
variable to zero, which is equivalent to starting from a new fixed
point.  Here we review this step and show that in the language of 
the Coulomb gas
analogy, this procedure may be understood 
as screening of charges of one type by 
proliferated charges of the other type. 
 
Let us start by considering the case of $J_2$ being relevant. 
The simplest approach to the strong coupling limit is to 
assume that $\Delta_2=0$. This can be done because if $J_2\gg 1$; 
then the term $J_2 \cos\Delta_2$ in the action (\ref{lowTaction}) 
constrains $\Delta_2$ to be $0$.
The kinetic part of the action becomes: 
\be
S_1=\int\frac{d\omega}{2\pi}\frac{R_Q}{2\pi}\Delta_1^2\frac{|\omega|\left(1+\frac{R_2}{r}\right)}
{R_1+R_2+\frac{R_1R_2}{r}}
\ee
and hence the flow for $J_1$ (or $J_+$) would become:
\be
\frac{dJ_1}{dl}=J_1\l(1-\frac{R_1+R_2+\frac{R_1R_2}{r}}{R_Q\l(1+\frac{R_2}{r}\r)}\r)
\ee
shifting the phase boundary S1-N2 and the FSC phase to:
\be
\frac{R_1+R_2+\frac{R_1R_2}{r}}{1+\frac{R_2}{r}}=R_Q
\label{j2rel}
\ee

The above calculation is a very straightforward way of obtaining 
the phase diagram, however, to understand the physics behind it 
let us take a step back. When $J_2$ is relevant, pair-tunnel 
events in junction 2 will proliferate. This means that any 
field felt by the gas particles of type 2 (corresponding to the 
pair-tunnel events in junction 2) will be {\it screened} by type-2 
particles attracted to the source of the field. So every type-1 
gas particle will acquire a {\it screening cloud} of particles 
of type-2, so that no field from the original type-1 particle 
is felt in junction 2. 
To make this a quantitative statement, a type-1 gas particle 
with charge $q_1$ exerts the field
\[
q_1 E^{(12)}_{(\tau)}=2 q_1 \frac{r}{R_Q} \log\l|\frac{\tau}{a}\r|
\]
on the type-2 particles. Type-2 particles will then form a 
screening cloud of charge $q_2$ so that:
\[
q_1 E^{(12)}_{(\tau)}=2 q_1 \frac{r}{R_Q} \log\l|\frac{\tau}{a}\r|=-q_2 E^{(22)}_{(\tau)}=2 q_2 \frac{(R_2+r)}{R_Q} \log\l|\frac{\tau}{a}\r|
\]
which leads to:
\[
q_2=q_1\frac{r}{R_2+r}
\]
now, the field that a test charge of type-1 would feel is:
\be
\ba{cc}
q_1 E'^{(11)}_{(\tau)}=q_2 E^{(12)}_{(\tau)}+q_1 E^{(11)}_{(\tau)}\vspace{2mm}\\
=-2 q_1\frac{1}{R_Q}\l(R_1+r-r\frac{r}{R_2+r}\r)\log\l|\frac{\tau}{a}\r|\vspace{2mm}\\
=-2 q_1\frac{1}{R_Q}\l(\frac{R_1+R_2+\frac{R_1 R_2}{r}}{1+\frac{R_2}{r}}\r)\log\l|\frac{\tau}{a}\r| ,
\ea
\ee
which we see gives exactly the same result as (\ref{j2rel}). 
Indeed this way is more complicated, however it could also 
be employed in more complicated setups, and gives some insight 
as to what physically happens to the system. In this case, the
charge tunneling from lead-1 to the grain, partially relaxes 
through the superconducting junction-2. The physical 
interpretation of the above results is also discussed in Sec. (\ref{ctcircuits}).    

Next, let us consider the case of relevant $J_+$. Here we 
need to set $\Delta_1=-\Delta_2\equiv\Delta$. This gives 
a free energy kinetic part:
\be
S_+=\int\frac{d\omega}{2\pi}\frac{R_Q}{2\pi}\Delta^2\frac{|\omega|\left(\frac{R_1}{r}+\frac{R_2}{r}\right)}
{R_1+R_2+\frac{R_1R_2}{r}},
\ee
and hence the flow for $J_1$ (or $J_2$) would become:
\be
\frac{dJ_1}{dl}=J_1\l(1-\frac{1}{R_Q}\l(r+\frac{R_1R_2}{R_1+R_2}\r)\r)
\ee
this shifts the phase boundary between the $SC^{\star}$ phase 
and FSC phase (Fig. \ref{Figure4}(b)) to:
\be
r+\frac{R_1R_2}{R_1+R_2}=R_Q.
\label{j+rel}
\ee
Here too, we can follow the screening principal to get the answer. 
The idea would be that a pair-tunnel event would acquire a pair-tunnel 
couple screening cloud so that other pair-tunnel couples won't feel any 
field. 

This method along with the Self-Consistent-Harmonic-Approximation \cite{Minnhagen} 
can be used to obtain more insight about the behavior of the system.

%----------------------------------------Coulomb gas for strong coupling

\section{Coulomb gas of phase-slips representation of the strong coupling case}

\subsection{Villain transformation - phase-slips \label{villain}}

To treat the strong coupling limit $J_1,\,J_2 \gg 1$, we need to 
derive a description of the two-junction system in terms of 
phase-slips: events in which the phase of one of a Josephson-junction 
tunnels from one trough of the Josephson $\cos$ potential into an 
adjacent trough. This event leads to a voltage drop across the 
junction (from $\hbar \dot{\phi}/2e=V$) and hence to dissipation. 
To derive this action, we make use of the villain-transformation. 
\cite{schoen,weiss-grabert,villain} 

Starting with the action (\ref{lowTaction})
\be
\ba{c}
S\approx \int\frac{d\omega}{2\pi}\frac{R_Q}{2\pi}\left(\Delta_1^2\frac{|\omega|\left(1+\frac{R_2}{r}\right)}
{R_1+R_2+\frac{R_1R_2}{r}}\right.\vspace{2mm}\\\left.+
 \Delta_2^2\frac{|\omega|\left(1+\frac{R_1}{r}\right)}
{R_1+R_2+\frac{R_1R_2}{r}}+
2\Delta_1\Delta_2\frac{|\omega|}
{R_1+R_2+\frac{R_1R_2}{r}}\right)+S_C+S_J
\ea
\ee
with:
\be
S_J=\int d\frac{\tau}{a}\left(-J \cos\Delta_1-J\cos\Delta_2\right)
\ee
Here too we modified the sum over frequencies into an integral, 
and introduced a high-frequency cut-off. 

% In the case of strong Josephson couplings, $J\gg 1$. This is obtained from considereing the relative importance of $S_C$ (\ref{S_C}) and $S_J$ (\ref{S_J}), however $S_C$ will not play a role in the next calculation, besides determining the high frequency cutoff, $a$. 

The assumption of strong J allows us to perform a Villain transformation:
\be
\ba{c}
\exp\left(\int d\tau J\left(\cos(\Delta_i)-1\right)\right)\vspace{2mm}\\
\approx\sum\limits_{{\eta^i_{(\tau)}}}\exp\left(-\int d\tau \frac{J}{2}\left(\Delta_i+2\pi\eta^i_{(\tau)}\right)^2\right)
\ea
\ee
where $\eta^i_{(\tau)}$ maps imaginary time to the integers, and the 
sum on the RHS is over all these functions. The function $\eta^i_{(\tau)}$ 
specifies in which trough of the potential $J_i\cos\Delta_i$ the i'th junction is. 
The essence of the villain transformation is that it completely eliminates the 
dynamics of intra-trough motion and only considers the tunneling between troughs. 
The intra-trough near-minimum motion is encoded into what will become the fugacity 
of a phase slip, $\zeta_i$. 

It is actually better to use the Fourier transform of the time derivative: 
$FT(\dot{\eta}^i_{(\tau)})=-i\omega\eta^i_{(\omega)}\equiv\rho^i_{(\omega)}$. 
Incorporating this allows us then to write:
\be
\ba{c}
\exp\left(-S_J\right)\approx\sum\limits_{{\rho^{1,\,2}_{(\tau)}}}\exp\left(-\int \frac{d\omega}{2\pi} \frac{J}{2}\left(\left|\Delta_1+2\pi\frac{\rho^1_{(\omega)}}{-i\omega}\right|^2\r.\r.\vspace{2mm}\\\l.\l.
+\left|\Delta_2+2\pi\frac{\rho^2_{(\omega)}}{-i\omega}\right|^2 \right)\right).
\ea
\ee
Expanding the square and putting it all in the action (\ref{lowTaction}) gives:
\be
\ba{c}
S\approx \int\frac{d\omega}{2\pi}\left(\Delta_1^2\left(\frac{J}{2}+\frac{|\omega|\left(1+\frac{R_2}{r}\right)}
{R_1+R_2+\frac{R_1R_2}{r}}\right)\right.\vspace{2mm}\\\left.+
 \Delta_2^2\left(\frac{J}{2}+\frac{|\omega|\left(1+\frac{R_1}{r}\right)}
{R_1+R_2+\frac{R_1R_2}{r}}\right)+
2\Delta_1\Delta_2\frac{|\omega|}
{R_1+R_2+\frac{R_1R_2}{r}}\r.\vspace{2mm}\\\l.
+\Delta_1J\frac{2\pi\rho_1}{i\omega}+\frac{J\left(2\pi\rho_1\right)^2}{\omega^2}+\Delta_2J\frac{2\pi\rho_2}{i\omega}+\frac{J\left(2\pi\rho_2\right)^2}{\omega^2}\right).
\ea
\label{villainaction}
\ee
Recalling: $Z=\sum\limits_{{\rho^{1,\,2}_{(\tau)}}}\int D[\Delta_1]D[\Delta_2]\exp(-S)$, 
we are ready to integrate out $\Delta_1,\,\Delta_2$ and get the partition 
function for the phase-slips gas. After doing this and taking the limit 
of large $J$ we get the following partition function:
%\be
%\ba{c}
%Z=\sum\limits_{{\rho^{1,\,2}_{(\tau)}}}\exp\left(-\int\frac{d\omega}{2\pi}\left\{\left(2\pi\rho_1\right)^2
%\left(\frac{-J}{2\omega^2}+\frac{J^2}{4\omega^2}\frac{\left(\frac{J}{2}+\frac{|\omega|\left(1+\frac{R_1}{r}\right)}
%{R_1+R_2+\frac{R_1R_2}{r}}\right)}{\left(\frac{J}{2}+\frac{|\omega|\left(1+\frac{R_1}{r}\right)}
%{R_1+R_2+\frac{R_1R_2}{r}}\right)\left(\frac{J}{2}+\frac{|\omega|\left(1+\frac{R_2}{r}\right)}
%{R_1+R_2+\frac{R_1R_2}{r}}\right)-\left(\frac{\omega}{R_1+R_2+\frac{R_1R_2}{r}}\right)^2}\right)\right.\right.\vspace{2mm}\\
%\left(2\pi\rho_2\right)^2
%\left(\frac{-J}{2\omega^2}+\frac{J^2}{4\omega^2}\frac{\left(\frac{J}{2}+\frac{|\omega|\left(1+\frac{R_2}{r}\right)}
%{R_1+R_2+\frac{R_1R_2}{r}}\right)}{\left(\frac{J}{2}+\frac{|\omega|\left(1+\frac{R_1}{r}\right)}
%{R_1+R_2+\frac{R_1R_2}{r}}\right)\left(\frac{J}{2}+\frac{|\omega|\left(1+\frac{R_2}{r}\right)}
%{R_1+R_2+\frac{R_1R_2}{r}}\right)-\left(\frac{\omega}{R_1+R_2+\frac{R_1R_2}{r}}\right)^2}\right)\vspace{2mm}\\
%\left.\left.
%2\l(2\pi\rho_2\r)\l(2\pi\rho_2\r)\left(\frac{-J}{2\omega^2}+\frac{J^2}{4\omega^2}\frac{\left(\frac{J}{2}+\frac{|\omega|}
%{R_1+R_2+\frac{R_1R_2}{r}}\right)}{\left(\frac{J}{2}+\frac{|\omega|\left(1+\frac{R_1}{r}\right)}
%{R_1+R_2+\frac{R_1R_2}{r}}\right)\left(\frac{J}{2}+\frac{|\omega|\left(1+\frac{R_2}{r}\right)}
%{R_1+R_2+\frac{R_1R_2}{r}}\right)-\left(\frac{\omega}{R_1+R_2+\frac{R_1R_2}{r}}\right)^2}\right)\right\}\right)
%\ea
%\ee
%considering the limit of large $J$ we get the following partition function:
\be
\ba{c}
Z=\sum\limits_{{\rho^{1,\,2}_{(\tau)}}}\exp\left(\frac{R_Q}{2\pi}\int\frac{d\omega}{2\pi}\left\{\left(2\pi\rho_1\right)^2\frac{1}{|\omega|}\frac{1+\frac{R_2}{r}}{R_1+R_2+\frac{R_1R_2}{r}}\right.\right.\vspace{2mm}\\
+\left(2\pi\rho_2\right)^2\frac{1}{|\omega|}\frac{1+\frac{R_1}{r}}{R_1+R_2+\frac{R_1R_2}{r}}\vspace{2mm}\\
\left.\left.+2\left(2\pi\rho_1\right)\left(2\pi\rho_2\right)\frac{1}{|\omega|}\frac{1}{R_1+R_2+\frac{R_1R_2}{r}}\right\}\right).
\label{gasZapp}
\ea
\ee
This is a partition function for a gas that consists 
of two kinds of particles, with $\rho_1,\,\rho_2$ being 
the densities of the two gasses. When carrying out the $\omega$ 
integrals we get the interaction energy between the gas particles. 
They are ($E^{(ij)}$ is the energy of interaction between two 
positive particles, one of species i and the other from species j):
\be
\ba{c}
E^{(11)}_{(\tau)}=-2\frac{1+\frac{R_2}{r}}{R_1+R_2+\frac{R_1R_2}{r}}\log\left(\frac{|\tau|}{a}\right)=-2\frac{1}{R_1+\frac{R_2 r}{R_2+r}}\log\left(\frac{|\tau|}{a}\right)\vspace{2mm}\\
E^{(22)}_{(\tau)}=-2\frac{1+\frac{R_1}{r}}{R_1+R_2+\frac{R_1R_2}{r}}\log\left(\frac{|\tau|}{a}\right)=-2\frac{1}{R_2+\frac{R_1 r}{R_1+r}}\log\left(\frac{|\tau|}{a}\right)\vspace{2mm}\\
E^{(12)}_{(\tau)}=-2\frac{1}{R_1+R_2+\frac{R_1R_2}{r}}\log\left(\frac{|\tau|}{a}\right).
\label{gasenergyapp}
\ea
\ee

As in the weak coupling limit, here too we derived a Coulomb 
gas description of the action (\ref{lowTaction}). However in 
the strong coupling limit the gas particles are phase slips, 
which produce a voltage drop over the junction.

\subsection{From the Coulomb gas to sine-Gordon}

The interacting gas of phase slips described in (\ref{villain}) 
can be encoded into a new sine-Gordon theory, conjugate to the 
original theory (\ref{lowTaction}). It is given by:
\be
\ba{c}
Z=\int D[\theta_1]\int D[\theta_2]\vspace{2mm}\\
\label{sgaction}
 \exp\left(-\int\frac{d\omega}{2\pi}\frac{|\omega|}{2\pi R_Q}\left(\left(r+R_1\right)\theta_1^2+\left(r+R_2\right)\theta_2^2-2r\theta_1\theta_2\right)\right.\vspace{2mm}\\
\left.+\int \frac{d\tau}{a}\left(\zeta_1\cos(\theta_1)+\zeta_2\cos(\theta_2)+\zeta_-\cos\left(\theta_1-\theta_2\right)\right)\right)\vspace{2mm}\\
=\int D[\theta_1]\int D[\theta_2]\vspace{2mm}\\
\exp\left(-\int\frac{d\omega}{2\pi}\vec{\theta}^\dagger{\bf G}\vec{\theta}+\int \frac{d\tau}{a}\left(\zeta_1\cos(\theta_1)+\zeta_2\cos(\theta_2)\right)\right)
\ea
\label{psactionapp}
\ee
where $\zeta_{1,\,2}$ play the role of fugacities of the phase-slips 
on junctions 1 and 2. By expanding this Sine-Gordon theory in the $\zeta's$ 
and following the steps of Appendix \ref{cb} we recover the Coulomb 
gas described in (\ref{gasenergyapp}).

\end{document}